\documentclass[twocolumn,tighten,twocolappendix]{aastex63}
\pdfoutput=1 
\usepackage{amsmath,amstext}
\usepackage[T1]{fontenc}
\usepackage{ae,aecompl}
\usepackage[utf8]{inputenc}
\usepackage{apjfonts} 
\usepackage[figure,figure*]{hypcap}
\usepackage{enumitem}
\usepackage{bm}
\usepackage{graphicx}
\input{hyperlink-year-only-natbib-patch}

\newcommand{\vect}[1]{\boldsymbol{#1}}

\newcommand{\cm}{\ensuremath{{\rm cm}}}
\newcommand{\km}{\ensuremath{{\rm km}}}
\newcommand{\pc}{\ensuremath{{\rm pc}}}
\newcommand{\kpc}{\ensuremath{{\rm kpc}}}
\newcommand{\mpc}{\ensuremath{{\rm Mpc}}}
\newcommand{\kms}{\ensuremath{{\rm km\,s^{-1}}}}

\newcommand{\Gyr}{\ensuremath{{\rm Gyr}}}

\newcommand{\K}{\ensuremath{{\rm K}}}
\newcommand{\s}{\ensuremath{{\rm s}}}

\newcommand{\msun}{\ensuremath{{M_{\odot}}}}

\shorttitle{Strong DM Self-interactions Diversify Halo Populations}
\shortauthors{Yang, Nadler, \& Yu}

\begin{document}

\title{Strong Dark Matter Self-interactions Diversify Halo Populations within and surrounding the Milky Way}

\author[0000-0002-5421-3138]{Daneng Yang}
\affiliation{Department of Physics and Astronomy, University of California, Riverside, California 92521, USA}

\author[0000-0002-1182-3825]{Ethan O.~Nadler}
\affiliation{Carnegie Observatories, 813 Santa Barbara Street, Pasadena, CA 91101, USA}
\affiliation{Department of Physics $\&$ Astronomy, University of Southern California, Los Angeles, CA, 90007, USA}

\author[0000-0002-8421-8597]{Hai-Bo Yu}
\affiliation{Department of Physics and Astronomy, University of California, Riverside, California 92521, USA}

\email{danengy@ucr.edu}

\email{enadler@carnegiescience.edu}

\email{haiboyu@ucr.edu}

\begin{abstract}
We perform a high-resolution cosmological zoom-in simulation of a Milky Way (MW)--like system, which includes a realistic Large Magellanic Cloud analog, using a large differential elastic dark matter self-interaction cross section that reaches $\approx 100~\mathrm{cm}^2\ \mathrm{g}^{-1}$ at relative velocities of $\approx 10~\mathrm{km\ s}^{-1}$, motivated by the diverse and orbitally dependent central densities of dwarf galaxies within and surrounding the MW. 
We explore the effects of dark matter self-interactions on satellite, splashback, and isolated halos through their abundance, central densities, maximum circular velocities, orbital parameters, and correlations between these variables.  
We use an effective constant cross section model to analytically predict the stages of our simulated halos' gravothermal evolution, demonstrating that deviations from the collisionless $R_{\rm max}$--$V_{\rm max}$ relation can be used to select  deeply core-collapsed halos, where $V_{\rm max}$ is a halo's maximum circular velocity, and $R_{\rm max}$ is the radius at which it occurs.  
We predict that a sizable fraction ($\approx 20\%$) of subhalos with masses down to $\approx 10^8~\msun$ is deeply core collapsed in our SIDM model. 
Core-collapsed systems form $\approx 10\%$ of the isolated halo population down to the same mass;
these isolated, core-collapsed halos would host faint dwarf field galaxies with extremely steep central density profiles.
Finally, most halos with masses above $\approx 10^9~\msun$ are core-forming in our simulation. 
Our study thus demonstrates how self-interactions diversify halo populations in an environmentally dependent fashion within and surrounding MW-mass hosts, providing a compelling avenue to address the diverse dark matter distributions of observed dwarf galaxies.
\end{abstract}

\keywords{\href{http://astrothesaurus.org/uat/353}{Dark matter (353)};
\href{http://astrothesaurus.org/uat/574}{Galaxy abundances (574)};
\href{http://astrothesaurus.org/uat/1083}{N-body simulations (1083)};
\href{http://astrothesaurus.org/uat/1880}{Galaxy dark matter halos (1880)};
\href{http://astrothesaurus.org/uat/1965}{Computational methods (1965)}}

\section{Introduction}

Self-interacting dark matter (SIDM) has long been considered a promising alternative to the cold, collisionless dark matter (CDM) paradigm, particularly due to its effects on small-scale structure. For example, SIDM has been studied in the context of cored halo inner density profiles, subhalo tidal evolution, the diversity of galactic rotation curves, gravitational lensing, and the interplay with baryonic feedback (e.g., see \citealt{Tulin170502358,Adhikari220710638} for reviews). 
A growing number of idealized and cosmological simulations show that dark matter (DM) self-interactions can yield diverse effects on halo properties as a function of halo mass (e.g., \citealt{Vogelsberger12015892,Rocha12083025,Zavala12116426,Dooley160308919,Robles190401469,Nadler200108754,Nadler210912120,Turner201002924,Ebisu210705967,Shirasaki220509920,Silverman220310104}). Moreover, these effects depend on halos' secondary properties, including concentration (e.g., \citealt{Kaplinghat150803339,Kamada161102716,Essig180901144,Correa220611298,Zeng211000259}).

Various SIDM simulations have been used to predict the impact of self-interactions on small-scale structure. These simulations suggest that a cross section $\sigma/m\gtrsim 1\rm ~cm^2~g^{-1}$ on relative velocity scales of $v\lesssim 200~\kms$ can address potential small-scale structure tensions associated with CDM (\citealt{Rocha12083025,Kamada161102716,Ren180805695}). 
Because observations of galaxy clusters imply that $\sigma/m\lesssim 0.1 \rm ~cm^2~g^{-1}$ on larger velocity scales ($v\approx 1000~\mathrm{km\ s}^{-1}$; e.g., \citealt{Peter12083026,Harvey150307675,Kaplinghat150803339,Andrade2020}), SIDM models that significantly affect small-scale structure therefore likely feature velocity-dependent scattering (e.g., \citealt{Correa200702958,Kim210609050,Silverman220310104}).
A velocity-dependent cross section is naturally realized in particle physics models that feature, e.g., Rutherford or M\o{}ller scattering \citep{Feng09053039,Feng09110422,Buckley09113898,Loeb10116374,Tulin13023898,Agrawal11611004611,Girmohanta220614395,Yang220503392}. 
Moreover, velocity-dependent SIDM models can be tuned to agree with cosmological observables over a wide range of scales and yield novel, scale-dependent effects on structure formation (e.g., \citealt{Kaplinghat150803339,Feng201015132,Gilman210505259,Gilman220713111,Yang210202375,Correa220611298,Loudas220913393,Lovell220906834,Meshveliani221001817}). 

In this context, it is interesting to consider how velocity-dependent self-interactions affect the properties of the low-mass halos that host the faintest galaxies. Self-interactions do not significantly affect halo evolution at early times (e.g., \citealt{Rocha12083025}), when the bulk of star formation occurs in low-mass galaxies \citep{Simon190105465}; thus, dwarf galaxy formation is expected to be similar in CDM and SIDM. Meanwhile, late-time (sub)halo properties can be altered due to the gravothermal and orbital evolution of SIDM (sub)halos, which may, in turn, affect the properties of the galaxies they host. In SIDM, halos with low stellar-to-halo mass ratios and concentrations are usually expected to have flat inner density profiles due to core formation, where core sizes depend on the SIDM model and halo properties (e.g., \citealt{Spergel9909386,Dave0006218,Vogelsberger12015892,Rocha12083025}). 
However, if the scattering rate is sufficiently large, some compact halos can evolve into a core-collapse phase where both the central density and velocity dispersion rise in time in a runaway fashion (e.g., \citealt{1968MNRAS.138..495L,Balberg0110561,Koda11013097,Feng210811967}). Tidal effects can further accelerate the gravothermal evolution of some subhalos (e.g., \citealt{Nishikawa190100499}), hence increasing the probability of observing satellite galaxies hosted by subhalos with high central densities. This mechanism has been demonstrated by simulating subhalos on the orbits chosen to resemble individual galaxies, e.g., for systems with properties similar to the MW satellite galaxies with high inner densities, such as Draco~\citep{Kahlhoefer190410539,Sameie190407872,Correa200702958}.

The halo of our own Milky Way (MW) galaxy harbors low-mass subhalos that host dwarf galaxies spanning several orders of magnitude in subhalo and stellar mass, down to the smallest ultra-faint dwarf galaxies with only hundreds of stars (for a review, see \citealt{Simon190105465}). Over $60$ candidate MW satellite galaxies are now known \citep{McConnachie12041562,Drlica-Wagner191203302}, and this population continues to motivate small-scale studies. For example, the ``too-big-to-fail'' problem (TBTF; \citealt{Boylan-Kolchin11030007,Boylan-Kolchin11112048}) suggests that the largest subhalos of MW-mass hosts have significantly higher maximum circular velocity ($V_{\mathrm{max}}$) values than inferred observationally for the brightest MW satellites. This problem can be alleviated by a variety of baryonic mechanisms, including tidal stripping by the central galactic disk and internal feedback (e.g., \citealt{Brook14103825,Garrison-Kimmel180604143}), or simply if the mass of the MW halo is lighter than commonly assumed (e.g., \citealt{Wang12034097,Jiang150802715}). Velocity-independent SIDM models with interactions at the $\sigma/m\gtrsim 1~\rm cm^2~g^{-1}$ level can also alleviate the tension because self-interactions significantly reduce the central densities of the corresponding halos in these scenarios (e.g., \citealt{Zavala12116426}); however, this may come at the cost of producing too few surviving subhalos, and too few high-density systems (e.g., \citealt{Kim210609050,Turner201002924,Silverman220310104}). Meanwhile, following the proper-motion measurements of MW satellites delivered by Gaia in recent years (e.g., \citealt{Simon180410230,Pace220505699}), an anticorrelation between satellites' inferred inner DM densities and pericentric distances has been reported (e.g., \citealt{Kaplinghat190404939}). Such an anticorrelation may be produced if only the densest subhalos survive in the inner regions of the MW, and potentially by observational selection effects. Nonetheless, SIDM scenarios in which low-mass subhalos efficiently core collapse may intrinsically predict such an anticorrelation if these extremely dense systems can survive the MW's tides. 

Going beyond the MW satellite population, intriguing properties of field dwarf galaxies have also been reported, including a potential TBTF problem for isolated Local Group dwarfs (e.g., \citealt{Garrison-Kimmel14045313,Papastergis14074665}). In addition to the low-density systems in question for TBTF-like tensions, recent observations also hint that some field dwarf galaxies have surprisingly high inner densities. For example, the Tucana dwarf galaxy is inferred to have a very-high-amplitude and steep central density profile (\citealt{Gregory190207228}; however, see \citealt{Taibi200111410}), and may be tidally affected as a splashback system of Andromeda (e.g., \citealt{Santos-Santos220702229}). Such overly dense systems are thought to be rare in CDM, particularly because feedback from galaxy formation primarily cores halos at these mass scales via supernova feedback (e.g., \citealt{Dutton2011111351}). Taken together, the existence of overdense and underdense outliers among the Local Group isolated dwarf population is similar to the well-studied ``diversity problem'' of galactic rotation curves (e.g., \citealt{deNaray09123518,Oman150401437,Ren180805695,Santos-Santos191109116}). Once again, strong, velocity-dependent SIDM models provide intriguing ways to address such tensions, because they can yield sizable populations of both core-collapsed and core-forming halos that may respectively correspond to overdense and underdense field galaxies, further beyond previous SIDM studies on this subject, which focus on the core-forming regime (e.g., \citealt{Creasey161203903,Kamada161102716,Ren180805695,Santos-Santos191109116,Zentner220200012}).

All of these considerations motivate a careful study of SIDM scenarios that produce (sub)halos in both the core-expansion and core-collapse phases, thereby yielding a large spread in their central densities. For halos with masses of $10^8$-$10^{10}~\msun$ and a median concentration, we estimate the collapse timescale to be $\sim 10~\Gyr$ for a cross section amplitude of $100~\rm cm^2~g^{-1}$, based on analytic predictions for the core-collapse timescale \citep{Essig180901144}. With such a large cross section, we expect a significant fraction of dwarf halos to be in the collapse phase. On the other hand, to avoid a substantial reduction of the number of subhalos due to evaporation associated with subhalo–-host halo interactions \citep{Nadler200108754}, the cross section must be suppressed at the velocity scale set by the MW host halo. Thus, we are led to consider a velocity-dependent cross section with $\sigma/m\lesssim 1~\rm cm^2~g^{-1}$ at the velocity scale $v\sim 200~\kms$ set by a $10^{12}~\msun$ host halo.

In this work, we study such a model by performing a high-resolution cosmological zoom-in simulation of an MW-like system, and we explore the effects of SIDM on halos both within and surrounding the MW-like host. 
Importantly, our simulation includes an analog of the Large Magellanic Cloud (LMC) system---i.e., a subhalo with a mass of~$\approx 10^{11}~\msun$ that accreted into the MW halo $\sim 1~\Gyr$ ago, and is currently at a distance of $\sim 50~\kpc$ from the center of the MW.
Following our estimate above, we consider a strong and velocity-dependent cross section that features Rutherford-like scatterings among DM particles and reaches a momentum transfer cross section $\sigma_T/m = 100~\rm cm^2~g^{-1}$ at relative velocity scales of $v\approx 10~\mathrm{km\ s}^{-1}$, which roughly correspond to the (sub)halos that host ultra-faint dwarf galaxies \citep{Jethwa161207834,Graus180803654,Nadler191203303,Silverman220310104}. We will show that this model yields a significant fraction of low-mass core-collapsed (sub)halos within and surrounding the MW, and that it naturally produces the anticorrelation between subhalo central density and pericentric distance hinted at by Gaia observations. 

Importantly, we explore the environmental dependence of self-interactions' effects on halos and subhalos using a large portion of the high-resolution region in our zoom-in simulations, out to $\sim 3~\mathrm{Mpc}$ from the center of the MW halo. This combination of resolution and volume allows us to analyze the SIDM effects on halos and subhalos both within and surrounding the MW. We specifically study subhalos of the MW, subhalos associated with its LMC analog, splashback subhalos of the MW, and isolated halos. Halos in each category are subject to distinct SIDM effects. For example, subhalos' core formation may be enhanced by tidal stripping due to the MW (e.g., \citealt{Vogelsberger12015892,Dooley160308919,Nadler200108754,Yang200202102}) or LMC \citep{Nadler210912120}, and their core collapse may be accelerated by these same tidal fields (e.g., \citealt{Kahlhoefer190410539,Nishikawa190100499,Sameie190407872,Yang210202375,Zeng211000259}). Meanwhile, splashback halos experience particularly strong tidal effects during their pericentric passages, and even isolated halos may core collapse if their initial concentration is sufficiently high. From an observational perspective, the environmentally dependent effects of SIDM on isolated halos, splashback halos, and subhalos are interesting because satellite and field dwarf galaxies with remarkable properties are rapidly being discovered in the Local Volume and beyond (e.g., \citealt{Caldwell161206398,Gregory190207228,Torrealba181104082,Li221014994}).
At face value, some of these galaxies' properties are difficult to explain in standard CDM scenarios (e.g., Crater II; see \citealt{Borukhovetskaya211201540}), motivating careful comparisons to full-fledged cosmological SIDM simulations that self-consistently predict the correlated properties of isolated halo and subhalo populations.

As part of our analysis, we apply a constant effective cross section model, which has recently been discussed in \cite{Yang220503392}, \cite{Yang220502957}, and \cite{Outmezguine220406568}, to analytically predict the gravothermal evolution stages of halos in our SIDM simulation. 
This model captures the leading-order effects of a generic velocity-dependent elastic cross section on halo dynamics. 
By quantifying halos' gravothermal evolution, we gain additional insights into the resulting SIDM effects on halo populations in comparison to CDM.
For example, we show that a combination of halos' maximum circular velocity, $V_{\mathrm{max}}$, and the radius at which it occurs, $R_{\mathrm{max}}$, can be used to accurately select deeply core-collapsed halos.
Furthermore, although this constant effective cross section approach has only been validated using isolated halos, we show that it is useful to explain the behavior of both halos and subhalos in our simulations. 
In future work, we plan to further explore this analytic approach to distinguish the effects of different SIDM models on halo and galaxy properties. 

The paper is organized as follows. 
Section \ref{sec:sims} introduces our cosmological zoom-in simulations, including our SIDM model implementation, simulation setup, and halo categorization.   
Section \ref{sec:core-collapse} presents detailed evolution histories of core-collapsed and core-forming halos, discusses the physics of gravothermal evolution, presents the $R_{\rm max}$--$V_{\rm max}$ relation for isolated halos and subhalos in our simulations, and proposes a criterion for selecting deeply core-collapsed systems based on their halo properties. 
Section \ref{sec:population_stats} compares the characteristics of halos within and surrounding the MW in our SIDM and CDM simulations, and places these results in the context of satellite and isolated dwarf galaxy observations. 
Section \ref{sec:discussion} discusses our findings in light of recent SIDM studies, as well as future work; Section \ref{sec:conclusions} summarizes and concludes. 

\section{Simulations}
\label{sec:sims}

\subsection{SIDM Implementation}
\label{sec:implementation}

\begin{figure}[t!]
  \centering
  \includegraphics[height=8.2cm]{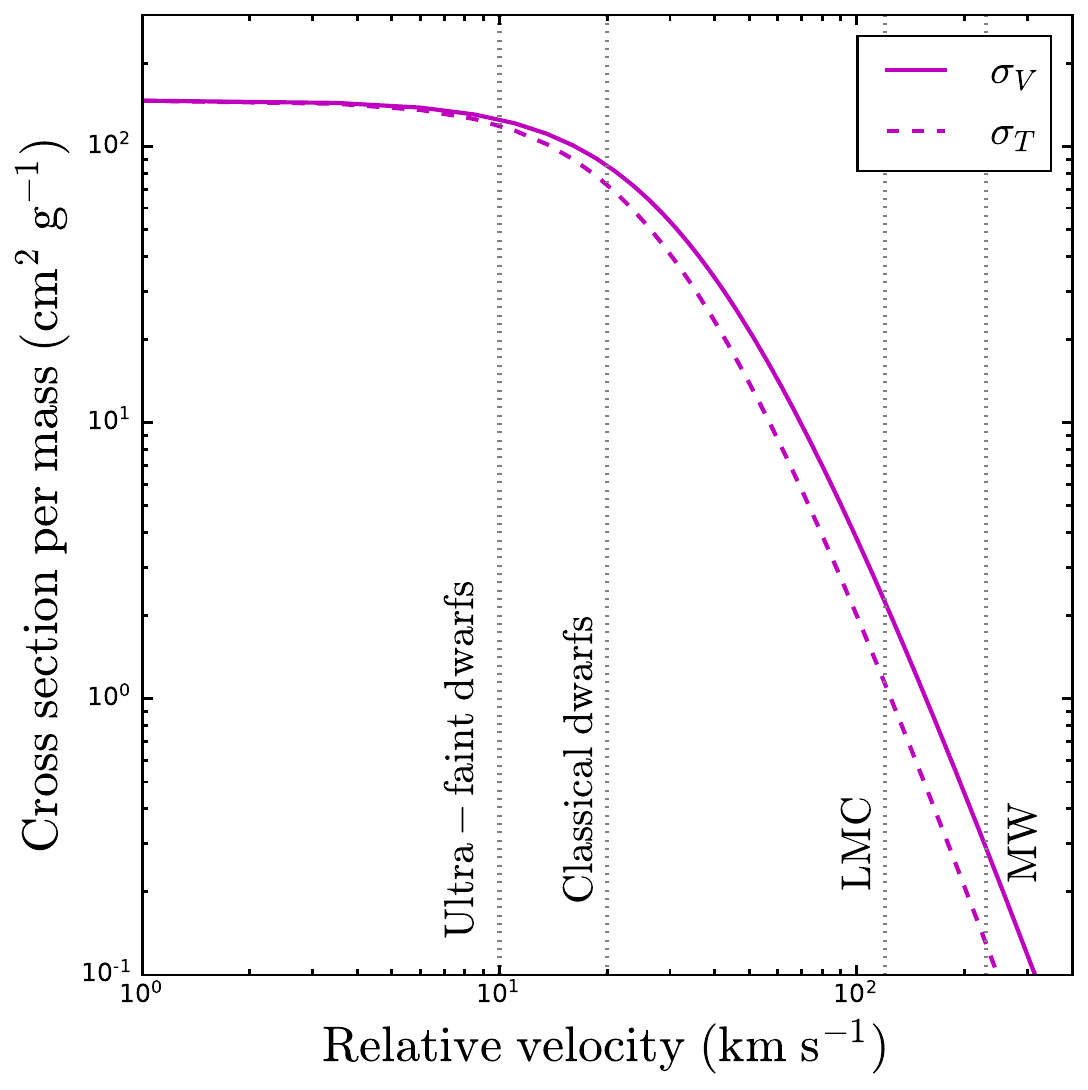}
  \caption{\label{fig:bmxs} The viscosity and momentum transfer cross sections, $\sigma_V$ (solid) and $\sigma_T$ (dashed), for the SIDM model simulated in this work. Vertical dotted lines show velocity scales approximately corresponding the the MW halo, LMC halo, classical dwarf, and ultra-faint dwarfs (e.g., \citealt{Nadler200108754}).}
\end{figure}

We consider a strong and velocity-dependent SIDM model that features Rutherford-like scatterings among DM particles, with a differential cross section
\begin{eqnarray}
\label{eq:xsr}
\frac{d\sigma}{d \cos\theta} = \frac{\sigma_{0}w^4}{2\left[w^2+{v^{2}}\sin^2(\theta/2)\right]^2 },  
\end{eqnarray}
where $w$ is a velocity parameter, and $\sigma_0$ controls the normalization \citep{Feng09110422,Ibe:2009mk}. 
Requiring the momentum transfer cross section $\sigma_T/m = 100~\rm cm^2~g^{-1}$, at $v=14~\kms$, and $\sigma_T/m = 2~\rm cm^2~g^{-1}$, at $v=100~\rm km~s^{-1}$, (equivalent to the ``vd100'' model in \citealt{Turner201002924}), we obtain $\sigma_0/m=147.1~\rm cm^2~g^{-1}$, $w=24.33~\rm km~s^{-1}$. 
Figure \ref{fig:bmxs} shows the velocity dependence of the viscosity and momentum transfer cross sections for this model (see \citealt{Yang220503392} for details). 
The large cross section and strong velocity dependence of our model is expected to cause a sizable fraction of low-mass halos and subhalos in our simulations to core collapse by $z=0$ (e.g., see \citealt{Turner201002924,Lovell220906834}).

We implement elastic DM self-interactions in the public \textsc{GADGET-2} code following the approach described in \cite{Yang220503392}, which in turn follows the procedure in \cite{Robertson160504307} with a few modifications. 
For isolated simulations, the details of our implementation are described in \cite{Yang220503392}; in this work, we extend the SIDM module to enable cosmological simulations. 
Briefly, our module allows a particle to scatter with multiple particles in each time step, where the scattering partners are selected based on a probabilistic approach, considering a neighbor search radius that equals the spline length, $\ell=2.8\epsilon=224~\mathrm{pc\ } h^{-1}$, where $\epsilon=80~\mathrm{pc\ } h^{-1}$ is the comoving Plummer-equivalent gravitational softening length in our simulations (see Section \ref{sec:simulation_details}).  
In Appendix~\ref{sec:appSIDM}, we provide more details on our implementation and validate our approach by showing that a cosmological simulation of an isolated halo with $\sigma/m=5~\rm cm^2~g^{-1}$ yields an equivalent density profile as in a Newtonian simulation. 

Our SIDM implementation fully accounts for the angular dependence of the differential scattering cross section, as in previous idealized \citep{Robertson160504307} and cosmological simulations (e.g., \citealt{Banerjee190612026,Nadler200108754,Nadler210912120}). Note that we do not distinguish the species of the incoming scattering particles in our simulation. 
Thus, our result is equivalent to a physical scenario of Rutherford scattering where the simulation particles are divided into two species of equal amounts, and the cross section is four times the simulated one \citep{Yang220503392}. 

\subsection{Cosmological Zoom-in Simulations}
\label{sec:simulation_details}

\begin{figure*}[htbp]
  \centering
  \includegraphics[width=\textwidth]{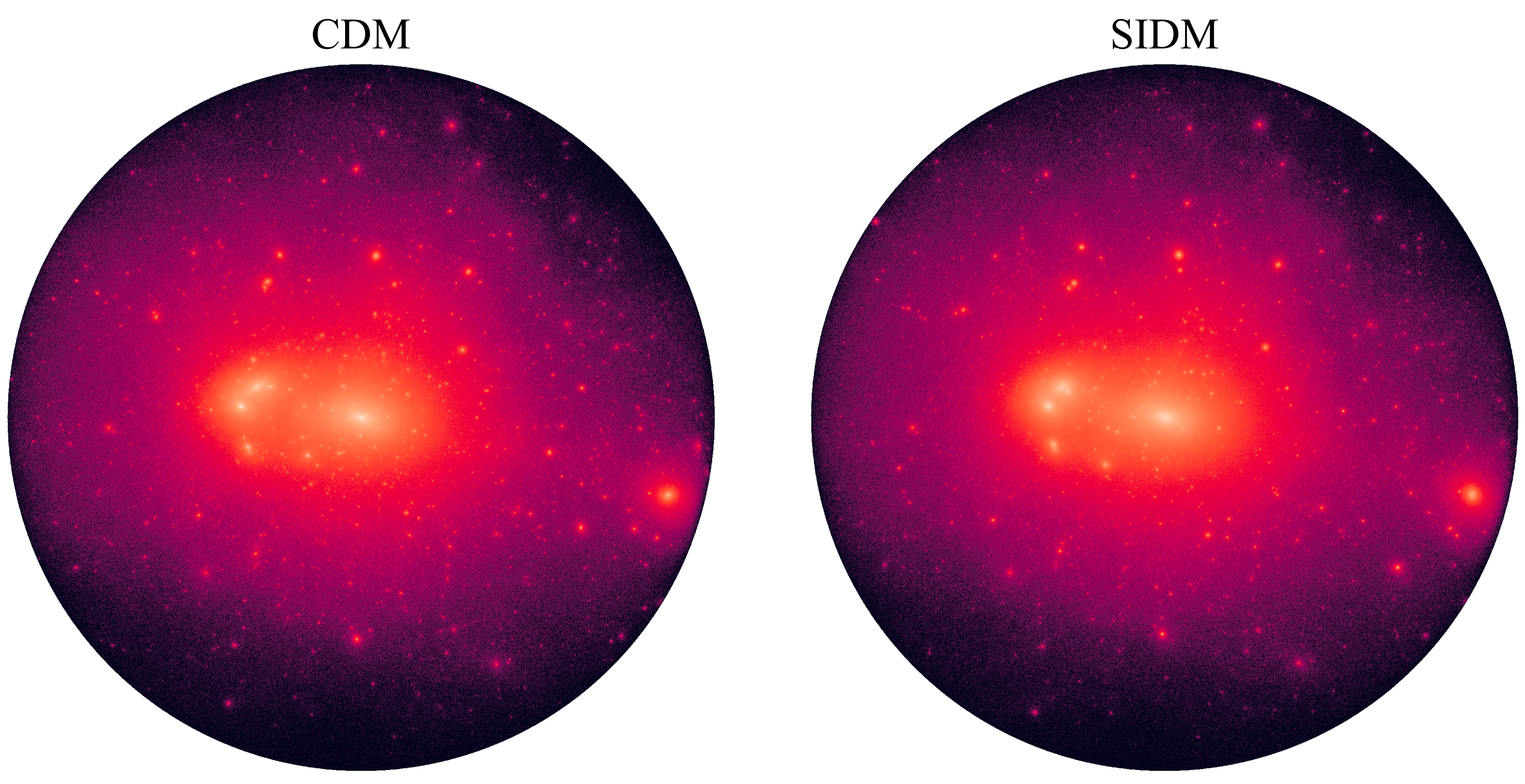}
  \caption{\label{fig:density} Projected DM density for our CDM (left) and SIDM (right) zoom-in simulations, centered on the MW host halo and spanning the region within its virial radius. Self-interactions visibly alter the properties of the MW's and its subhalos' DM structure. Moreover, the DM structure of the LMC system (to the left of the main halo in each panel) as it orbits within the MW is affected by self-interactions. Visualizations were created using \textsc{meshoid} (\url{https://github.com/mikegrudic/meshoid}). 
 }
\end{figure*}

We perform a cosmological zoom-in simulation of an MW-like system using the SIDM model described above, and we compare our results to a corresponding CDM simulation with identical initial conditions. A lower-resolution version of this MW-like simulation was originally presented in \cite{Mao150302637}, used in \cite{Nadler191203303,Nadler200800022}, \cite{Mau220111740} to model the MW satellite galaxy population, and resimulated in a velocity-independent SIDM model to study the effects of self-interactions on subhalos associated with the LMC in \cite{Nadler210912120}. The higher-resolution CDM simulation that we use here was first presented in \cite{Nadler191203303}, and our high-resolution SIDM simulation with a velocity-dependent cross section is presented here for the first time.

Our CDM and SIDM zoom-in simulations have a high-resolution particle mass of $4\times 10^4~M_{\mathrm{\odot}}~h^{-1}$, with a  Plummer-equivalent gravitational softening of $80~\mathrm{pc}~h^{-1}$. The simulations are performed using \textsc{GADGET-2} \citep{Springel0505010}; halos are identified using \textsc{ROCKSTAR} \citep{Behroozi11104372}, and merger trees are constructed using \textsc{consistent-trees} \citep{Behroozi11104370}. The cosmological parameters for both runs are $\Omega_M=0.286$, $\Omega_\Lambda=0.714$, $n_s=1$, $h=0.7$, and $\sigma_8=0.82$ \citep{WMAP9}.\footnote{This choice of $n_s$ yields an LMC analog that falls into the MW at the correct lookback time (see the discussion in Appendix A of \citealt{Mau220111740}).} We report virial masses using the \cite{Bryan_1998} definition, which corresponds to $\Delta_{\mathrm{vir}}\approx 99$ times the critical density of the universe at $z=0$ given our cosmological parameters.

At $z=0$, the mass of the MW host halo in both our CDM and SIDM simulation is $1.6\times 10^{12}~M_{\mathrm{\odot}}$. This mass is consistent with many recent MW mass measurements (e.g., see \citealt{Shen211109327} for a recent compilation). Furthermore, the virial concentration of both our CDM and SIDM host halos is $c_{\mathrm{NFW}}=11.4$, which is also broadly consistent with the inferred concentration of the MW halo (e.g., \citealt{Callingham180810456}). As discussed in \cite{Nadler200800022}, we emphasize that our host halo's mass and concentration measurements are performed in the absence of baryons, which alter the MW's inferred mass and concentration (e.g., \citealt{Cautun191104557}).

In both runs, an analog of the LMC with a $z=0$ mass of $1.8\times 10^{11}~M_{\mathrm{\odot}}$ accretes into the MW (i.e., first crosses within the MW's virial radius) $\sim 1~\mathrm{Gyr}$ ago. These properties are consistent with recent estimates of the LMC's mass (e.g., based on its dynamical impact on stellar streams; \citealt{Erkal181208192}), and the orbits of our LMC analogs are consistent with the LMC's measured orbital properties (e.g., \citealt{Kallivayalil13010832}). In addition to these LMC analogs, our hosts experience major mergers between $1 < z < 2$ with $\sim 10^{11}~M_{\mathrm{\odot}}$ halos that are broadly similar to Gaia-Enceladus (see \citealt{Nadler191203303}). Our hosts have quiescent merger histories after this event, and therefore have formation histories that resemble the MW's reconstructed merger history (e.g., see \citealt{Helmi200204340} for a review).

Finally, we emphasize that the high-resolution zoom-in regions in both our CDM and SIDM simulations extend to $\sim 10$ times the virial radius of the MW host halo, corresponding to $3\ \mathrm{Mpc}$.\footnote{In particular, greater than $90\%$ of the particles within $\sim 3\ \mathrm{Mpc}$ of the MW host halo center are high-resolution particles \citep{Wang210211876}. Thus, we only study halos within this radius from the center of the MW host.} Thus, there is a substantial population of halos \emph{outside} the virial radius of the MW host halo in both simulations. This population includes isolated ``field'' halos that do not reside within the virial radius of larger systems, subhalos of lower-mass hosts in the region surrounding the MW, and ``splashback'' halos of the MW---i.e., halos that previously orbited inside of the MW but are outside of its virial radius today (e.g., \citealt{Adhikari14094482,Diemer:2014xya,More150405591}). Here, we study four halo populations: (1) MW subhalos, (2) LMC-associated subhalos (i.e., MW subhalos that accreted into the MW as subhalos of the LMC), (3) splashback subhalos of the MW, and (4) isolated halos. We defer a study of subhalos (and splashback halos) of lower-mass hosts throughout the zoom-in region to future work.

Figure \ref{fig:density} shows projected DM density fields within the virial radius of our CDM (left) and SIDM (right) hosts, and Figure \ref{fig:MWLMC} shows the CDM (black) and SIDM (magenta) MW host halo density profiles. 
The MW host halo exhibits a $\sim 1~\mathrm{kpc}$ core in our SIDM simulation, which is expected given that our cross section is greater than $\sim 0.1~\mathrm{cm}^2\ \mathrm{g}^{-1}$ at the MW halo velocity scale (e.g., \citealt{Nadler200108754}).
Note that this core is not expected to remain in a full hydrodynamic setting because SIDM efficiently responds to a baryon-dominated central gravitational potential, thus regenerating a cusp \citep{Kaplinghat13116524,Sameie180109682,Robles190401469,Rose220614830}.

Visually, our SIDM model has little effect on the spatial distribution and abundance of MW subhalos.  
This is qualitatively expected; due to the strong velocity dependence of our SIDM model (Figure~\ref{fig:bmxs}), the scattering cross section decreases as $v^{-4}$ at large velocities, suppressing evaporation due to subhalo--host halo interactions, which may reduce the subhalo masses and potentially completely unbind them at late times \citep{Vogelsberger12015892,Nadler200108754,Slone210803243,Zeng211000259}.
We return to this point in detail when discussing the subhalo mass function in Section \ref{sec:population_stats}.
However, due to the mass and concentration dependence of the velocity profile in a halo, we also expect the SIDM effects to be stronger in inner regions of more compact and lower-mass halos.  
Thus, although the projected density fields in our CDM and SIDM simulations are similar, the details of subhalos' positions and internal density structure are clearly affected by SIDM on small scales. As we will demonstrate, our SIDM model facilitates both core formation (in massive, low-concentration halos) and core collapse (in low-mass, high-concentration halos), for halos both within and surrounding the MW.

\subsection{Categorizing Halos}
\label{sec:categorization}

We use the \textsc{ROCKSTAR}~\citep{Behroozi11104372} halo finder to identify halos in our simulation snapshots, and we use the \textsc{consistent-trees}~\citep{Behroozi11104370} merger tree code to reconstruct merger histories of these halos.  
Our simulations' resolution allows us to robustly resolve both isolated halos and tidally evolved subhalos with present masses greater than $10^8~\msun~h^{-1}$, corresponding to systems with~$>2500$ particles at $z=0$. 
In particular, this cut is well above the convergence limit for subhalo abundance measurements in zoom-in simulations \citep{Nadler220902675}, and is comparable to the particle count threshold necessary to robustly resolve the tidally stripped density profiles of CDM subhalos \citep{Errani201107077}.

In Appendix \ref{sec:convergence}, we demonstrate that---even for tidally evolved subhalos at our resolution cut---gravothermal evolution in our cosmological simulation is in good agreement with idealized simulations that pass stringent convergence tests for halos above our fiducial mass threshold. 
We will also show that the core-collapse timescales of halos in our SIDM simulation are in reasonable agreement with those predicted by an empirical SIDM fluid model. 
Nonetheless, we note that the gravothermal evolution of the lowest-mass halos in our SIDM simulation may potentially suffer from convergence issues because detailed convergence tests have not been carried out for core-forming or core-collapsed systems in a cosmological setting.
Because insufficient resolution in low-mass halos generally leads to less efficient collisional relaxation than desired, the population of core-collapsed halos we identify is conservative, in the sense that an even higher-resolution simulation is expected to yield a larger fraction of core-collapsed systems down to a fixed halo mass. 

Figure \ref{fig:demo} illustrates the halo populations we study, highlighting the objects that show signs of core collapse based on the analyses discussed below. In particular, we define the following categories of halos: 
\begin{figure}[t!]
  \centering
  \includegraphics[width=7.2cm]{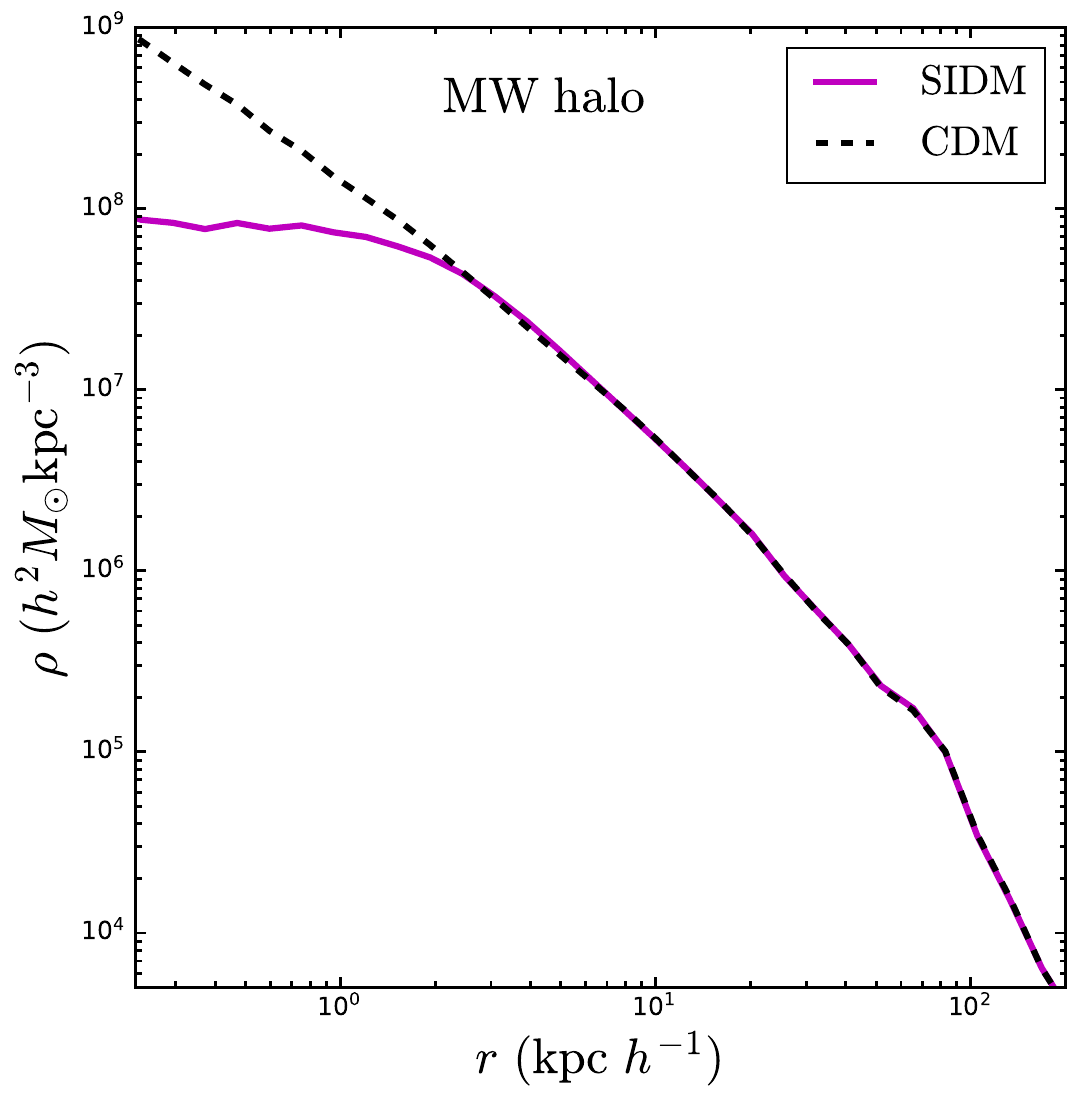}
  \caption{\label{fig:MWLMC} Density profile of the MW host halo in our CDM (dashed black) and SIDM (solid magenta) simulations, plotted down to our spatial resolution limit of $\ell=2.8\epsilon=224~\mathrm{pc\ } h^{-1}$.}
\end{figure}

\begin{enumerate}
    \item \emph{MW subhalos} are halos that reside within the virial radius of the MW at $z=0$ and are not associated with the LMC, including higher-order subhalos (specifically, the \texttt{upid} of these systems must equal the \texttt{id} of the MW host halo at $z=0$ in the \textsc{ROCKSTAR} halo list).
    \item \emph{LMC-associated subhalos} are objects that were within the virial radius of the LMC halo at the time of LMC infall into the MW ($\sim 1~\mathrm{Gyr}$ ago, in our simulation), following the ``fiducial definition'' in \cite{Nadler191203303}. Many of these objects still reside within the virial radius of the LMC at $z=0$, and thus are higher-order subhalos of the MW because the LMC itself is an MW subhalo.
    \item \emph{Splashback halos} are halos that have orbited within the virial radius of the MW halo at $z>0$, but reside outside the MW virial radius today. These systems are usually characterized by a single, strong tidal stripping and heating event along their evolution history, corresponding to a pericentric passage around the MW. 
    \item \emph{Isolated halos} reside outside of the virial radius of any larger halo at $z=0$ (specifically, the \texttt{upid} of these systems must equal $-1$ in the $z=0$ \textsc{ROCKSTAR} halo list), have never passed within the virial radius of the MW halo, and are within $3~\mpc$ of the MW halo center.  
    These systems are subject to much less significant tidal effects than those of subhalos; thus, the effects of SIDM on these halos are expected to exhibit less scatter than for tidally affected systems. 
\end{enumerate} 
According to these definitions, there are 647 (626) isolated halos within, 110 (96) MW subhalos, 15 (19) LMC-associated subhalos, and 38 (35) splashback halos above our fiducial mass threshold of $10^8~\msun~h^{-1}$ in our CDM (SIDM) simulation.

In Figure \ref{fig:demo}, we see by eye that a significant population of core-collapsed halos exists both within and surrounding the MW. In the following sections, we explore the signatures of SIDM on each class of halos defined above.

\begin{figure*}[t!]
  \centering
  \includegraphics[trim={0 1.5cm 0 0},width=0.9\textwidth]{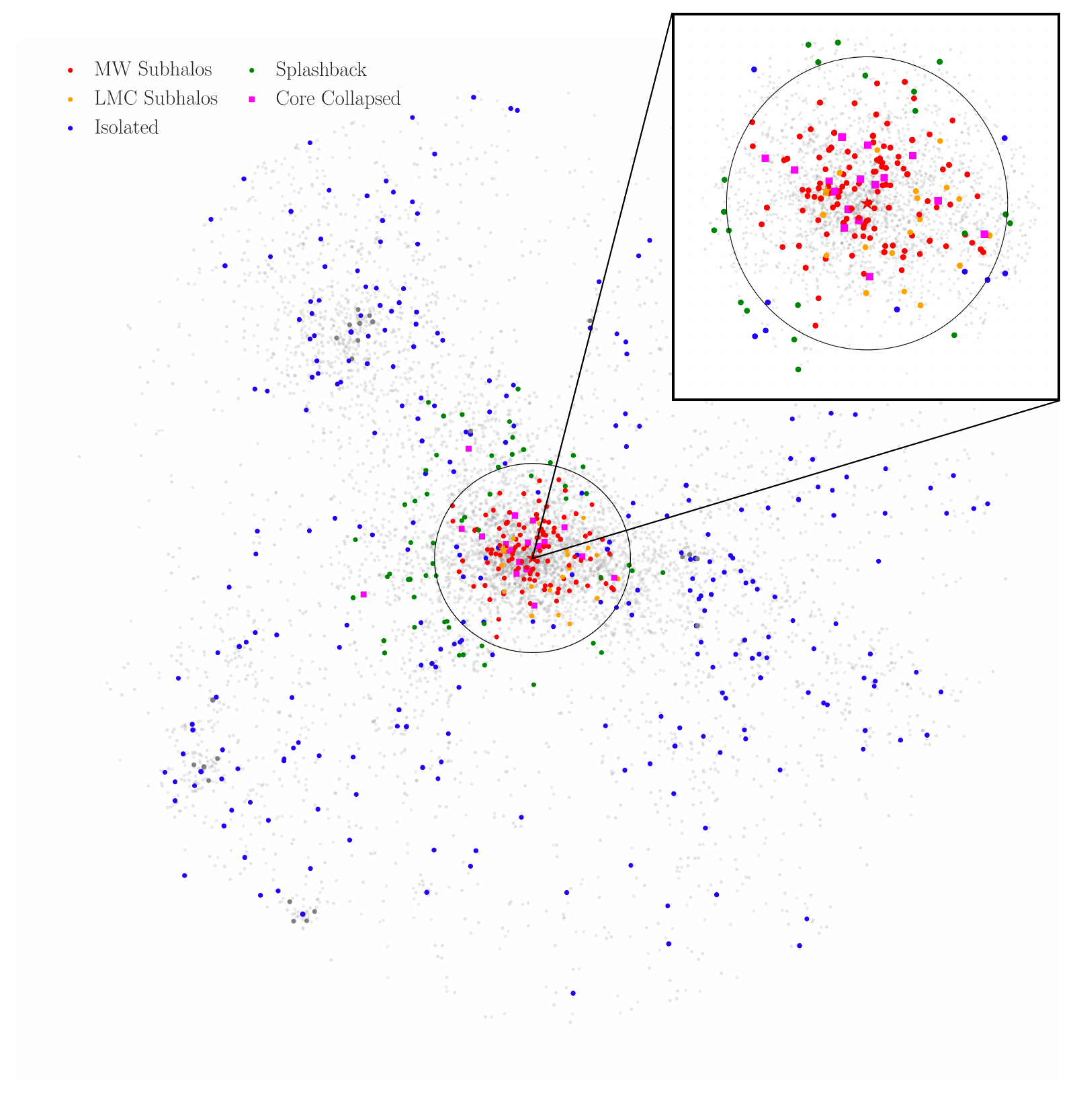}
  \caption{\label{fig:demo} Scatter plot of halos within $5R_{\rm vir}$ of the MW host halo in our SIDM simulation. The red (orange) star indicates the center of the MW (LMC) host halo. Halos with $M_{\rm vir}(z=0) > 10^8~\msun$ (i.e., $> 2000$ high-resolution particles) are shown in color, with MW subhalos shown in red, LMC-associated subhalos shown in orange, splashback halos of the MW (i.e., halos that were formerly subhalos of the MW host) shown in green, and isolated halos shown in blue. Halos that show signs of core collapse according to our criteria in Section~\ref{sec:categorization} are shown by magenta squares. The black circle shows the MW host's virial radius, and the zoom-in panel highlights the halo populations within the virial radius, in projection. Solid gray points show subhalos that are outside of the MW but reside within another host, and transparent gray points show halos and subhalos with $5\times 10^6~{\msun} < M_{\rm vir}(z=0) < 10^8~\msun$ (i.e., halos below our $2000$ particle limit, down to objects with 100 high-resolution particles), which are not used in our main analyses.
  }
\end{figure*}

\begin{deluxetable*}{{c@{\hspace{0.07in}}c@{\hspace{0.07in}}c@{\hspace{0.07in}}c@{\hspace{0.07in}}c@{\hspace{0.07in}}c@{\hspace{0.07in}}c}}[t!]
\centering
\tablecolumns{6}
\tablecaption{Properties of Selected Benchmark Halos in the SIDM (CDM) Simulation at Redshift Zero}
\tablehead{
\colhead{Name} & \colhead{$M_{\rm vir}$ ($\msun~h^{-1}$)} & \colhead{$R_{\rm vir}~\rm (kpc~h^{-1})$} & \colhead{$V_{\rm max}~\rm (km~s^{-1})$} & \colhead{$R_{\rm max}~\rm (kpc~h^{-1})$} & \colhead{$c_{\mathrm{eff}}$}
}
\startdata
LMC analog & $1.29\times 10^{11}~ (1.21\times 10^{11})$  & $104~(102)$  &  $104~(103)$   &  $18.0~(19.3)$ & $12.5~ (11.4)$\\
\hline
MW subhalo & $2.35\times 10^{8}~(2.73\times 10^{8})$   & $12.7~(13.3)$ &  $24.8~(20.2)$  &  $0.278~(0.929)$ & $98.7~(31.0)$ \\ 
\hline
Isolated        & $4.36\times 10^{8}~(4.53\times 10^{8})$   & $15.6~(15.8)$ &  $21.9~(18.9)$  &  $0.439~(0.878)$ & $76.8~(38.9)$ \\
\hline
Splashback & $1.78\times 10^{8}~(1.85\times 10^{8})$   & $11.5~(11.7)$ &  $19.6~(17.3)$  &  $0.419~(0.761)$ & $59.6~(33.3)$ \\
\hline
\enddata
{\footnotesize \tablecomments{The first column lists the identity of each benchmark halo, the second (third) column lists virial mass (virial radius) at $z=0$, the fourth (fifth) column lists maximum circular velocity (the radius at which the maximum circular velocity is achieved), and the sixth column lists effective concentration (Equation~\ref{eq:eff}). The LMC analog resides in the core-forming phase and all other benchmark halos reside in the core-collapsed phase; these halos have significantly larger $c_{\rm eff}$ in SIDM compared to CDM.}\vspace{-5mm}}
\label{tab:bm}
\end{deluxetable*}

\section{Halos in the Core-collapse and Core-formation Phases}
\label{sec:core-collapse}

In this section, we explore the effects of self-interactions on the gravothermal evolution of halos---in both the core-formation and core-collapse regions---in our SIDM simulation. To motivate the halo properties that we study, we present evolution histories of representative core-forming and core-collapsed halos in Section \ref{sec:cbm}; we then compare analytic estimates of SIDM effects on halo structure in terms of maximum circular velocity, $V_{\mathrm{max}}$, and the radius at which this maximum occurs, $R_{\mathrm{max}}$, in Section~\ref{sec:sidm_features}; with these predictions and measurements in hand, we define conservative criteria for selecting core-collapsed and core-forming systems based on these halo properties in Section~\ref{sec:sidm_identification}.

\subsection{Halo Evolution Histories}
\label{sec:cbm}
 
By visually inspecting the density profiles of halos in our SIDM simulation, we find that core-forming halos and deeply core-collapsed halos exhibit clear changes in their density profiles relative to CDM. 
To motivate our subsequent population analyses, which will be focused on the most extreme core-collapsed and core-forming halos, we therefore begin by studying the structure and evolution history of a representative halo in each regime, in detail. Specifically, we choose the LMC analog halo as a clear (and particularly well-resolved) example of a core-forming system, and we choose a representative low-mass, core-collapsed MW subhalo. We also identify representative, low-mass core-collapsed isolated and splashback halos; we present their evolution histories in Appendix \ref{sec:additional_benchmark}. 
Table~\ref{tab:bm} lists the $z=0$ properties for each of our four benchmark systems, both in our SIDM simulation and for their matched counterpart in CDM. 

In the following examples, and throughout the paper, we measure density profiles of halos above our resolution cut as follows: we compute the kinetic and potential energies of particles near each halo and use bound particles to compute halo properties. 
To compute the potential energy more precisely, we iterate the bound particle selection five times, each time removing some unbound particles; we find that the number of bound particles stabilizes after five iterations.

\subsubsection{Benchmark Core-forming LMC Halo}

The LMC analog halo in our SIDM simulation, which has a mass of $\sim 10^{11}~\msun$, has formed a prominent $\mathcal{O}(\mathrm{kpc})$ core by $z=0$. Given this mass, our analytic effective cross section model described in Section~\ref{sec:sidm_identification} predicts a core-collapse timescale of $t_{\rm c} \approx 2\times 10^3~\Gyr$. 
Hence, the LMC analog is expected to reside well within the core-forming phase. Note that we do not attempt to compare our LMC analog's predicted density profile to observations; although the LMC is relatively well observed, inferring its inner kiloparsec DM distribution is challenging because of its proximity and central bar.

\begin{figure*}[htbp]
  \centering
  \includegraphics[height=5cm]{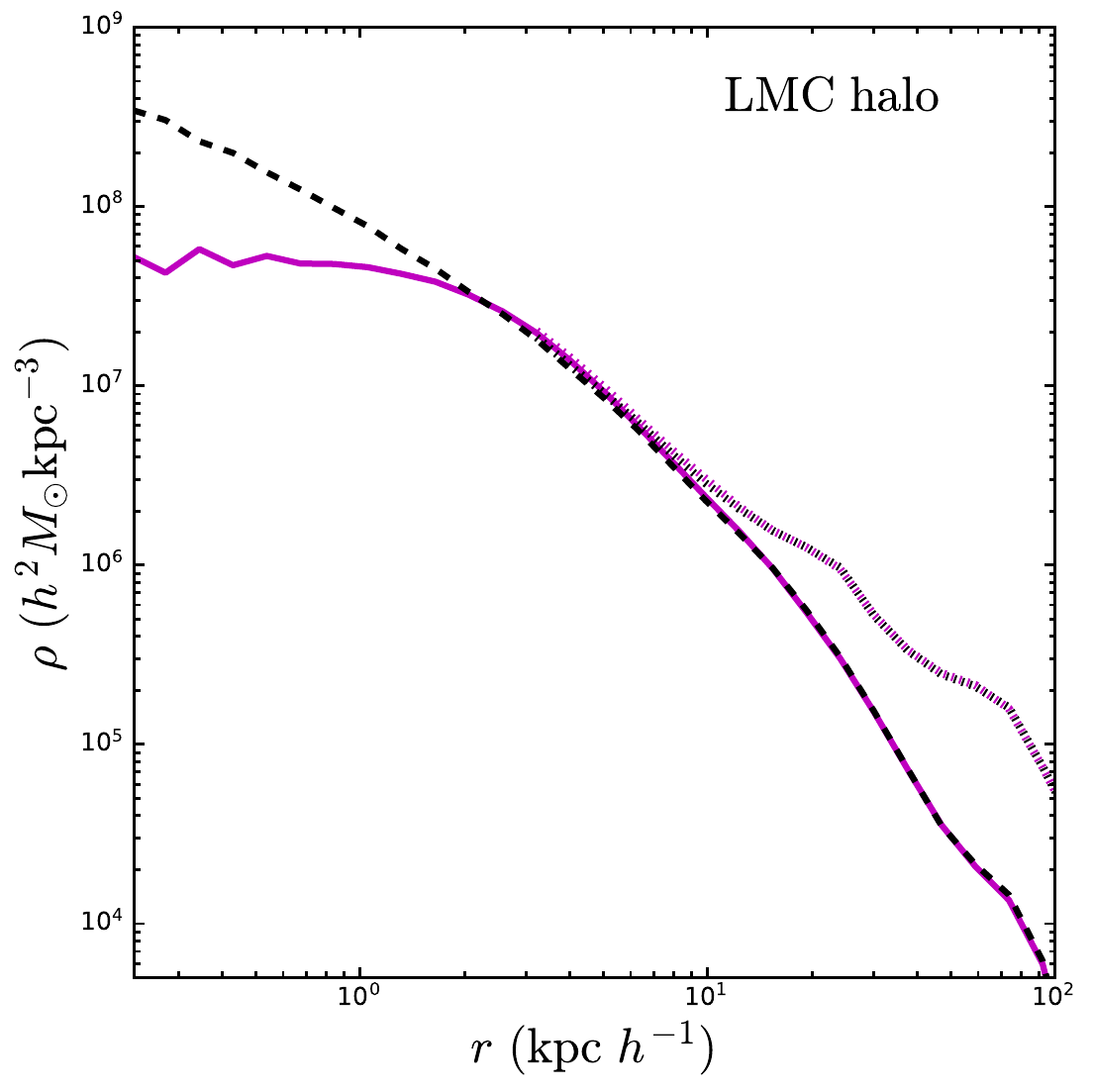}
  \includegraphics[height=5cm]{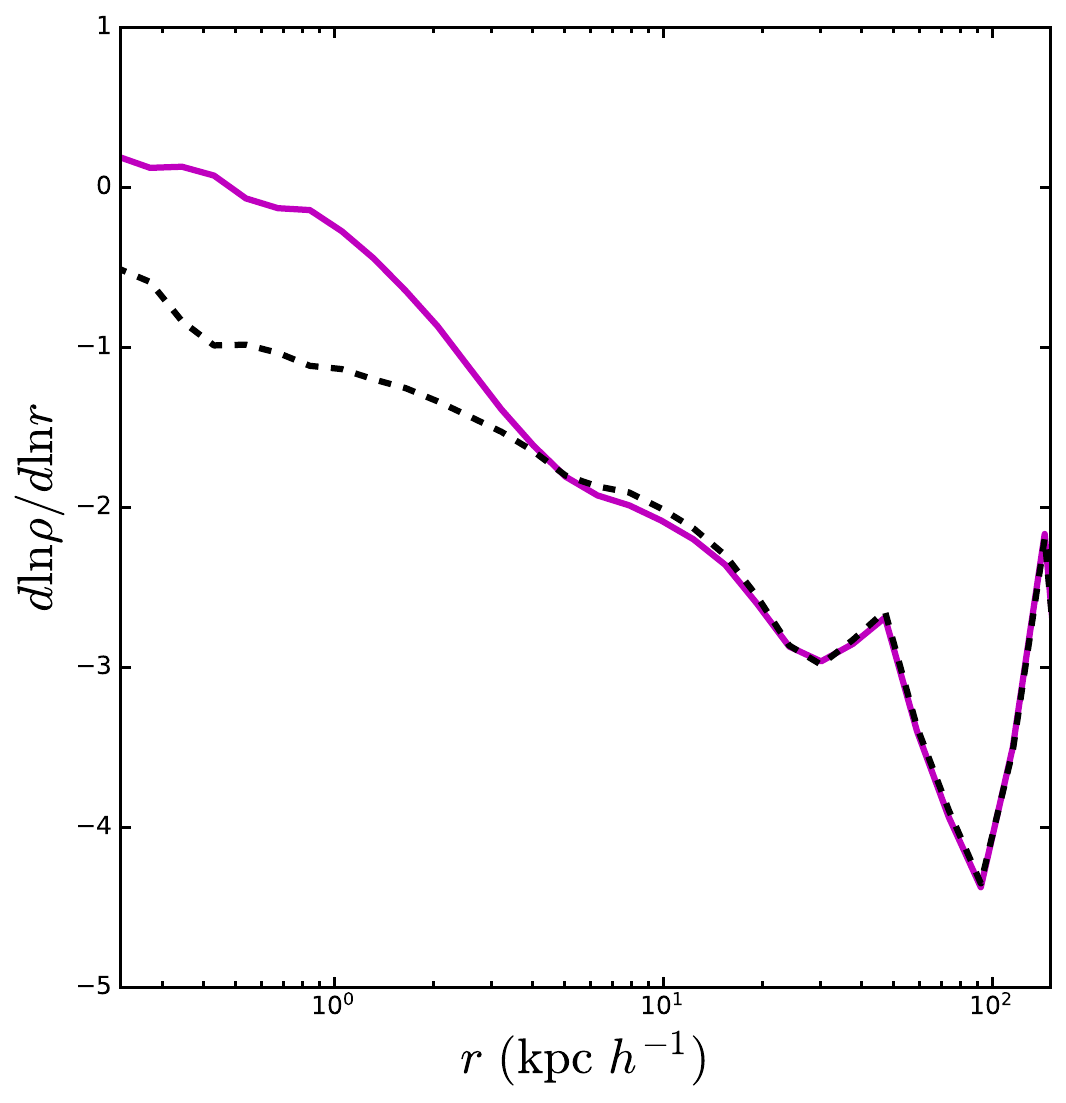}
  \includegraphics[height=5cm]{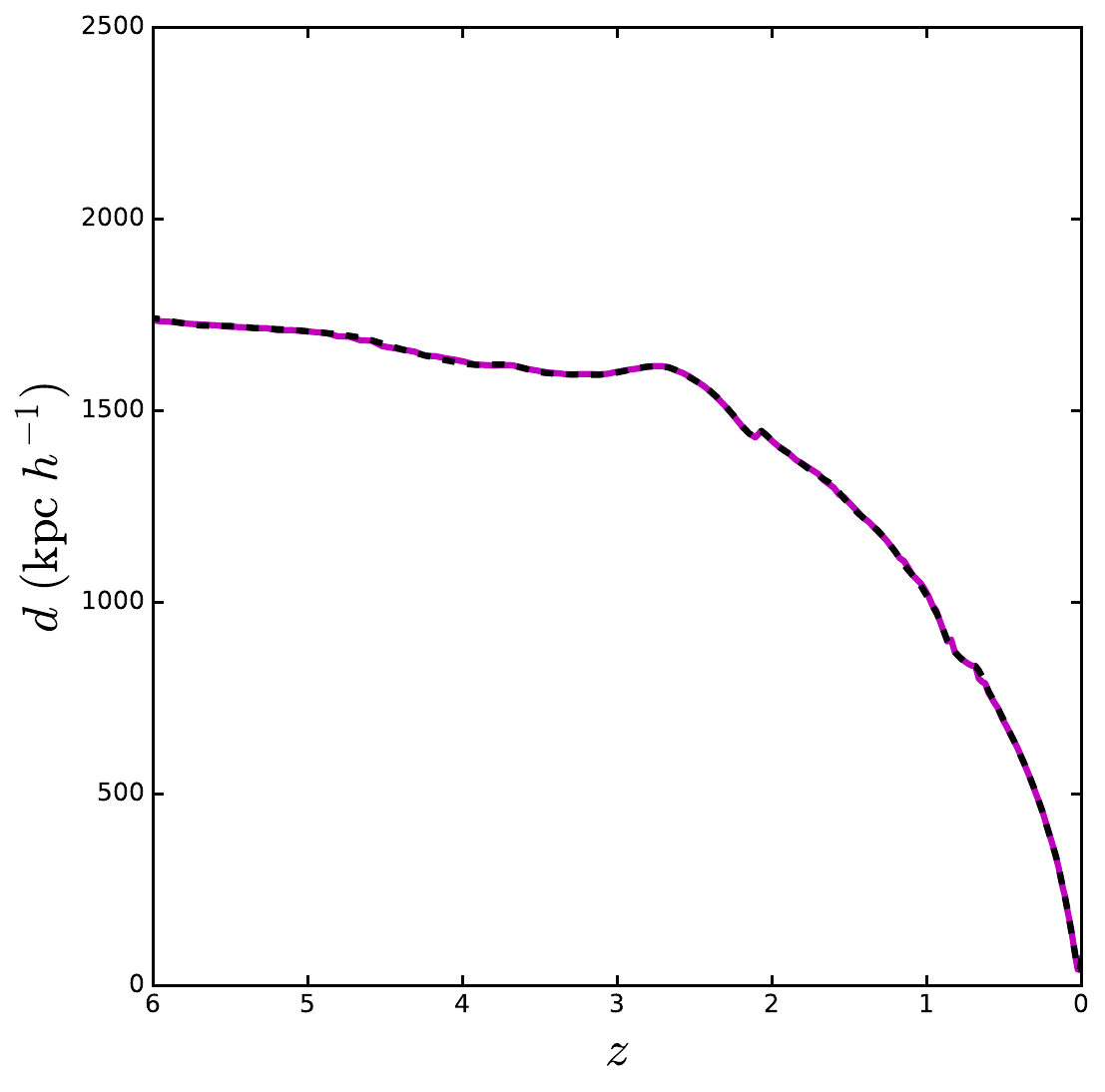}\\
  \includegraphics[height=5cm]{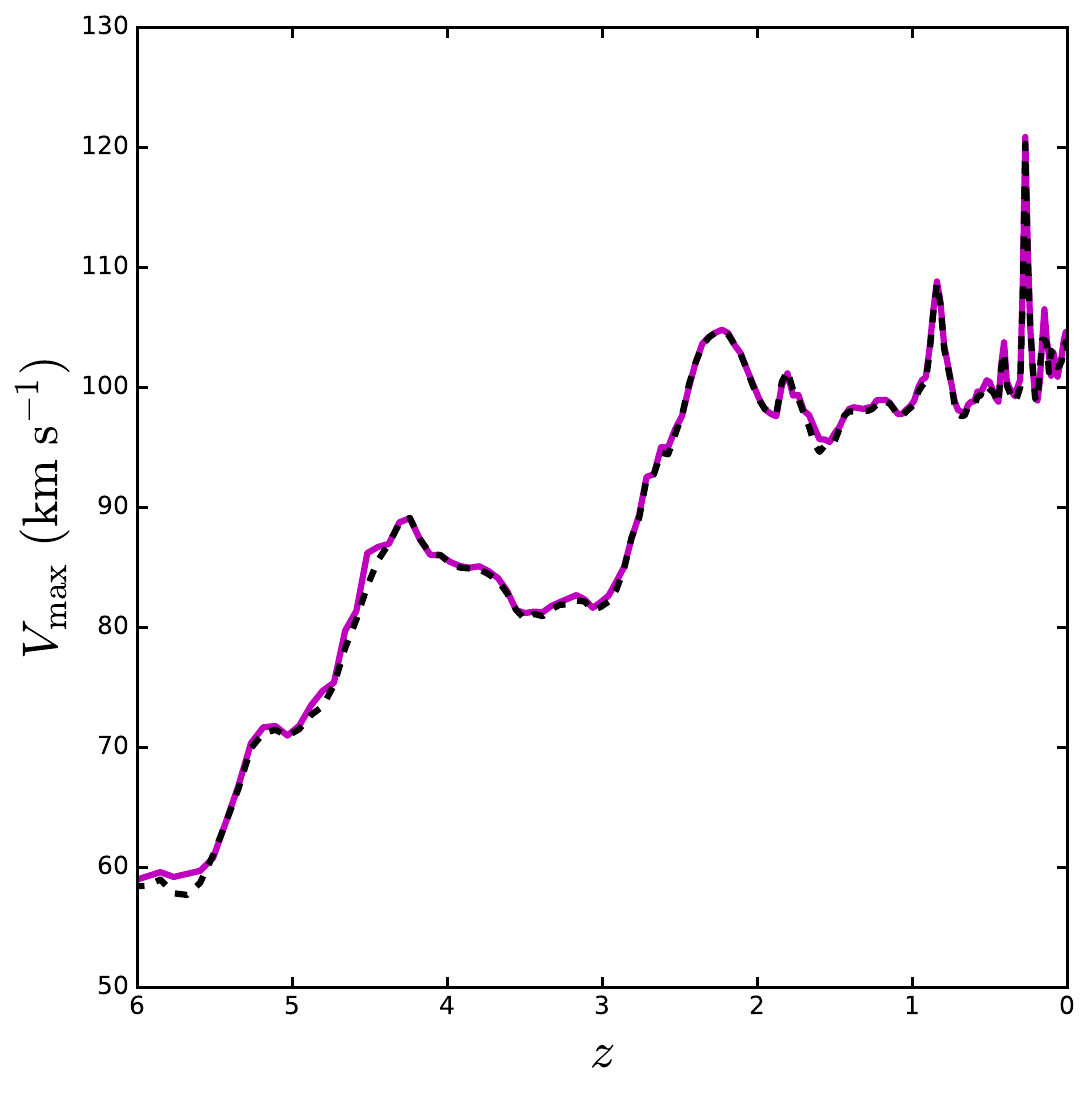} 
  \includegraphics[height=5cm]{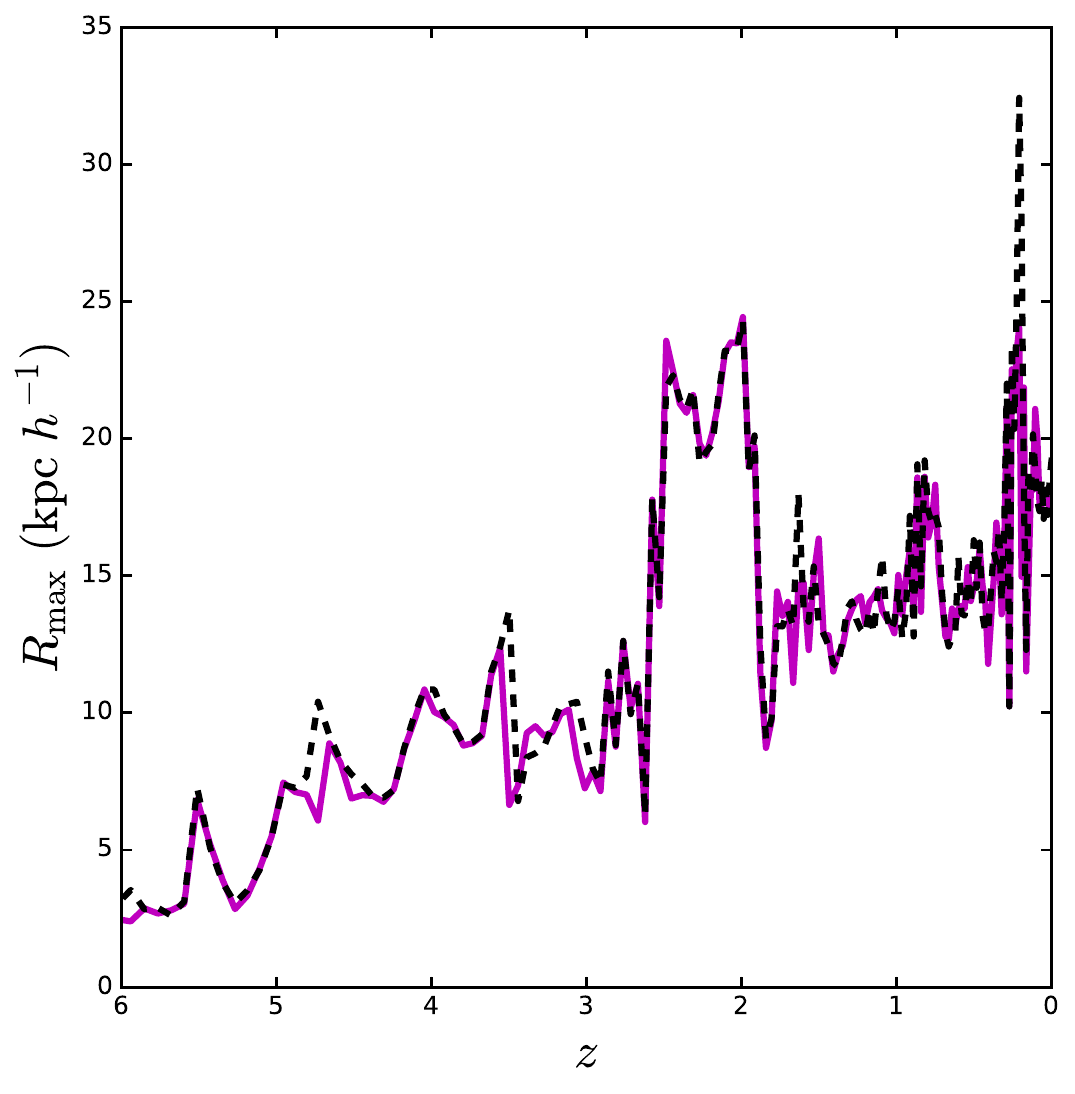}
  \includegraphics[height=5cm]{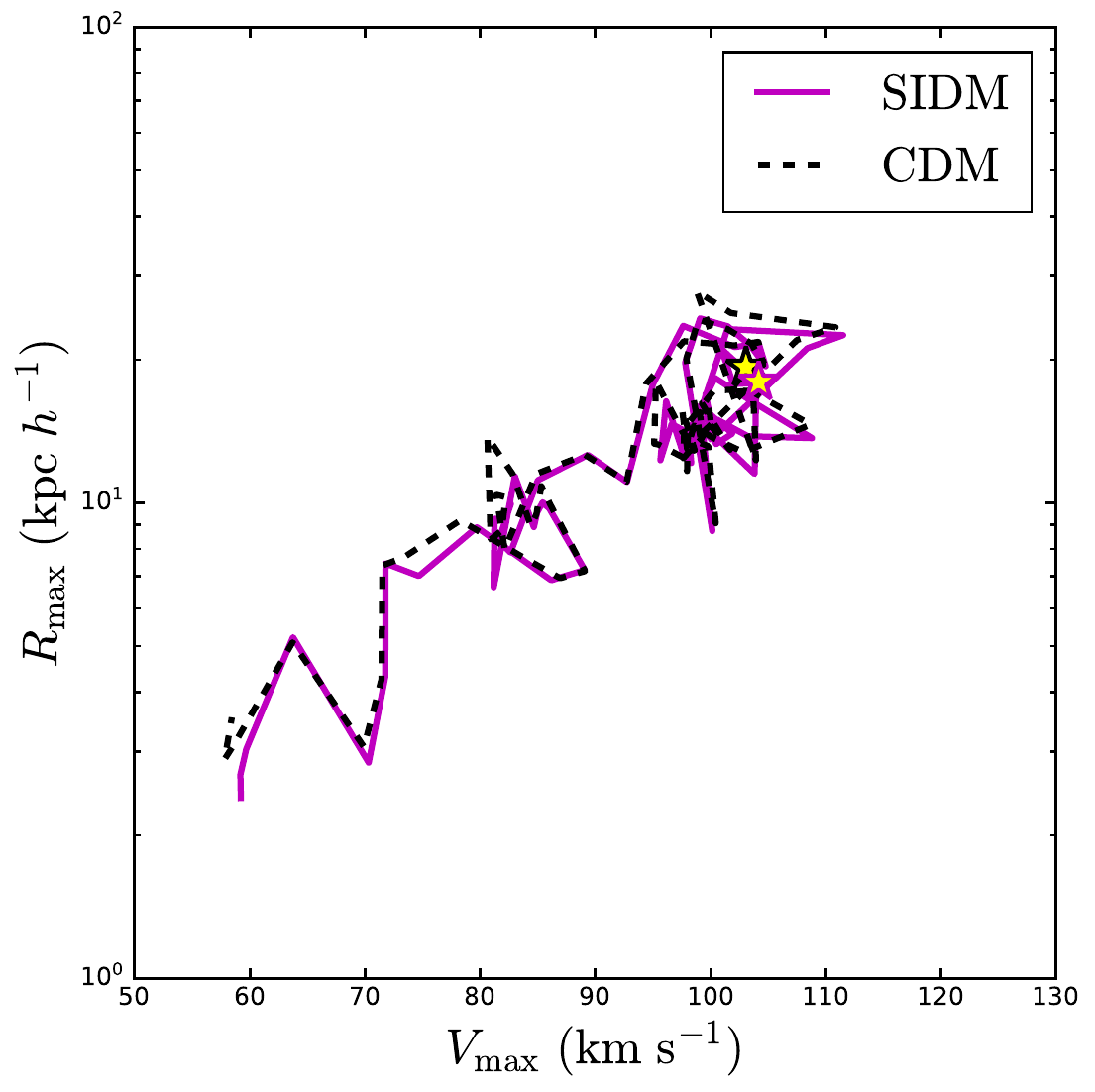}
  \caption{\label{fig:bmlmc}
Profiles (top) and evolution histories (bottom) of the LMC analog halo in CDM (black) and SIDM (magenta). On the top row, the three panels show the density profile, logarithmic density profile slope, and distance from the MW center as a function of redshift, from left to right. 
On the bottom row, the three panels (from left to right) show the evolution of $V_{\rm max}$ and $R_{\mathrm{max}}$ as functions of redshift, and the evolution trajectory in the $R_{\rm max}$--$V_{\rm max}$ plane. In both simulations, the LMC analog halo enters the MW virial radius $\approx 1~\mathrm{Gyr}$ ago, at $z\approx 0.1$.
}
\end{figure*}

We match LMC analog halos in our CDM and SIDM simulations based on their orbital and $V_{\rm max}$ evolution histories, which are illustrated in the bottom left and bottom middle panels of Figure~\ref{fig:bmlmc}. 
We observe that the LMC accreted into the MW at $z\sim 0.2$, or $\sim 1~\mathrm{Gyr}$ ago---consistent with the LMC's inferred infall time based on proper-motion measurements (e.g., \citealt{Kallivayalil13010832})---coinciding with the peak in its $V_{\rm max}$ history before it begins to be tidally stripped by the MW.

The top left and top middle panels of Figure~\ref{fig:bmlmc} respectively show the density profile and density profile logarithmic slope of the LMC analog halo in our CDM and SIDM simulations. 
SIDM thermalizes the inner halo region, reducing the central density; we have also checked that self-interactions increase the inner velocity dispersion, as expected. 
In our SIDM simulation, the density profile slope starts to become shallower than that in CDM at $r\sim 1~\mathrm{kpc}$, and approaches zero in the subkiloparsec regions. 
At large radii, both the CDM and SIDM LMC analogs have nonconstant density slopes. 
This is reasonable, because many subhalos are better fit by Einasto profiles with rolling logarithmic density slopes at large radii rather than Navarro--Frenk--White (NFW) profiles~\citep{2008MNRAS.391.1685S}.  
We analyze density profiles using only bound particles, with results shown as solid lines for SIDM and dashed lines for CDM; for demonstration, we also plot density profiles using all the particles in dotted lines. At large radii, the contribution of unbound particles to the density profiles becomes significant.

\begin{figure*}[htbp]
  \centering
  \includegraphics[height=4.6cm]{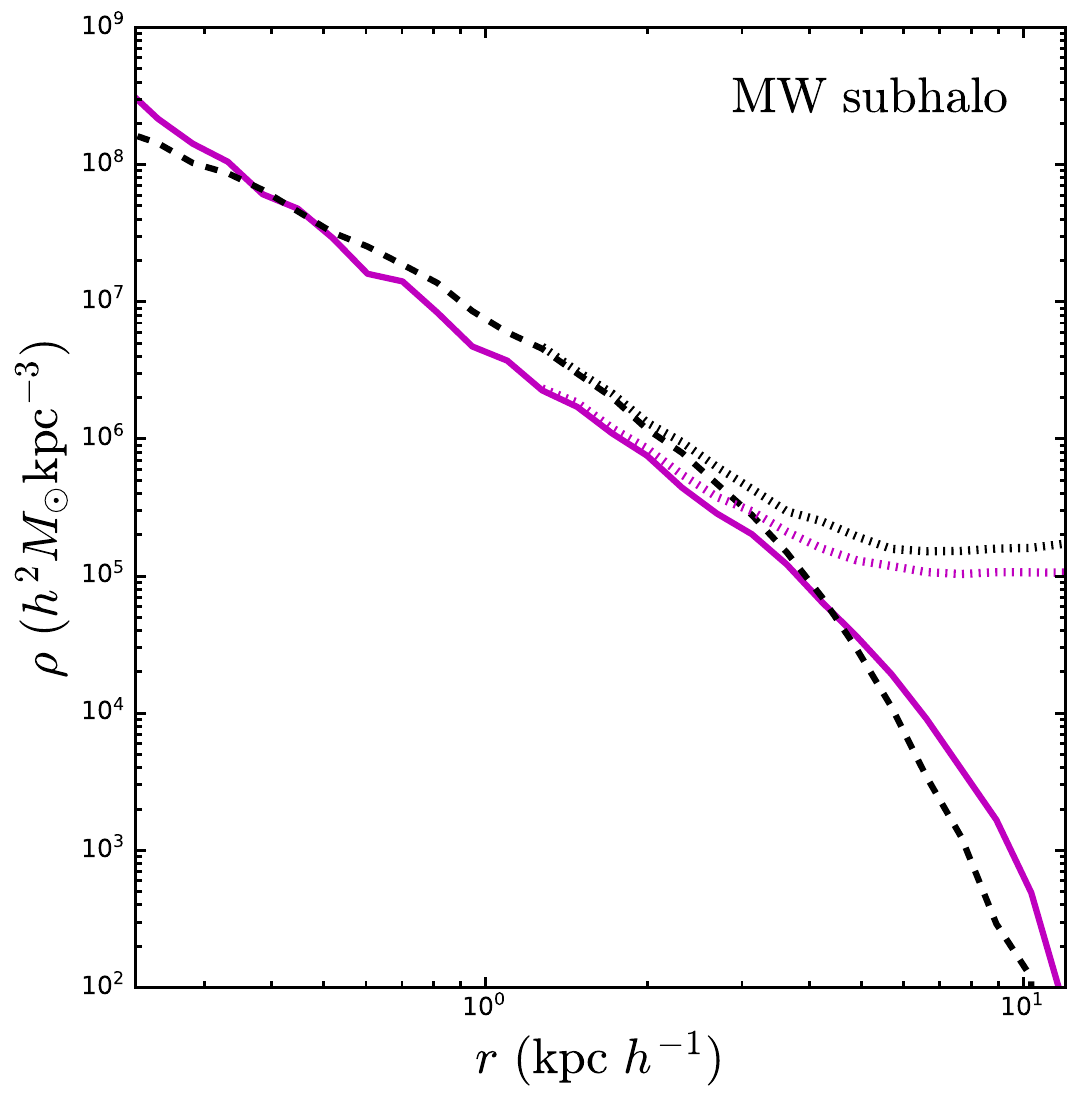}
  \includegraphics[height=4.6cm]{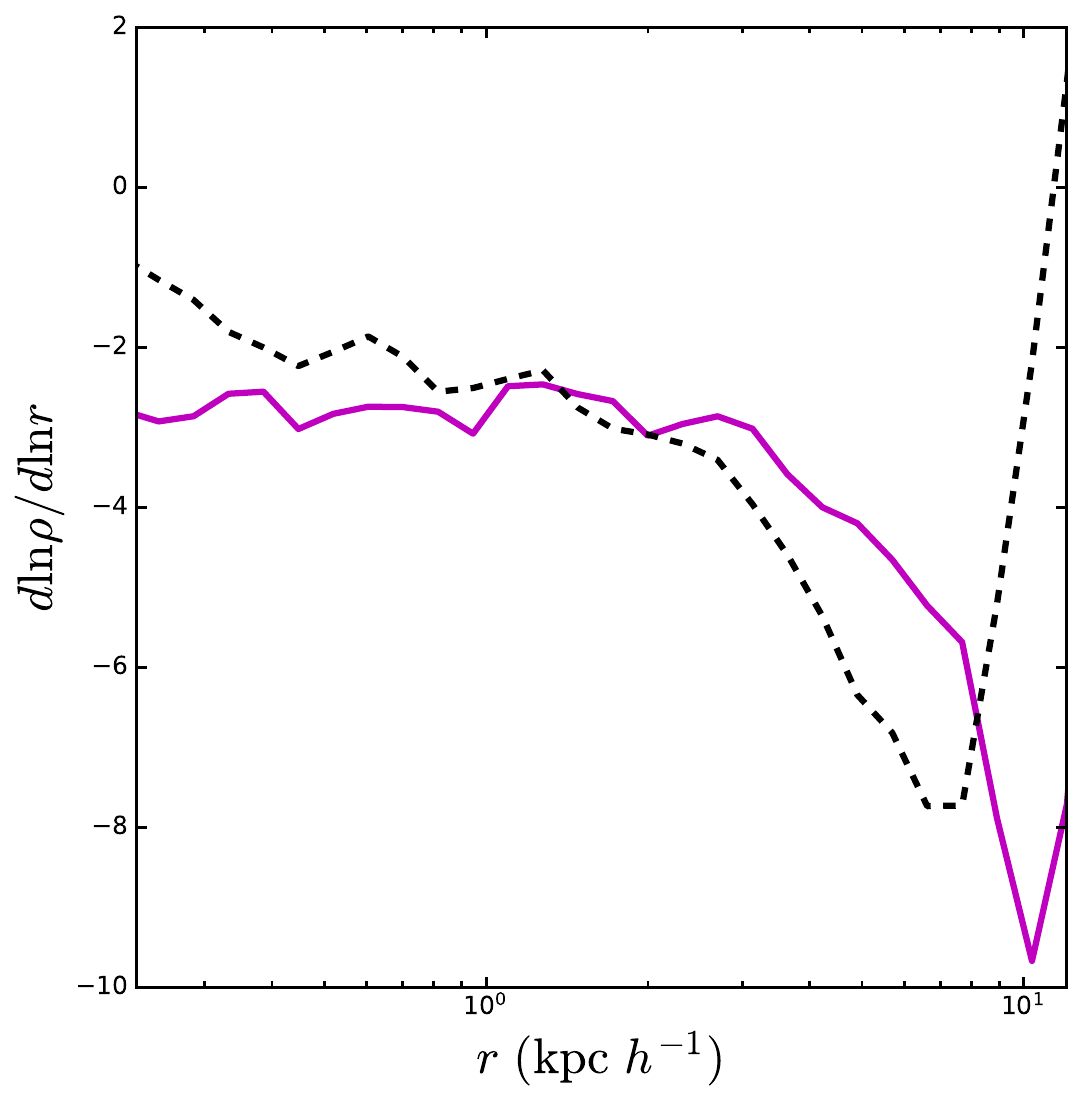}
   \includegraphics[height=4.6cm]{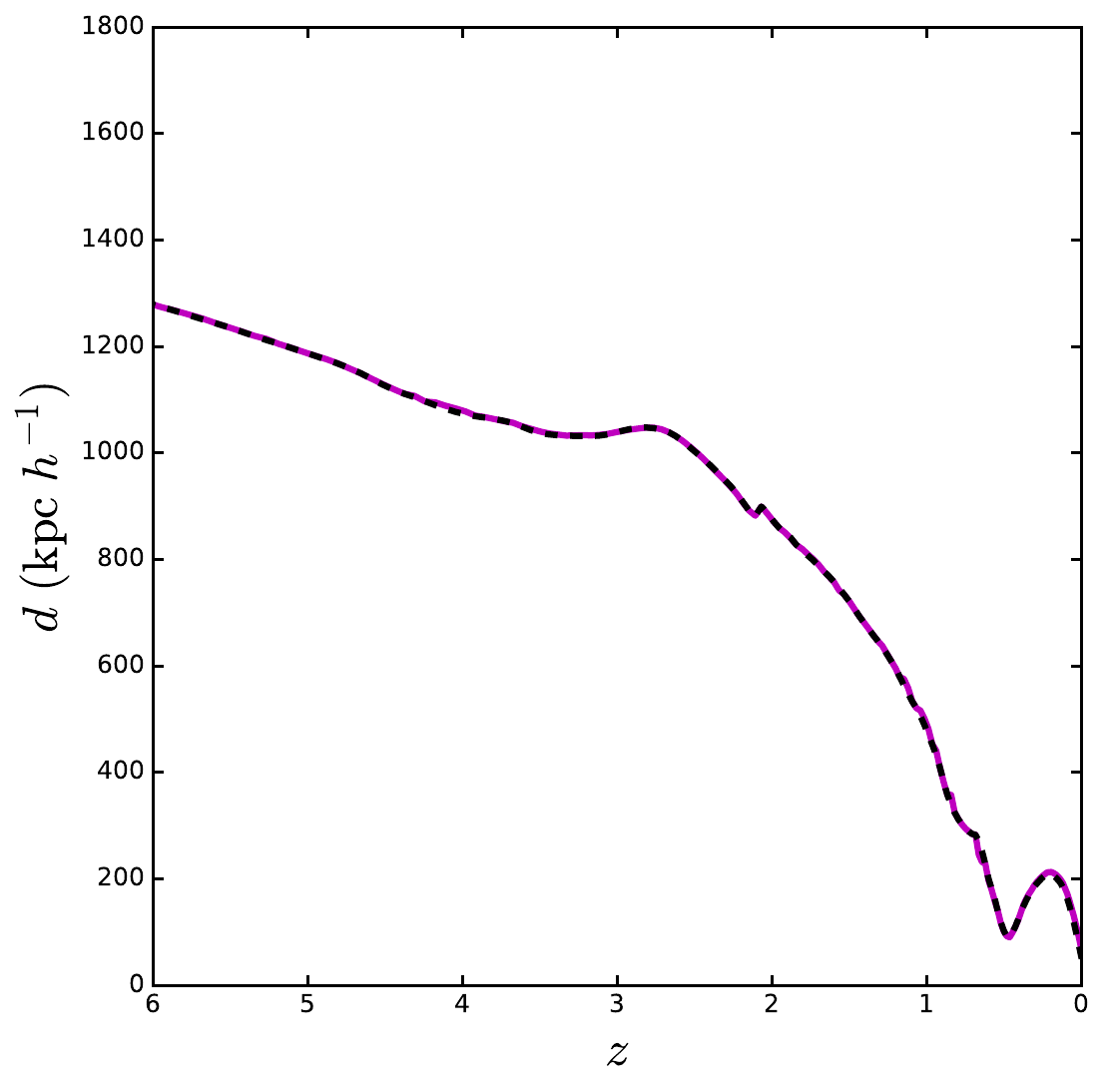}\\
  \includegraphics[height=4.6cm]{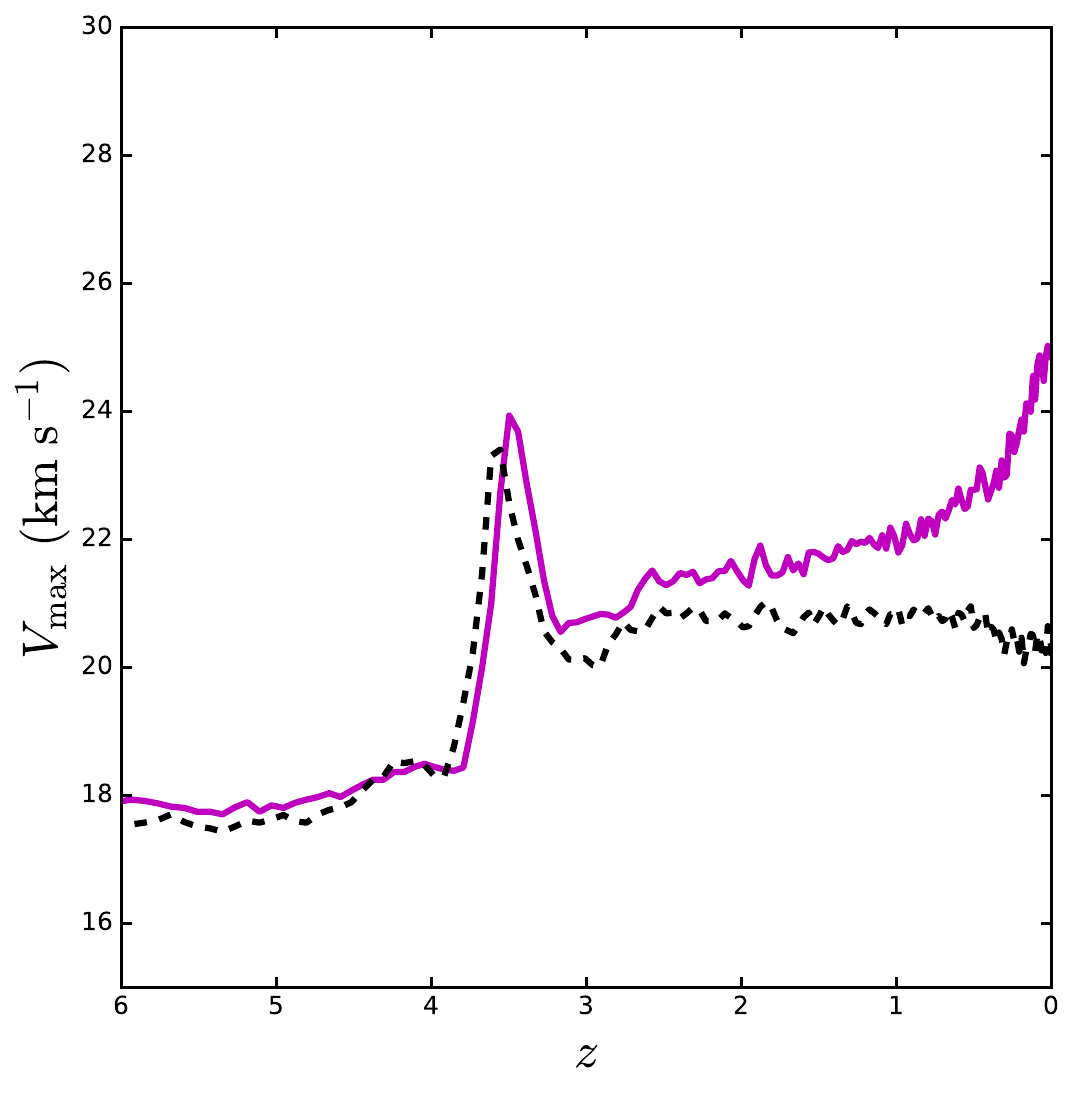}
  \includegraphics[height=4.6cm]{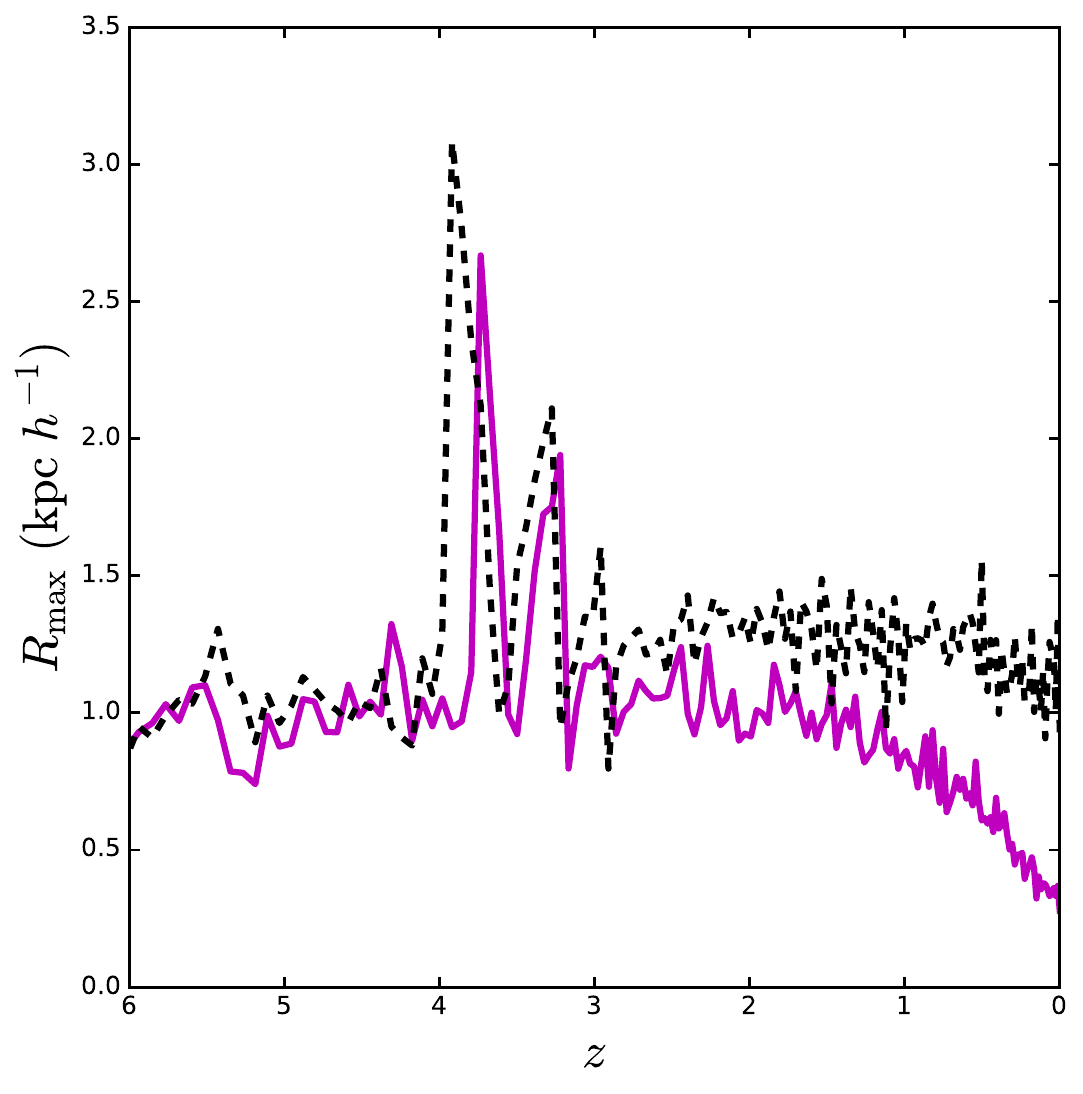}
  \includegraphics[height=4.6cm]{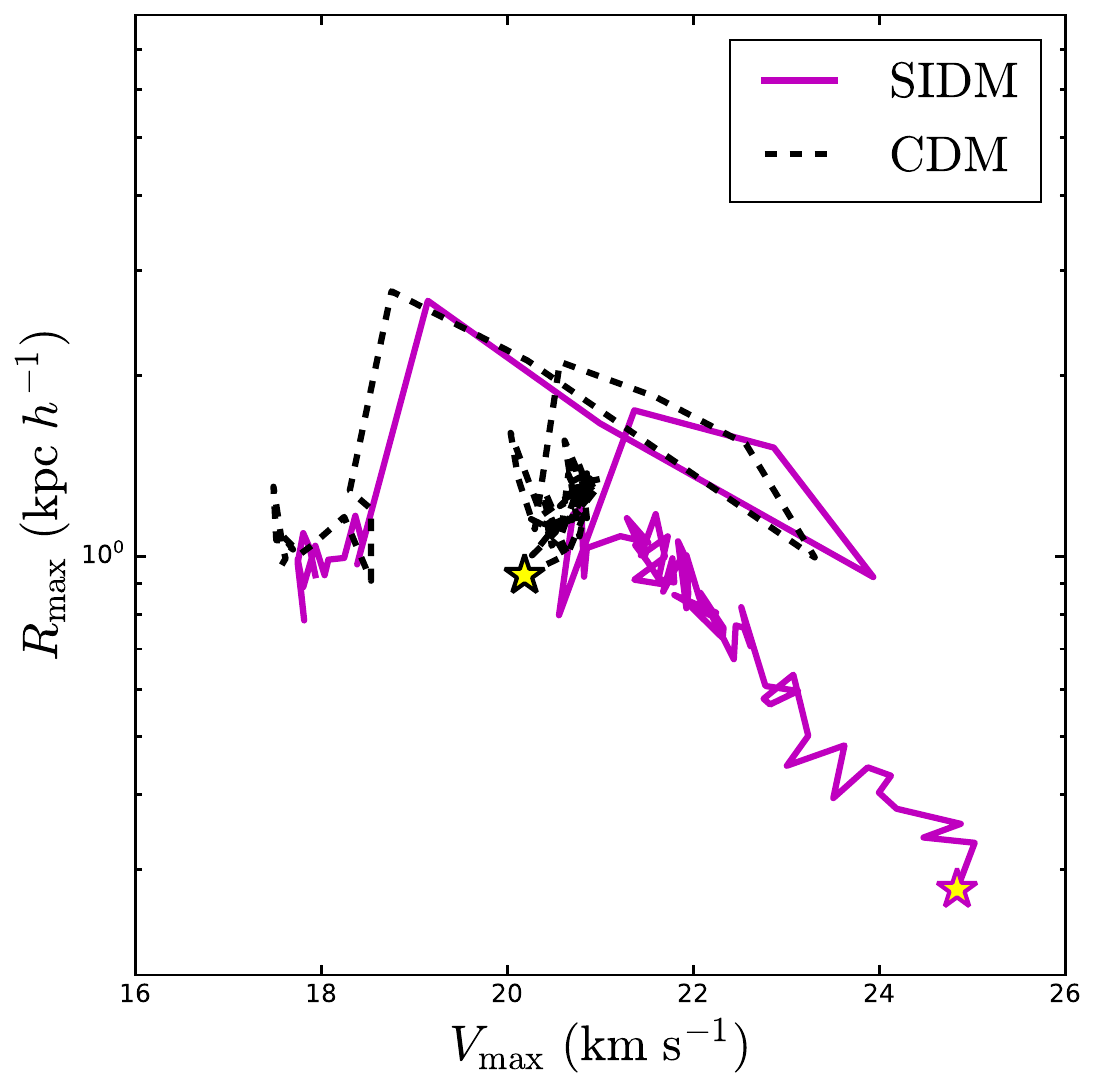}
  \caption{\label{fig:bmmw}
Same as Figure \ref{fig:bmlmc}, for a benchmark MW subhalo that shows signs of core collapse at $z=0$ in our SIDM simulation. In both simulations, this subhalo enters the MW virial radius at $z\approx 0.62, d\approx 0.212~\mpc~h^{-1}$. 
}
\end{figure*}

In the bottom panels, we show the evolution of the LMC analog halo's maximum circular velocity, $V_{\rm max}$, and the radius at which this maximum occurs, $R_{\rm max}$, as well as the relation between these two variables over time.
The core-forming LMC analog halo in our SIDM simulation exhibits a very similar $V_{\rm max}$ and $R_{\rm max}$ evolution compared to its CDM counterpart.
By visually inspecting many pairs of matched profiles, we find that core-forming halos in SIDM do not appreciably differ from their matched CDM counterparts in terms of their $V_{\rm max}$ or $R_{\rm max}$ evolution, in general.
This is expected because core formation mainly affects the circular velocity profile inside the core radius, which is in general much smaller than $R_{\rm max}$ until the final stages of halos' gravothermal evolution. 
We note that a small number of halos are actively transitioning from the core-forming to core-collapse phase at $z=0$; these halos display moderate changes in $V_{\rm max}$ and $R_{\rm max}$ compared to those of CDM. 
In our analyses of the core-collapsed halos below, we focus on systems that show \emph{unambiguous} signs of core collapse at $z=0$, in order to avoid incorrectly identifying a larger fraction of core-collapsed halos than is warranted by our simulation results.

\subsubsection{Benchmark Core-collapsed MW Subhalo}

Figure~\ref{fig:bmmw} shows similar plots as Figure~\ref{fig:bmlmc}, but for a benchmark MW subhalo that shows clear signs of core collapse at $z=0$. 
For this subhalo, we find that the CDM density profile is again better fit by an Einasto profile, where the density slope runs with radius, even at large radii, and that its logarithmic density slope approaches $\sim -1$ in the inner regions, as expected.
Unlike core-forming halos, core-collapsed halos have cuspy central density profiles, with an inner logarithmic density slope between $\sim -2$ and $-3$ \citep{Balberg0110561}; both of these features are clearly visible in the benchmark subhalo. 

We find that a clear, consistent indicator of core collapse is an enhancement in $V_{\rm max}$ and a coincident decrease in $R_{\rm max}$ at late times, as the gravothermal cusp forms and collapses to smaller radii. 
In the 2D $R_{\rm max}$--$V_{\rm max}$ planes, we mark the points at redshift zero with stars filled with yellow; near $z=0$, and particularly after infall into the MW, the SIDM subhalo sharply diverges from its CDM counterpart in this parameter space. 
Thus, the core-collapse process is likely accelerated after infall by tidal effects (e.g., \citealt{Kahlhoefer190410539,Nishikawa190100499,Sameie190407872}).

In addition, by visually inspecting the evolution histories of many halos, we consistently observe that core-collapse signatures in $V_{\mathrm{max}}$ and $R_{\mathrm{max}}$ begin prior to infall, often after major merger events in a halo's growth history (e.g., the event at $z\sim 3.5$ for our benchmark MW subhalo), and are then accelerated by tides at late times. This is perhaps surprising, because major mergers are canonically thought to reset the timescale for gravothermal evolution \citep{Yoshida0006134,Dave0006218}. However, realistic halos in a cosmological setting experience upward fluctuations in their concentration following mergers (e.g., \citealt{Wang200413732}), which may facilitate rapid subsequent gravothermal evolution even if core collapse is not dynamically linked to merger events; this mechanism may also increase the core-collapse rates of subhalos relative to isolated halos, as subhalos tend to experience more active early accretion histories due to their environments. Exploring the relationship between major mergers and core-collapse dynamics is an interesting area for future study.

In Appendix \ref{sec:additional_benchmark}, we present similar evolution histories for benchmark isolated and splashback halos that show signs of core collapse. 
Many isolated halos that show signs of core collapse, including the benchmark case studied in Appendix \ref{sec:additional_benchmark}, also undergo major mergers and/or tidal events (around host halos other than the MW) at early times, which manifest as rapid increases in their $V_{\rm max}$ histories.
Thus, major mergers and tidal evolution can play an important role in facilitating core collapse, even for halos that end up isolated today. This is particularly relevant for splashback halos, which usually experience a single pericentric passage around the MW before reaching their apocenter at $r>R_{\mathrm{vir}}$ today.
We find that SIDM splashback halos' evolution differs from its CDM evolution significantly only at late times, after these pericentric passages.

\begin{figure*}[t!]
  \centering
  \includegraphics[width=8cm]{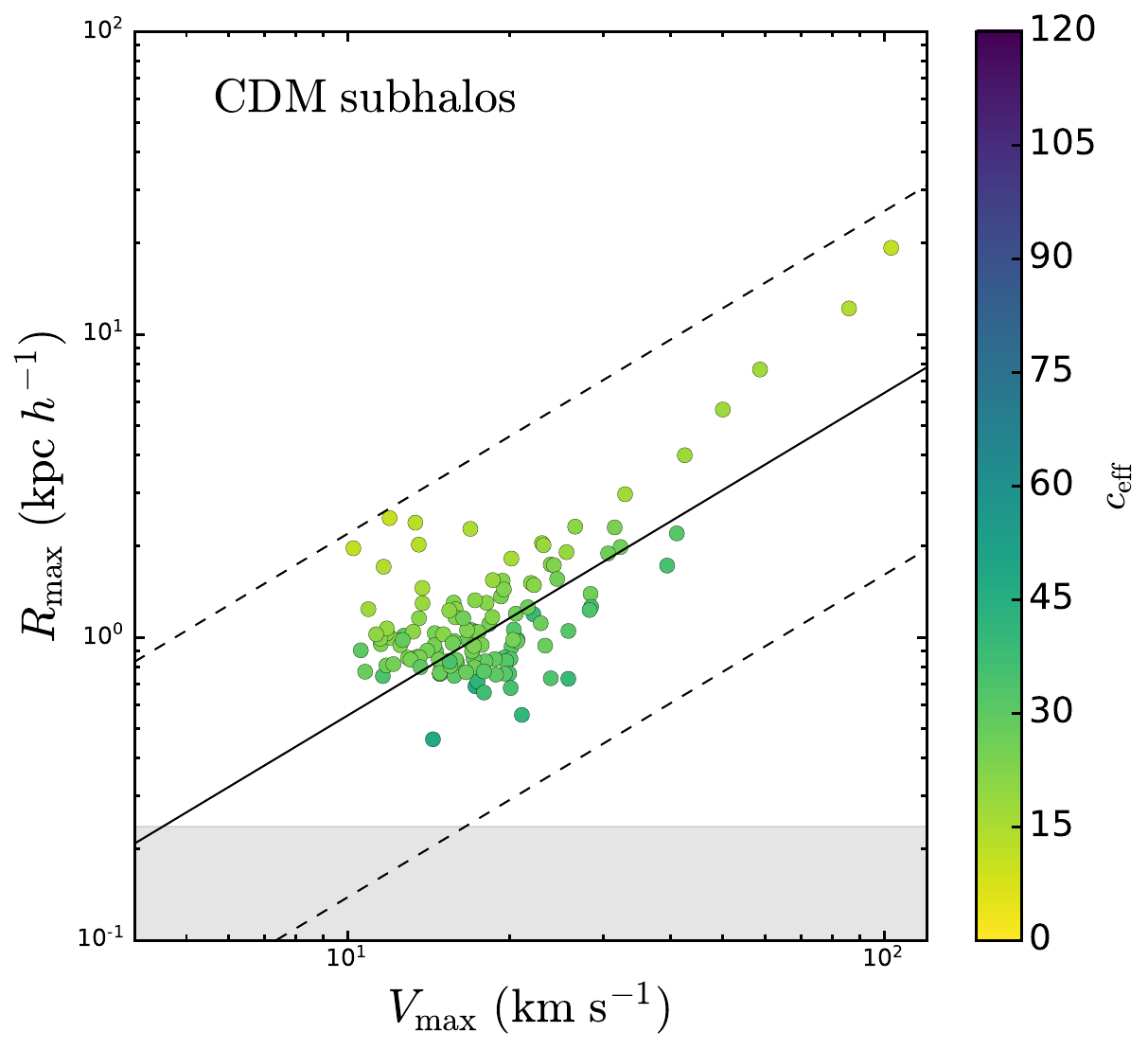}
  \includegraphics[width=8cm]{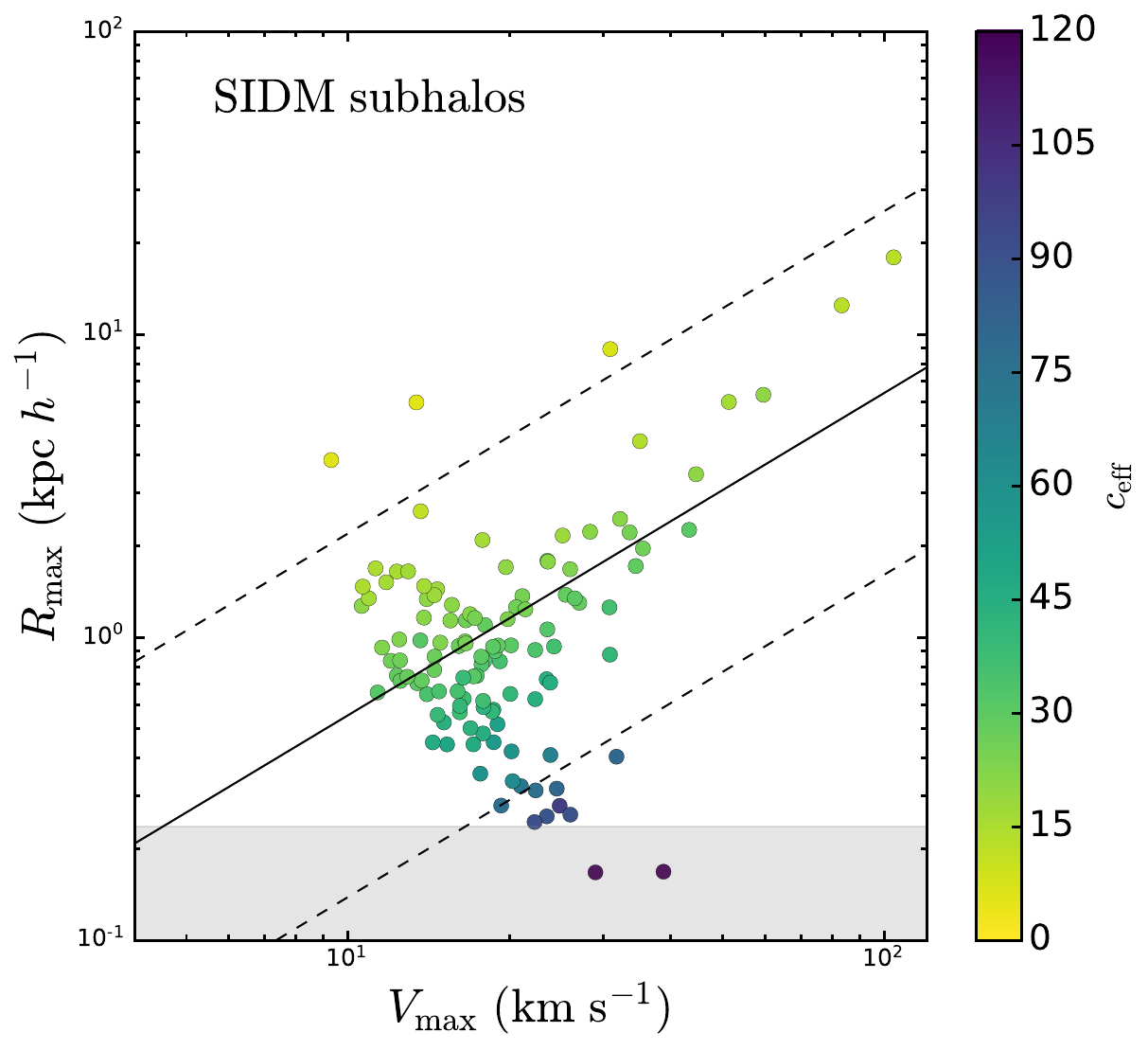}
  \\
  \includegraphics[width=8cm]{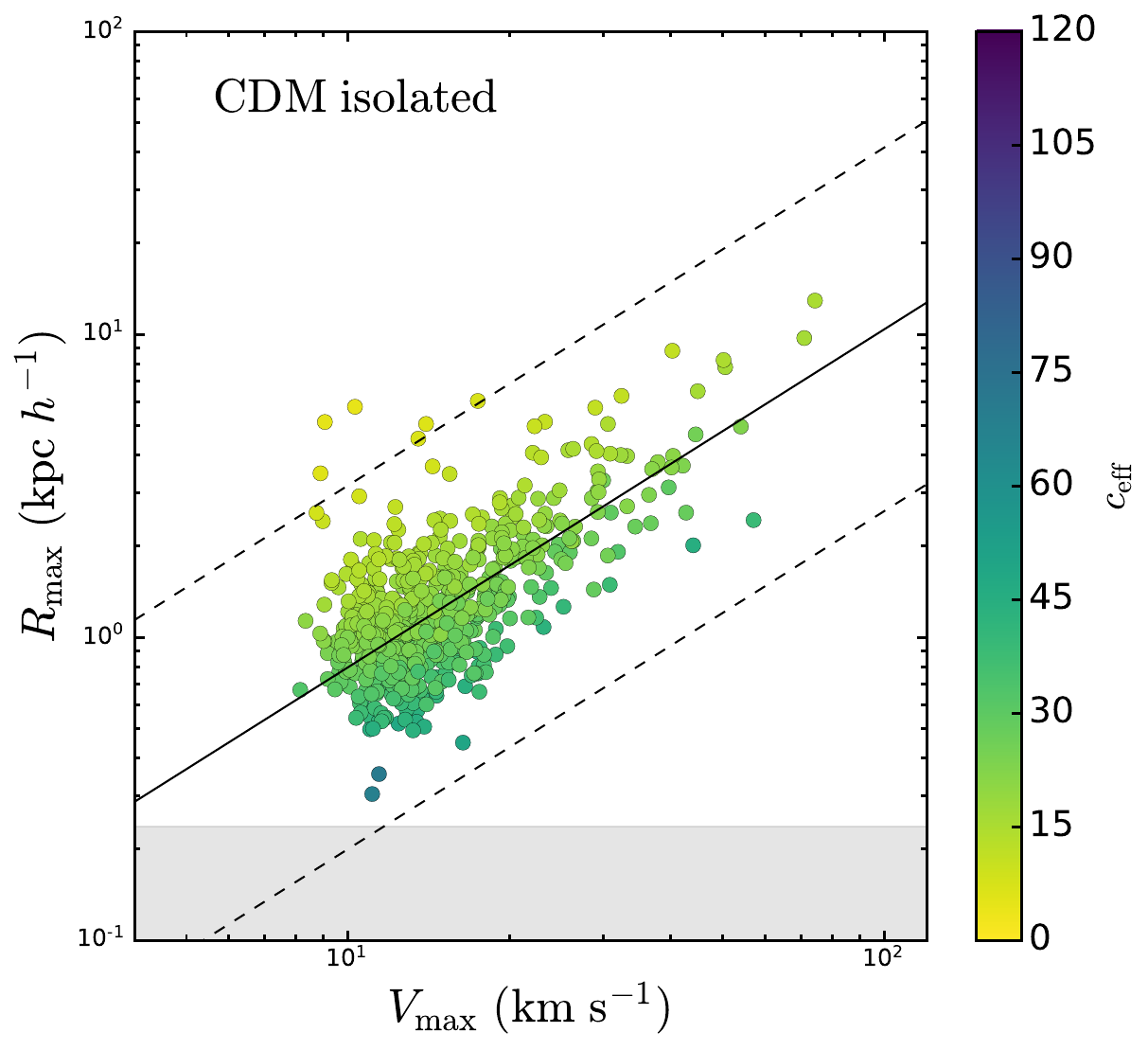}
  \includegraphics[width=8cm]{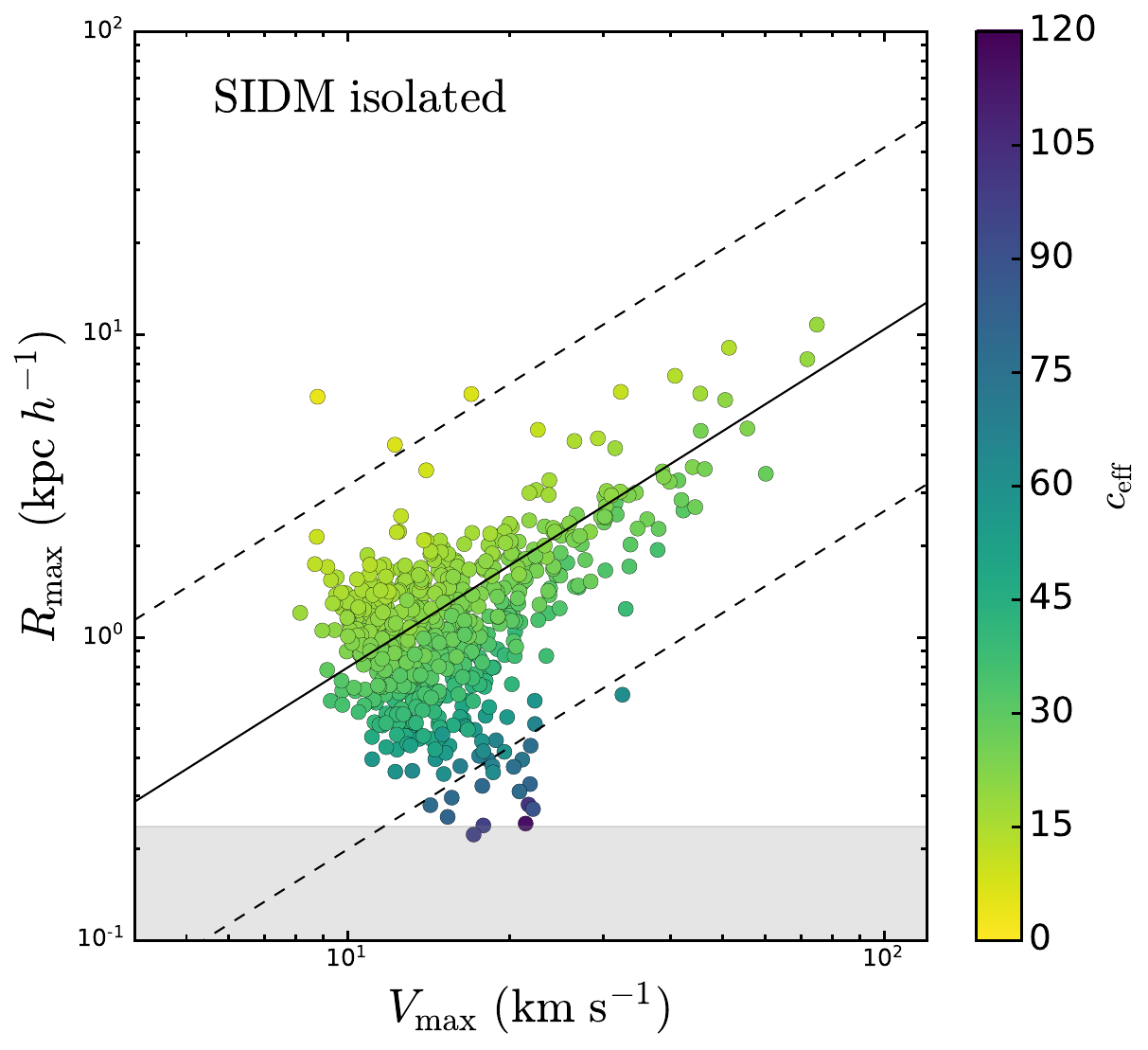}
  \caption{\label{fig:vmaxRvmax1} Top: the $R_{\rm max}$--$V_{\rm max}$ distribution for MW subhalos in our CDM (top left) and SIDM (top right) simulations. The median CDM subhalo $R_{\rm max}$--$V_{\rm max}$ relation, together with a symmetric $\pm 0.6$~dex band is shown to delineate diffuse and compact halo populations. Data points are colored according to the effective concentration $c_{\rm eff}$ of each halo. Bottom: same as the top panels, for isolated halos in our CDM (bottom left) and SIDM (bottom right) simulations. Note that the median $R_{\rm max}$--$V_{\rm max}$ relation is shown accordingly for isolated halos in these panels.} 
\end{figure*}

Thus, our SIDM simulation predicts the existence of a population of isolated, core-collapsed halos that are expected to host faint, dense galaxies. Interestingly, only $\approx 20\%$ of all isolated halos we consider previously (before $z=0$) orbited inside larger hosts; thus, even relatively quiescent formation histories are sufficient to drive core collapse in our SIDM model, although tidal interactions accelerate its onset. Observational searches for such isolated dwarf galaxies with extremely steep DM density profiles (potentially reminiscent of the Tucana dwarf galaxy) are particularly interesting in this context.

\subsection{SIDM Features in the $R_{\rm max}$--$V_{\rm max}$ plane}
\label{sec:sidm_features} 

\begin{figure*}[t!]
  \centering
  \includegraphics[width=7.2cm]{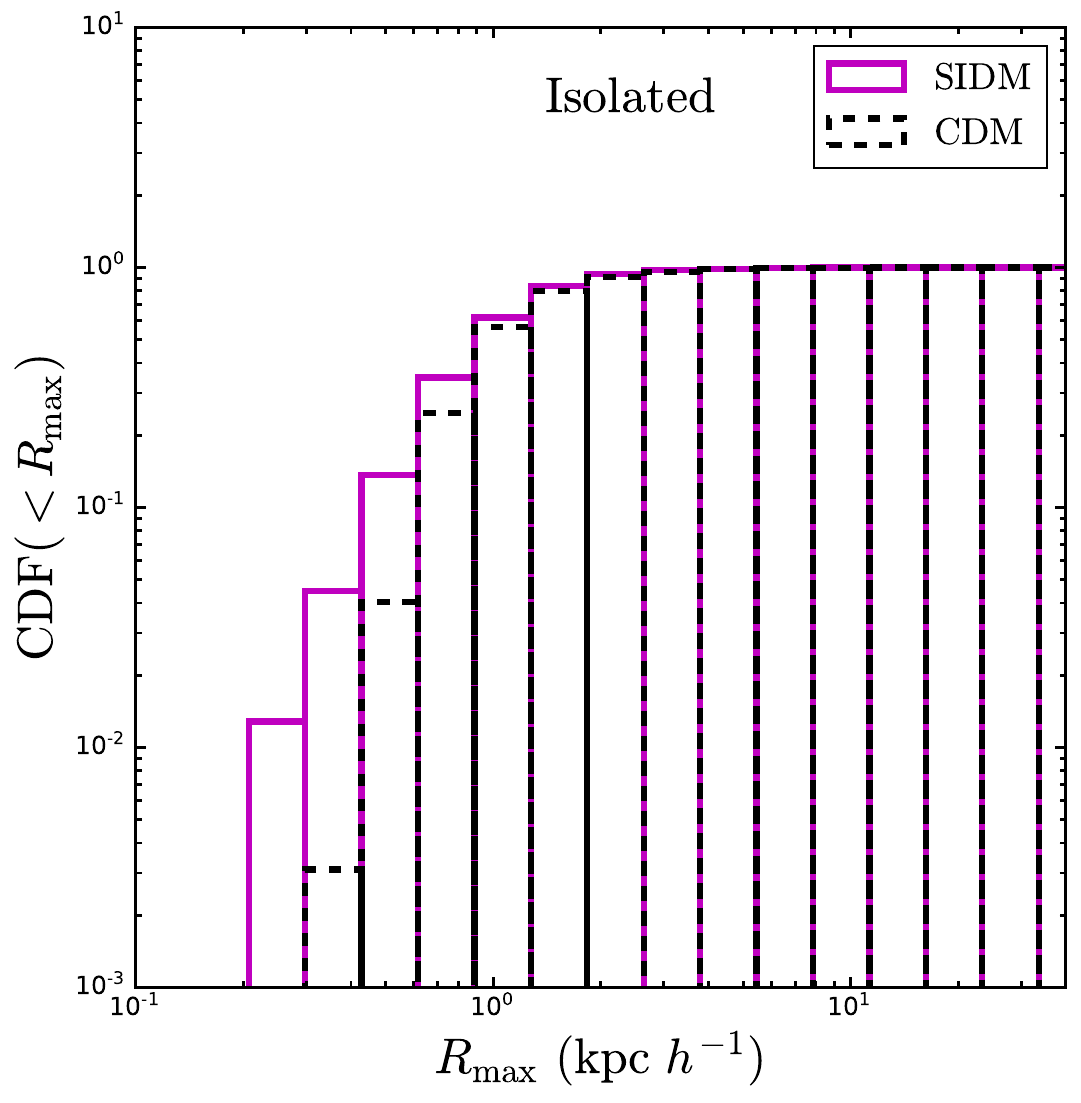}
  \includegraphics[width=7.2cm]{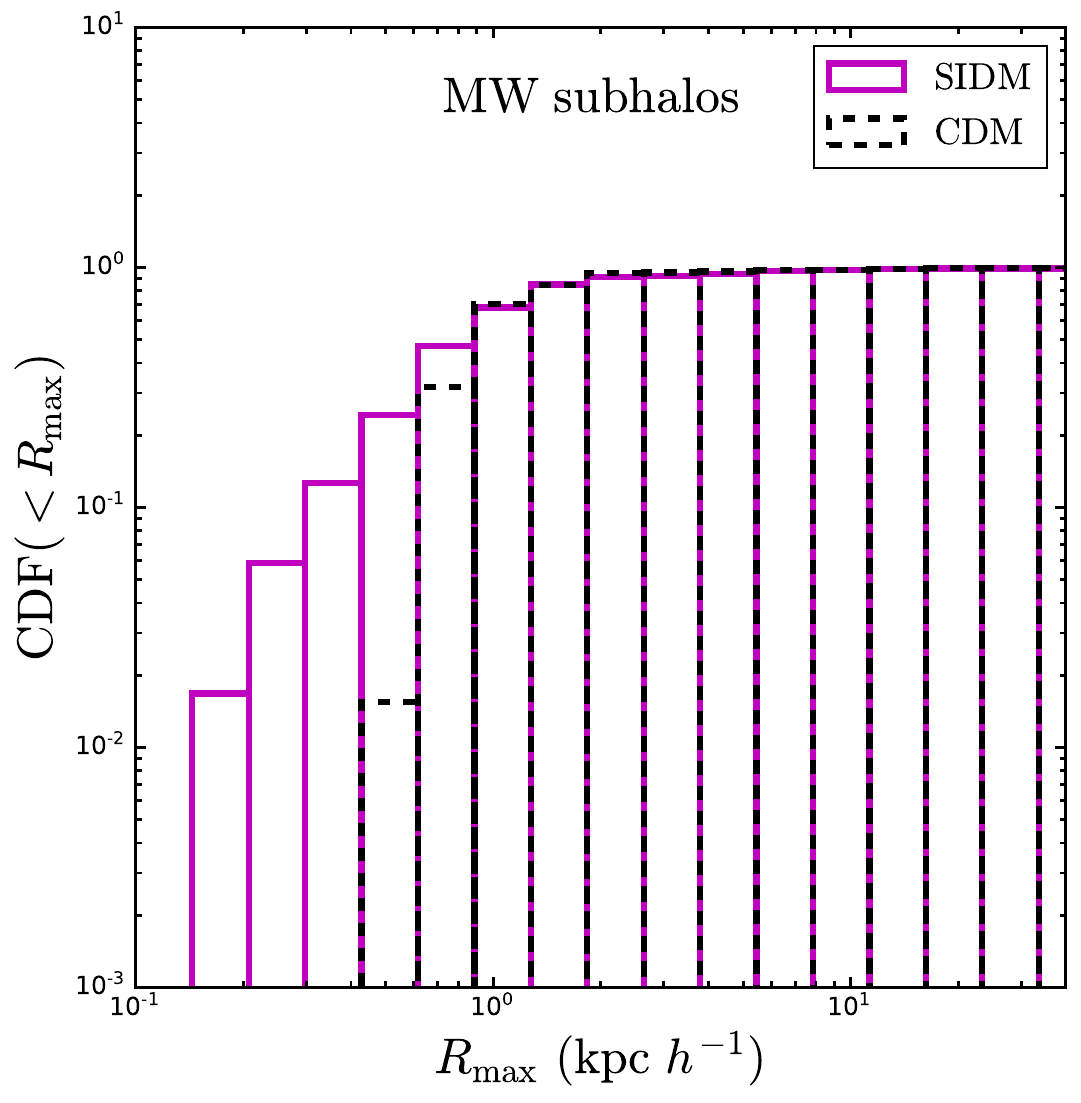}
  \caption{\label{fig:hmaxcum} The cumulative $R_{\rm max}$ distribution function of isolated halos (left) and MW subhalos (right).
  The SIDM results (magenta), compared to the CDM results (black), show excesses at smaller $R_{\rm max}$. These small-$R_{\rm max}$ halos are expected to have core collapsed, as confirmed by our case studies and analytic effective cross section model.}
\end{figure*}

As suggested by Figure~\ref{fig:bmmw}, core-collapsed halos are shifted toward larger $V_{\mathrm{max}}$ and smaller $R_{\mathrm{max}}$ than their CDM counterparts, with a more significant shift in $R_{\mathrm{max}}$. 
Thus, to study the statistical impact of SIDM on halo properties, we consider the distribution of halos in the $R_{\rm max}$--$V_{\rm max}$ plane. 
In Figure~\ref{fig:vmaxRvmax1}, we show the $R_{\rm max}$--$V_{\rm max}$ distribution of all MW subhalos (top) and isolated halos (bottom) that have masses higher than $10^8~\msun~h^{-1}$ in our CDM (left) and SIDM (right) simulations. For both isolated halos and subhalos in SIDM, a population of objects moves below the lower dashed curve, indicating that they are likely core collapsed. 
Isolated core-forming halos shift to slightly lower $R_{\mathrm{max}}$ relative to CDM, as suggested by Figure~\ref{fig:bmlmc}, while some diffuse MW subhalos shift toward \emph{larger} $R_{\mathrm{max}}$, indicating enhanced tidal stripping of these systems compared to their CDM counterparts.

To better study the difference between the CDM and SIDM halo populations, we plot the median relation in the CDM cases, together with symmetric $\pm 0.6$~dex bands (see Appendix \ref{sec:fitting} for the numerical details of our $R_{\rm max}$--$V_{\rm max}$ fit). 
We define the core-collapsed candidates as halos that reside below the $-0.6$~dex curve from the median relation. 
Applying this selection, we find 18 isolated core-collapsed halos (two of which are splashback halos) and
10 core-collapsed MW subhalos (one of which is an LMC-associated subhalo). In Figure~\ref{fig:hmaxcum}, we show the cumulative probability distribution functions (CDF) of $R_{\rm max}$ for isolated halos (left) and subhalos (right). The excess of SIDM over CDM at low $R_{\rm max}$ approximately indicates the rate of isolated and MW subhalo core-collapsed candidates.
In particular, using a conservative threshold of $R_{\mathrm{max}}<300~\pc~h^{-1}$ to pick out core-collapsed candidates (see Figure~\ref{fig:vmaxRvmax1}), we can read off that roughly $10\%$ ($20\%$) of the isolated (MW subhalo) population has potentially core collapsed. The core-collapse rate is higher for subhalos because tidal interactions with the MW accelerate gravothermal evolution (e.g., \citealt{Kahlhoefer190410539,Nishikawa190100499,Sameie190407872}), as suggested by our benchmark study in Figure~\ref{fig:bmmw}.

In Figure~\ref{fig:vmaxRvmax1}, we also color code the points by their corresponding effective concentrations, defined as 
\begin{equation}c_{\rm eff} \equiv \frac{R_{\rm vir}}{R_{\rm max}/2.1626}.
\end{equation}
This quantity can be evaluated for both the CDM and SIDM halos and reduced to the common definition $c_s=R_{\rm vir}/R_s$ for NFW halos. 
From the gradient in the $c_{\rm eff}$, we see that SIDM effects are strongly correlated with halo concentration, in agreement with many previous studies \citep{Kaplinghat150803339,Kamada161102716,Essig180901144,Zeng211000259}. In particular, the $c_{\mathrm{eff}}$ gradient in the $R_{\rm max}$--$V_{\rm max}$ plane is almost perpendicular to the underlying relation, approximately pointing in the direction that differentiates core-collapsed and core-forming SIDM halos. 

Aside from the core-collapsed candidates discussed above, we find a few isolated halos and MW subhalos residing to the upper left of the $+0.6$~dex band from the median $R_{\rm max}$--$V_{\rm max}$ relation in both our CDM and SIDM simulations. 
Due to their large sizes and slowly rising rotation curves, the isolated halos in this region of parameter space with $V_{\mathrm{max}}\gtrsim 30~\km~\mathrm{s}^{-1}$ can potentially host ultra-diffuse field dwarf galaxies (UDGs; e.g., see \citealt{Jiang181110607}). 
However, due to the $(3~\mpc)^3$ volume of the high-resolution region that we study halos within, we cannot statistically probe the observed UDG populations in the current study; in future work, it will therefore be interesting to consider larger zoom-in volumes in different environments to compare with halo properties inferred for samples of observed UDGs in the field (e.g., \citealt{Kong220405981,PinaMancera211200017}). 

At the low-$V_{\mathrm{max}}$ end, our simulations predict that a handful of diffuse low-mass halos may host faint, low-density isolated dwarf galaxies in the Local Volume. 
These systems may be field analogs of the lowest surface brightness MW satellite galaxies (e.g., Crater 2 and Antlia 2; \citealt{Torrealba160107178,Torrealba181104082}). 
We do not observe significant differences between the isolated CDM and SIDM halo populations in the diffuse, high-$R_{\rm max}$ region of the $R_{\rm max}$--$V_{\rm max}$ plane; this is expected, as those halos have low concentrations and suppressed self-interaction rates~\citep{Kong220405981}.

On the other hand, we find that several SIDM subhalos remain in (or evolve to) very diffuse states, whereas no comparably diffuse CDM subhalos exist in our simulation.
This may result from mass loss due to ram pressure stripping as a consequence of subhalo--host halo interactions in our SIDM simulation (e.g., \citealt{Dooley160308919,Nadler200108754}).
At the same time, cored SIDM subhalos may be more efficiently tidally heated, particularly because our MW host has a slightly higher density at intermediate radii compared to CDM to compensate for its central core \citep{Banerjee190612026,Slone210803243}; all of these effects underscore the diversity and environmental dependence of SIDM (sub)halo populations.
We investigate SIDM effects on diffuse halos further in Section~\ref{sec:udg}. 

\subsection{Analytic Predictions for the Core-collapsed Population}
\label{sec:sidm_identification}

To provide theoretical context for the core-collapsed candidate halos identified with large $V_{\mathrm{max}}$ and small $R_{\mathrm{max}}$ above, we analytically estimate the expected number of core-collapsed halos expected in our SIDM model as follows. 
First, we reduce our differential cross section into an effective constant cross section following~\cite{Yang220503392}, \cite{Yang220502957}, and \cite{Outmezguine220406568}: 
\begin{align}
\label{eq:eff}
\sigma_{\rm eff} &= \frac{1}{512 \nu_{\rm eff}^8 } \int v^2 d v d\cos\theta \frac{d \sigma}{d\cos\theta} v^5 \sin^2\theta \exp \left[-\frac{v^2}{4\nu_{\rm eff}^2}\right],&
\end{align}
where $\nu_{\rm eff} =0.64 V_{\rm max, NFW} \approx 1.05 r_{\rm eff} \sqrt{G\rho_{\rm eff}}$ is a characteristic velocity dispersion scale, with $r_{\rm eff} = R_{\rm max}/2.1626$, and $\rho_{\rm eff} = (V_{\rm max}/(1.648 r_{\rm eff}))^2/G$. 
Both $r_{\rm eff}$ and $\rho_{\rm eff}$ reduce to the $r_s$ and $\rho_s$ parameters for NFW halos.
With $\sigma_{\rm eff}$ calculated for every halo in our CDM simulation, we then estimate core-collapse timescales according to
\begin{eqnarray}
\label{eq:tc}
t_{\rm c}  &=& \frac{150}{\beta} \frac{1}{(\sigma_{\rm eff}/m) \rho_{\rm eff} r_{\rm eff}} \frac{1}{\sqrt{4\pi G \rho_{\rm eff}}}, \\ \nonumber
\end{eqnarray}
where $\beta\approx 0.75$ for an NFW halo (see \citealt{Pollack150100017,Essig180901144}).

\begin{figure*}[t!]
  \centering
  \includegraphics[width=7.2cm]{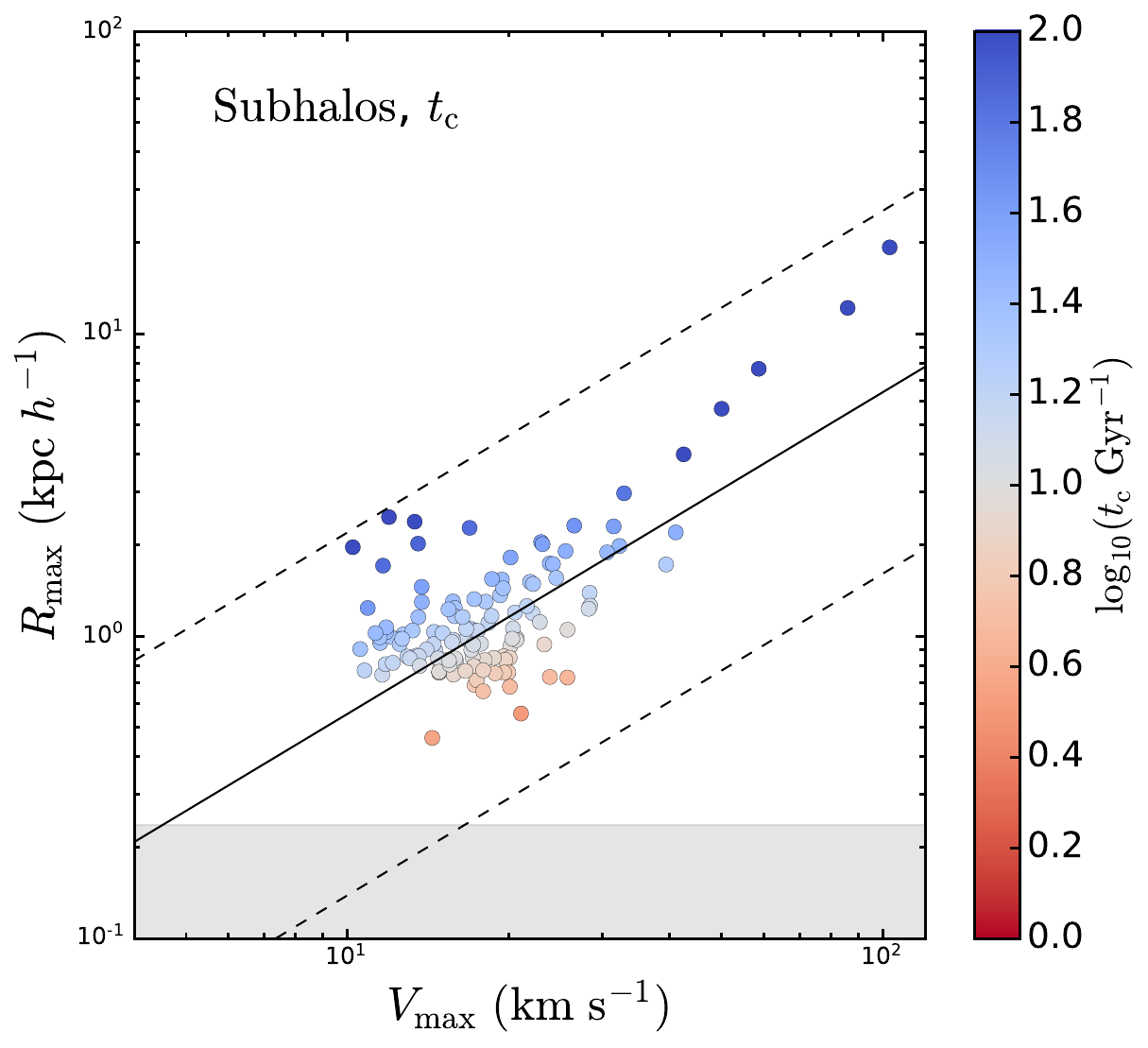}
  \includegraphics[width=7.2cm]{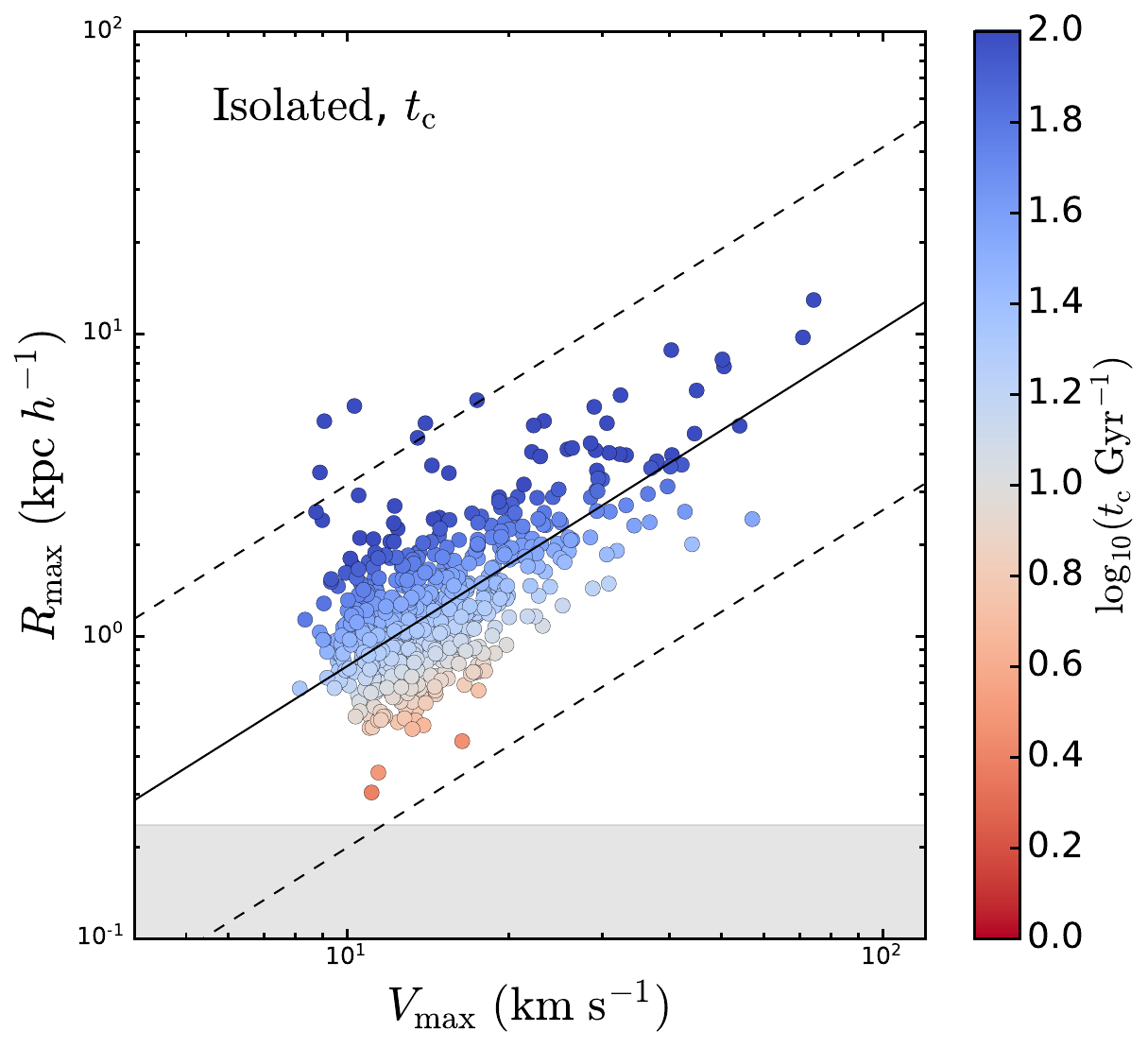}
  \caption{\label{fig:tcbarCDM} Core-collapse timescales estimated for CDM subhalos (left) and isolated halos (right) using their $z=0$ halo properties. Halos with $t_c$ shorter than the age of the universe (in the lower right areas of these distributions) are expected to core collapse by $z=0$, consistent with our SIDM simulation results shown in Figure~\ref{fig:vmaxRvmax1}. 
}
\end{figure*}

We use this model to estimate the distribution of $\sigma_{\rm eff}$ for our isolated halo and subhalo CDM populations. 
We find that the $\sigma_{\rm eff}$ distributions peak at large values, $\sigma_{\mathrm{eff}}\gtrsim 10~\cm^2~\mathrm{g}^{-1}$, because they are dominated by low-mass halos (for which our cross section is large; see Figure~\ref{fig:bmxs}) due to the underlying halo mass function. 
The resulting predictions for $t_{\rm c}$ are shown in Figure~\ref{fig:tcbarCDM}.\footnote{Note that we compute $t_{\rm c}$ using CDM (rather than SIDM) halo properties because Equation~\ref{eq:tc} is calibrated on $\rho_{\rm eff}$ and $r_{\rm eff}$ values derived from NFW profiles, which do not accurately describe SIDM halos.}
We find that $t_{\rm c}$ is shorter than the age of the universe for CDM halos with high concentration and small $R_{\mathrm{max}}$, implying that these halos' SIDM counterparts should core collapse by $z=0$. 
In particular, counting the number of halos with $t_{\rm c}<10~$Gyr, we find $58$ isolated halos, $31$ MW subhalos, $1$ LMC subhalo, and $5$ splashback halos. 
Thus, the number of core-collapsed halos predicted by our analytic model is roughly double of that found in our SIDM simulation using the $V_{\mathrm{max}}$--$R_{\mathrm{max}}$ criteria described in the previous section. 
For this reason, we argue that our fiducial core-collapsed criteria based on $V_{\mathrm{max}}$ and $R_{\mathrm{max}}$ is conservative, in that it selects the most extreme core-collapsed objects. 

\section{Population Statistics of Halos within and surrounding the Milky Way}
\label{sec:population_stats}

Having identified the effects of self-interactions on specific halo properties, we now compare various observationally relevant properties of MW subhalos and isolated halos in our CDM and SIDM simulations. We also discuss the general trends displayed by our subhalos and isolated halos, in the context of observed MW satellite galaxies and field dwarf galaxies, respectively. We note that our predictions should not be interpreted directly in the space of observables; such predictions would require forward-modeling that accounts for observational selection effects and systematic uncertainties. Furthermore, our simulations do not include baryons or a central MW-like galaxy, both of which affect such comparisons, particularly for MW subhalos. Nevertheless, our study provides a benchmark case for highlighting the effects of SIDM with respect to the CDM limit in light of observations. Note that, for isolated halos toward the lower-mass end, we expect that the baryonic feedback has a negligible effect on the halo density profile. 

\subsection{Mass Functions}

\begin{figure*}[t!]
  \centering
  \includegraphics[width=7.2cm]{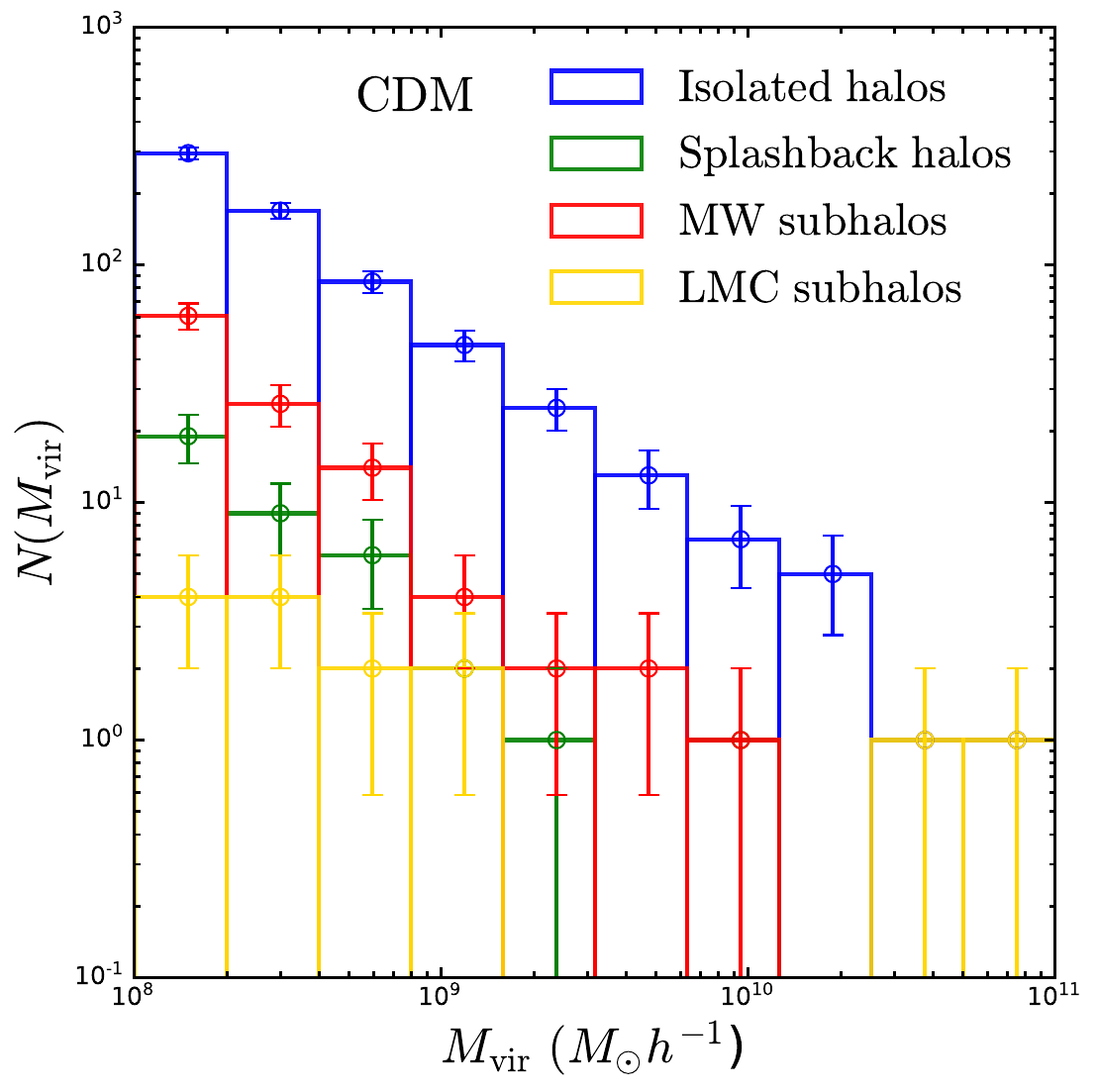}
  \includegraphics[width=7.2cm]{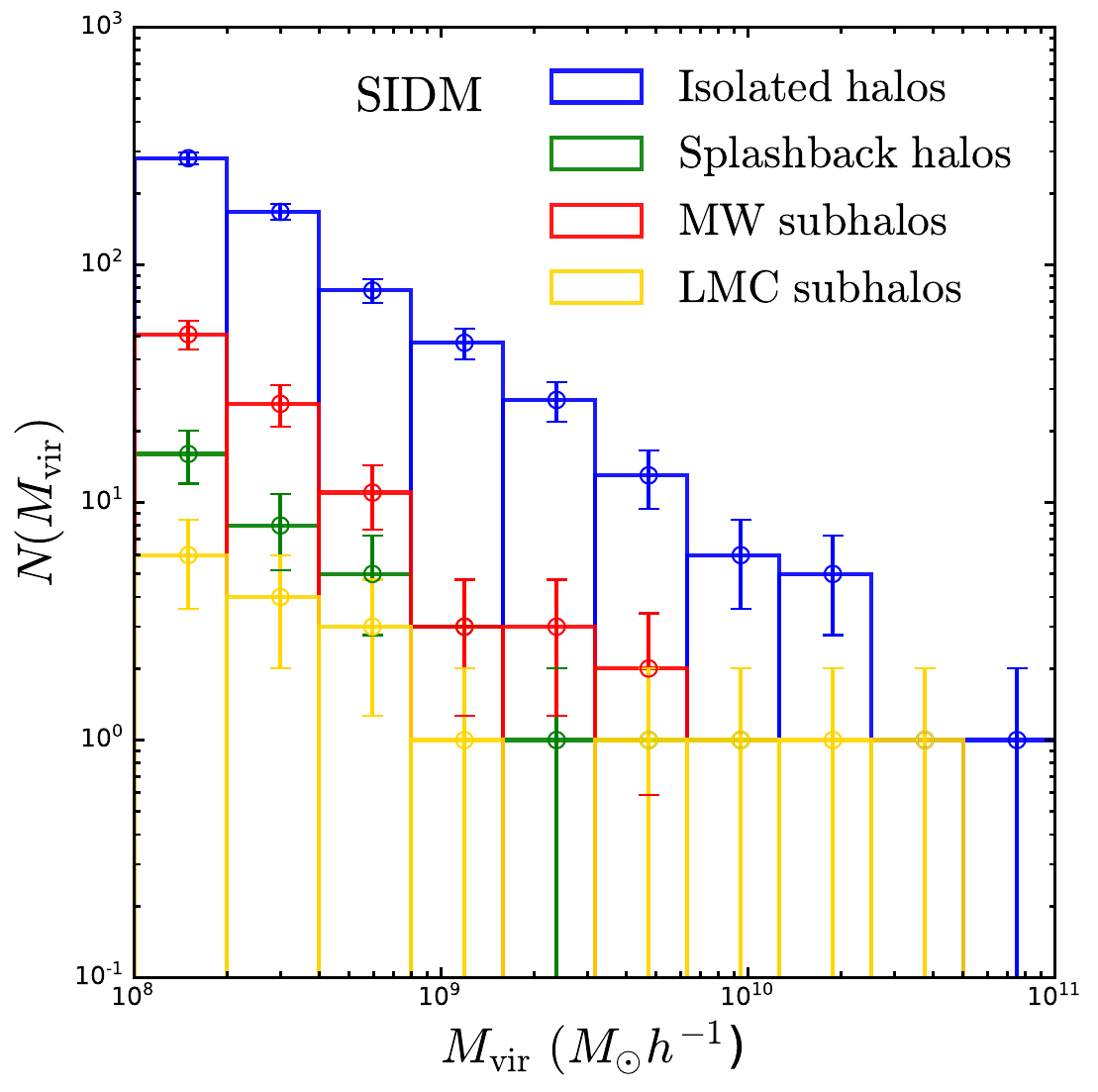}
  \caption{\label{fig:shmf} Mass functions for isolated halos (blue), MW subhalos (red), and LMC-associated subhalos (magenta) in our CDM (left) and SIDM (right) simulations.
  }
\end{figure*}

Figure~\ref{fig:shmf} compares the CDM and SIDM mass functions for halos in our four main categories.
Because of the volume of our high-resolution region, which extends to $\approx 10R_{\mathrm{vir}}$ of the MW host halo, there are significantly more isolated halos than subhalos.
All mass functions rise toward low halo masses, with slopes consistent with previous measurements in zoom-in simulations (e.g., \citealt{Nadler220902675}).
Interestingly, for halos down to our fiducial mass resolution limit of $10^8~\msun~h^{-1}$, the SIDM mass function is nearly unchanged compared to the CDM case. 
This is expected for isolated halos, which are only affected by internal self-interactions and are thus not subject to mass loss or disruption. 

On the other hand, the similarity between our CDM and SIDM \emph{subhalo} mass functions is nontrivial. As discussed extensively in \cite{Nadler200108754}, self-interactions can potentially affect the subhalo mass function at a significant level due to ram pressure stripping caused by self-interactions between subhalo and host halo particles. 
This mechanism depends on the amplitude of the cross section at the typical relative velocity between subhalo and host halo particles, $\approx 200~\km~\mathrm{s}^{-1}$ for the MW host we consider here. At this velocity scale, our momentum transfer cross section is much smaller than $1~\cm^2~\mathrm{g}^{-1}$ (see Figure~\ref{fig:bmxs}), so it is plausible that evaporation is a small effect for typical subhalos in our simulations, consistent with other previous studies (e.g., \citealt{Dooley160308919,Slone210803243,Zeng211000259}). Another contributing factor may be that a fraction of low-mass subhalos in our SIDM simulation is core-collapsed (or at least in the core-collapse phase), and is thus more resilient to tidal disruption than its core-forming counterpart. In this context, it is reassuring that our comparison between the CDM and SIDM mass functions agrees reasonably well with those of \cite{Turner201002924}, who performed a zoom-in simulation of the same SIDM model.

\subsection{MW Subhalo Rotation Curves}

MW satellite galaxies are expected to reside in halos with sufficiently high peak virial temperatures, e.g., above the atomic hydrogen cooling limit of $10^4~$K, which corresponds to a peak maximum circular velocity of $V_{\rm peak}\approx 16.3~\rm km/s$ (\citealt{Graus180803654}; also see \citealt{Nadler191203303} and references therein).
Meanwhile, the stellar velocity dispersion measurements suggest that many of the observed MW satellites occupy halos with $V_{\rm max}<30~\rm km/s$ today, although there are large observational uncertainties in the halo properties for many of these systems (e.g., see \citealt{Bullock170704256,Sales220605295} for reviews). We use these ranges of halo properties to hone in on the subhalos expected to host observable satellite galaxies in our simulations.

\begin{figure*}[htbp]
  \centering
  \includegraphics[width=7.2cm]{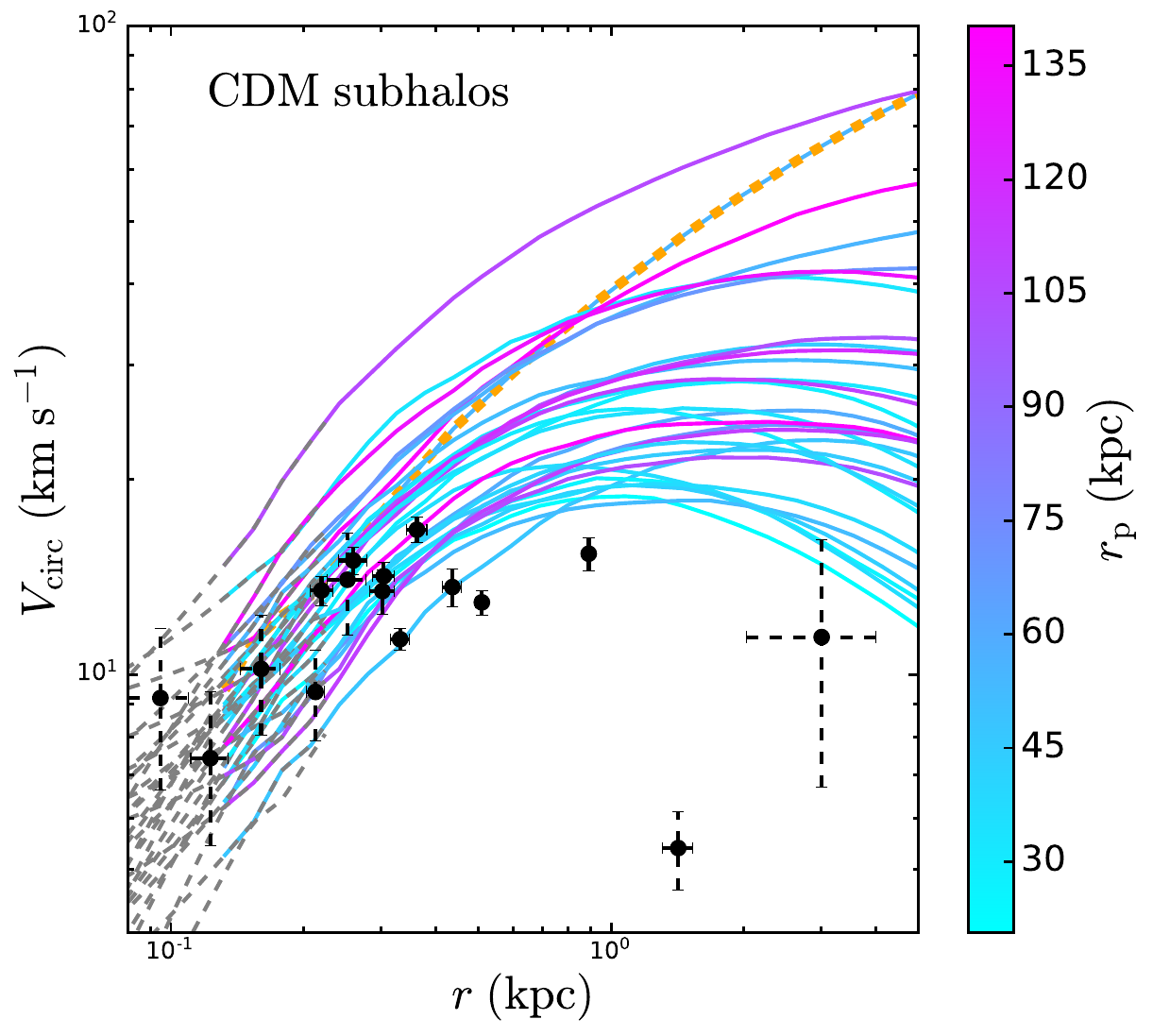}
  \includegraphics[width=7.2cm]{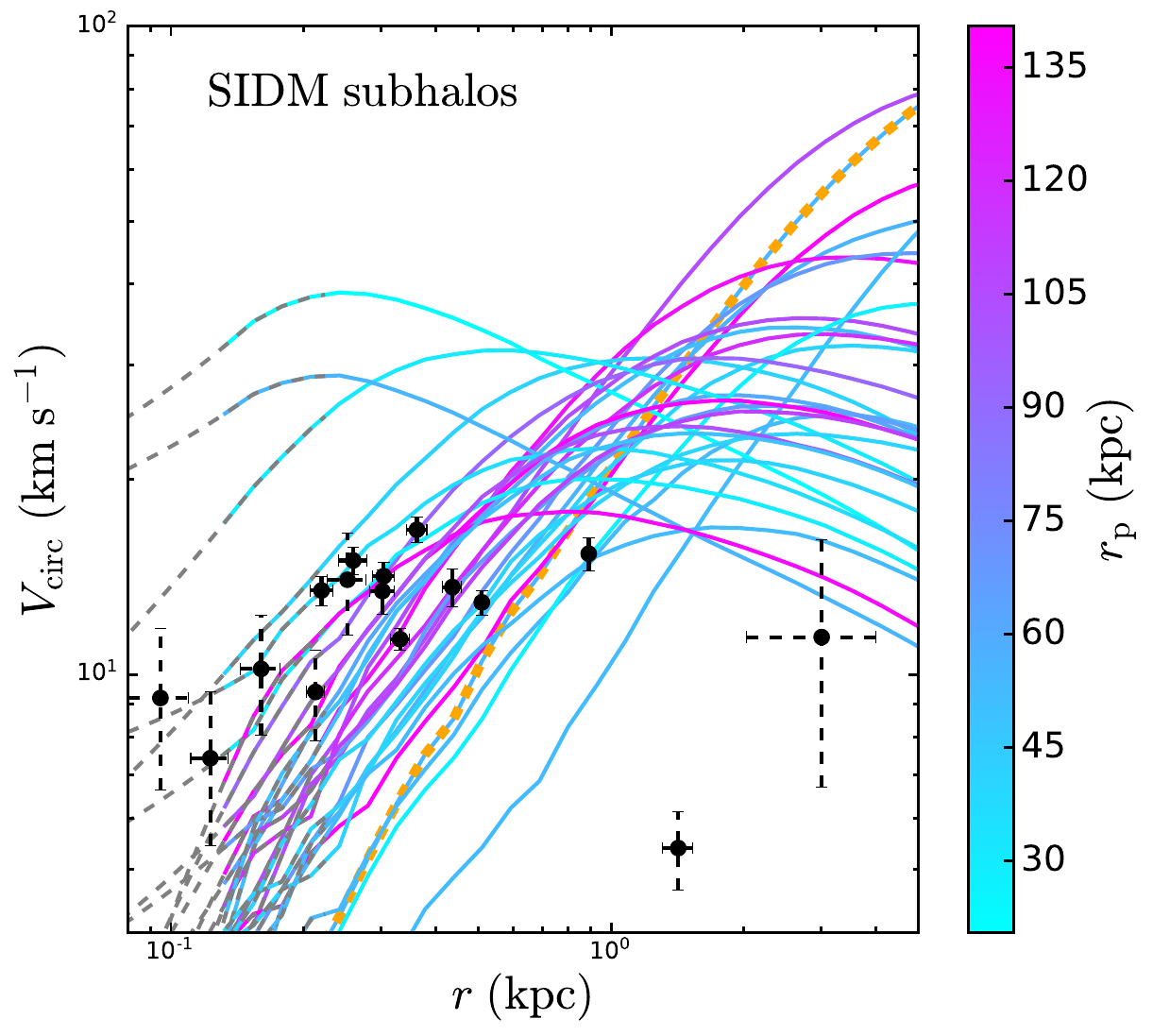}
  \\
  \includegraphics[width=7.2cm]{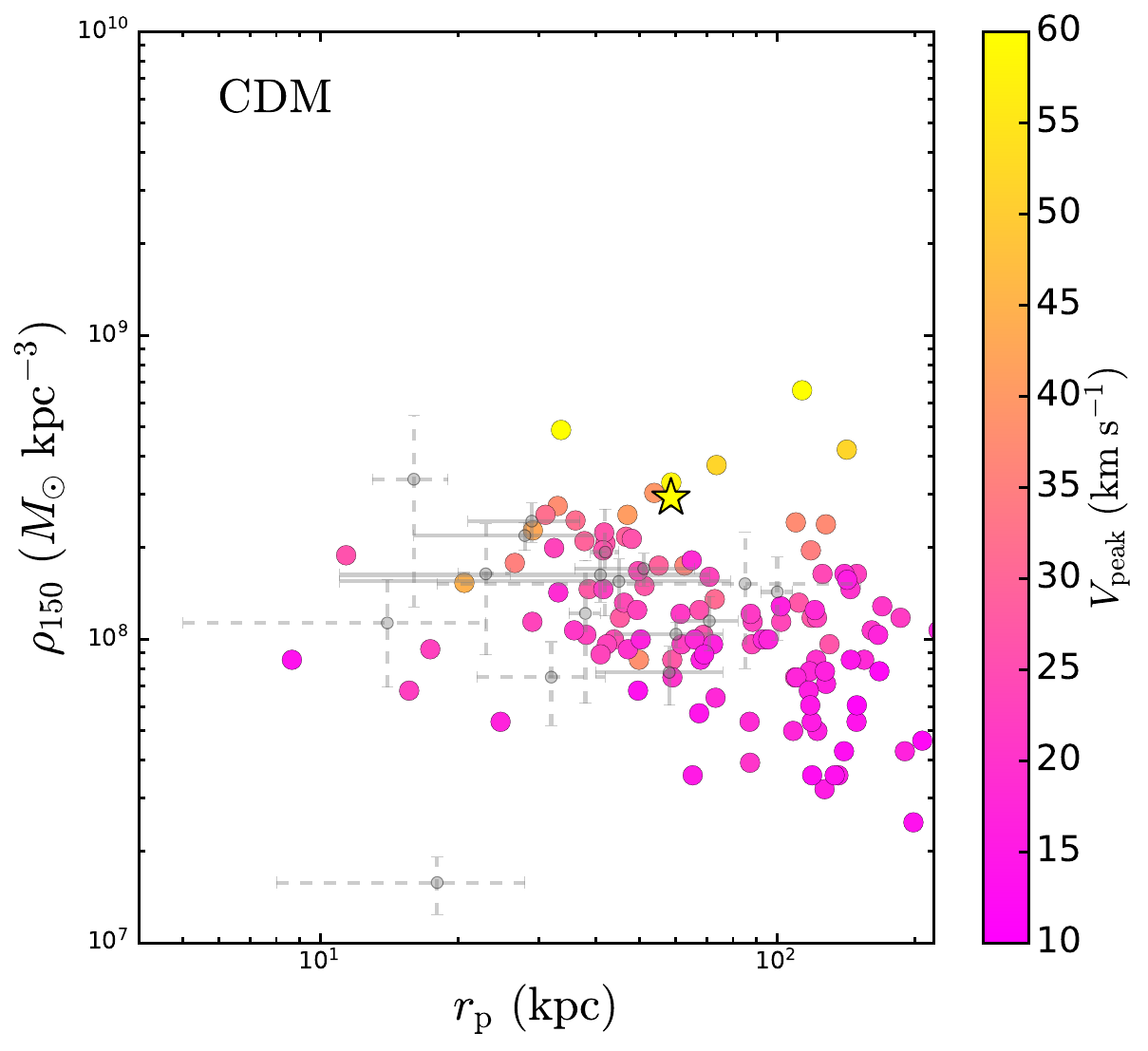}
  \includegraphics[width=7.2cm]{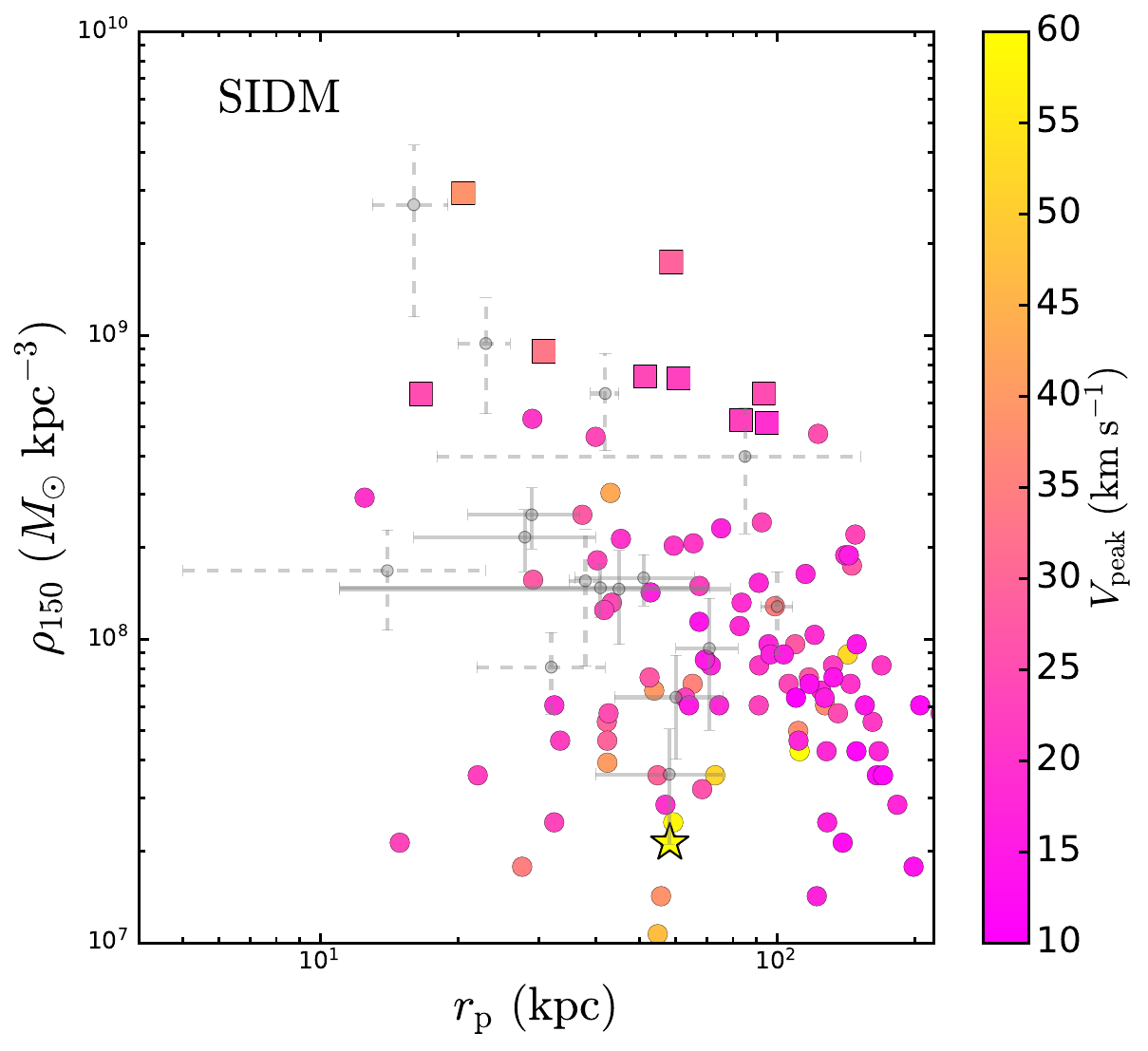}
  \caption{\label{fig:vcirc} Top: circular velocities of $30$ simulated MW CDM (left) and SIDM (right) subhalos of the highest $V_{\rm peak}$ that reside at least $30$ kpc from the MW center; the LMC halo is shown by an orange dashed curve in each panel. Subhalos are colored according to their pericenter distance. The segments with only gray dotted lines denote the extrapolation below our convergence radius. For comparison, circular velocity measurements at the half-light radius for observed MW classical and fainter satellites are shown in solid and dashed black error bars, respectively, with the data compiled in~\cite{Silverman220310104}. Bottom: the distribution of pericentric distance, $r_{\rm p}$, versus inner DM density, $\rho_{\rm 150}$, for MW subhalos that have masses higher than $10^8~M_{\odot}~h^{-1}$ in our CDM (left) and SIDM (right) simulation. Subhalos are colored according to their peak maximum circular velocity. Gray error bars show pericenter versus inferred central density for a sample of observed MW satellite galaxies from \cite{Kaplinghat190404939}; the data differ in the two panels because an NFW (isothermal) halo profile is assumed to extrapolate the observationally inferred density profile to $150~\pc$. The LMC analog halo is shown as a yellow star in each panel, and squares denote deeply core-collapsed halos in the right panel.
  }
\end{figure*}

In the top panels of Figure~\ref{fig:vcirc}, we plot the circular velocities of the 30 MW subhalos with the highest values of $V_{\rm peak}$, and that reside at least 30 kpc from the MW center, in our CDM and SIDM simulations. 
For comparison, the circular velocity measurements at the half-light radius for the observed MW satellites are shown for the nine classical MW satellite galaxies (i.e., satellites with $V$-band luminosity $M_V<-8.8$, excluding the LMC and SMC; solid black error bars), and for fainter dwarf galaxies (including a handful of ultra-faint dwarfs; dashed black error bars), using the data compiled in~\cite{Silverman220310104}. Many of the subhalo rotation curves in our CDM simulation are higher than the observed points, even if we exclude our LMC analog when comparing to the data, thus reproducing the canonical TBTF problem~\citep{Boylan-Kolchin11030007,Boylan-Kolchin11112048}. 

Several baryonic mechanisms may alleviate the TBTF tension in SIDM~\citep{Zolotov12070007,Sawala151101098,Wetzel160205957}, including enhanced tidal stripping or complete disruption due to the Galactic disk, along with internal feedback for sufficiently massive systems. In particular, the former could be a dominant effect~\citep{Sawala151101098,Garrison-Kimmel170103792}. For example,~\cite{Robles190401469} showed that the inclusion of a disk can reduce the circular velocities of subhalos at $0.3~\kpc$ by $20\textup{--}30$\%, bringing CDM predictions into better agreement with the data. However, these authors also found that the least dense subhalos (among the most massive ones) tend to have the smallest pericentric distances from the Galactic center, inconsistent with the trend observed for the brightest MW satellites~\citep{Kaplinghat190404939}; we discuss this point further below. 

Our SIDM simulation demonstrates a much larger scatter for the most massive subhalos' rotation curves, underscoring the diversity of halo populations resulting from self-interactions. For SIDM subhalos with large pericentric distances of $r_{\rm p}\gtrsim75~{\rm kpc}$, the circular velocity profiles are systematically shifted lower compared to CDM as a result of core formation and tidal stripping. For those with small pericentric distances of $r_{\rm p}\lesssim75~{\rm kpc}$, the subhalos with a relatively large mass have lower circular velocities as well, but many low-mass subhalos have \emph{enhanced} velocity profiles due to core collapse (note that there are 9 deeply core-collapsed MW subhalos in our sample). Since our SIDM simulation does not include a Galactic potential, we cannot make a concrete comparison with the data. Nevertheless, our SIDM model produces both core-forming and core-collapsed subhalos, resulting in a more diverse DM distribution than that predicted in CDM, hence reproducing certain features of the data that are difficult to explain otherwise. 

We further comment on the SIDM solution to the TBTF problem. Previous DM-only simulations of MW-like systems showed that the TBTF problem could be resolved due to SIDM core formation~\citep{Vogelsberger12015892,Zavala12116426}. However, detailed kinematic analyses of the classical MW satellites found that their inner densities are so diverse that the self-interacting cross section inferred from individual dwarfs can vary by $1$ order of magnitude for core-forming subhalos~\citep{Read180506934,Valli171103502,Hayashi200712780}. More recently, SIDM simulations with an MW disk potential showed that the TBTF problem can actually be {\em oversolved} due to core formation and additional tidal stripping from the disk, if all subhalos are in the core-forming phase~\citep{Silverman220310104}. Our SIDM simulation predicts that some core-collapsed subhalos can have inner densities comparable to or higher than their CDM counterparts (see also~\citealt{Turner201002924,Correa220611298}), a necessary condition for the model to pass the test in the presence of an MW disk potential.

Interestingly, the impact of the baryonic mechanisms discussed above may differ for SIDM subhalos in comparison to CDM. For example, SIDM subhalos that already have prominent cores before a pericentric passage may be disrupted more easily by the MW disk than their cuspy CDM counterparts, while SIDM subhalos approaching core collapse may be accelerated in this process due to tides. Dedicated simulations will be needed to test whether such core-collapsed subhalos can withstand tidal disruption by the MW. At the same time, SIDM halo profiles may be more robust due to internal baryonic feedback processes than those in CDM due to thermalization~\citep{Kaplinghat13116524,Ren180805695}, implying that halo responses may further distinguish the models~(e.g., \citealt{Robles170607514,Robertson171109096}). 

\subsection{MW Subhalo Central Densities and Pericentric Distances}

In addition to internal halo properties, the correlation between MW satellites' orbital kinematics and their inferred density profiles may further help distinguish between CDM and SIDM scenarios (e.g., \citealt{Nadler200108754}). Indeed, \cite{Kaplinghat190404939} reported an anticorrelation between DM density at $150~$pc ($\rho_{\rm 150}$) and pericentric distance ($r_{\rm p}$) for classical MW satellite galaxies. This anticorrelation is unexpected if tidal stripping is the dominant mechanism that reduces subhalos' central densities during pericentric passages. On the other hand, ``survivor bias''---i.e., the fact that high-density subhalos are less prone to tidal disruption in the host's inner regions---may counteract this effect and bring the central density--pericenter relation into agreement with that inferred observationally, even in a CDM context \citep{Kaplinghat190404939,Hayashi200712780}, although a reassessment is needed when the tidal effects of the stellar disk are included \citep{Robles190401469}. A similar mechanism is known to bias the radially dependent mass--concentration relation for subhalos (e.g., \citealt{Moline160304057,Moline211002097}). Observational selection effects can also potentially influence this relation because fainter satellites, which are expected to occupy lower-mass and thus (initially) more concentrated subhalos, are easier to detect near the pericenter. We therefore focus on the \emph{correlation} between these properties, which responds nontrivially to SIDM physics.

In the bottom panels of Figure~\ref{fig:vcirc}, we show the $\rho_{\rm 150}\textup{--}r_{\rm p}$ distribution predicted in our CDM (left) and SIDM (right) simulations, compared to data points for the classical (solid gray) and ultra-faint (dashed gray) dwarfs from~\cite{Kaplinghat190404939}. The data points differ in the two panels because of the assumptions used to extrapolate measurements from large radii to $150~$pc. 
In particular, an NFW profile is used in the CDM case, while a cored-isothermal profile is used in the SIDM case. 
An anticorrelation between $\rho_{150}$ and $r_{\rm p}$ is observed in both cases; however, it is only statistically significant for the SIDM density profile extrapolation according to a Spearman correlation test. Interestingly, both our CDM and SIDM simulations yield statistically significant anticorrelations, with Spearman coefficients of $-0.37$ and $-0.18$, respectively. It is important to note that, for CDM, this anticorrelation is mainly driven by low-$V_{\rm peak}$ subhalos. In fact, the trend vanishes in our CDM simulation if we only consider the $15$ subhalos with the largest values of $V_{\rm peak}$ (consistent with the results of \citealt{Kaplinghat190404939}), but it remains for SIDM. Since the most massive subhalos are expected to host the brightest satellite galaxies, our result suggests that the observed anticorrelation for the brightest MW satellites cannot be reproduced in our CDM simulation. We also caution that comparing these predictions to data requires more realistic simulations (e.g., including a Galactic disk) and detailed forward-modeling (e.g., accounting for observational selection effects); we leave this analysis to future work.

There are some additional subtleties regarding the differences between the central density--pericenter relations in CDM and SIDM. 
First, SIDM subhalos show a much larger scatter in their inner densities. 
Core-forming subhalos, which tend to have large $V_{\rm peak}$, are shifted lower compared to their CDM counterparts, while core-collapsed subhalos are shifted higher and are marked by squares. 
In CDM, the halos of higher $V_{\rm peak}$ have systematically higher inner densities measured at a fixed radius, as expected. 
This trend is \emph{reversed} in our SIDM model, where higher-$V_{\rm peak}$ subhalos tend to have lower inner densities due to core formation, and vice versa for lower-$V_{\rm peak}$ subhalos due to core collapse. We contrast this with the results in \cite{Ebisu210705967}, which suggest that velocity-independent self-interactions of $\sigma/m=1$ and $3~\mathrm{cm^2\ g^{-1}}$ tend to \emph{erase} the anticorrelation between central density and pericentric distance, because in such SIDM models all subhalos are in the core-expansion phase, and their central densities are reduced relative to CDM (e.g., see \citealt{Robles190401469,Silverman220310104}).

\begin{figure*}[t!]
  \centering
  \includegraphics[width=7.2cm]{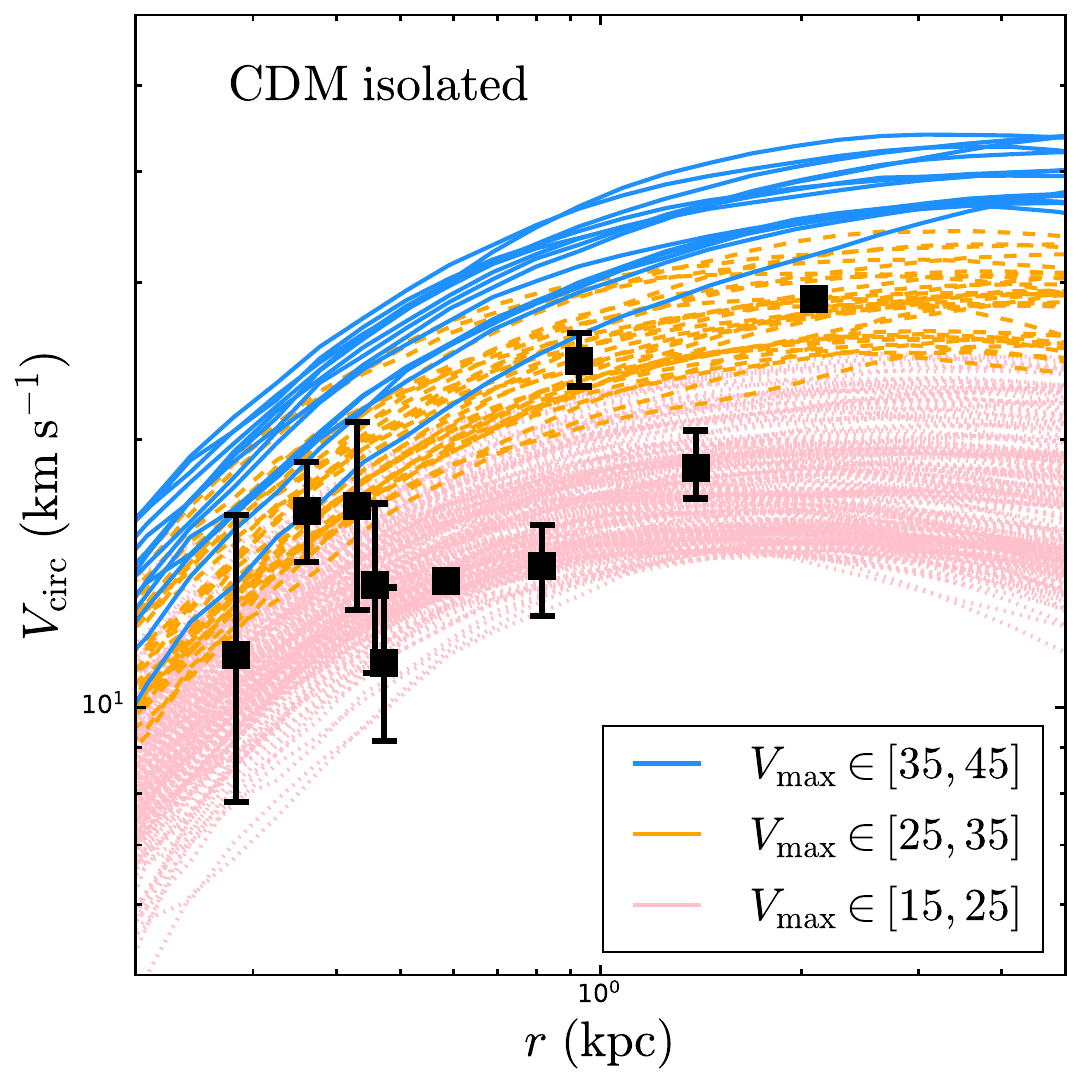}~~~~~
  \includegraphics[width=7.2cm]{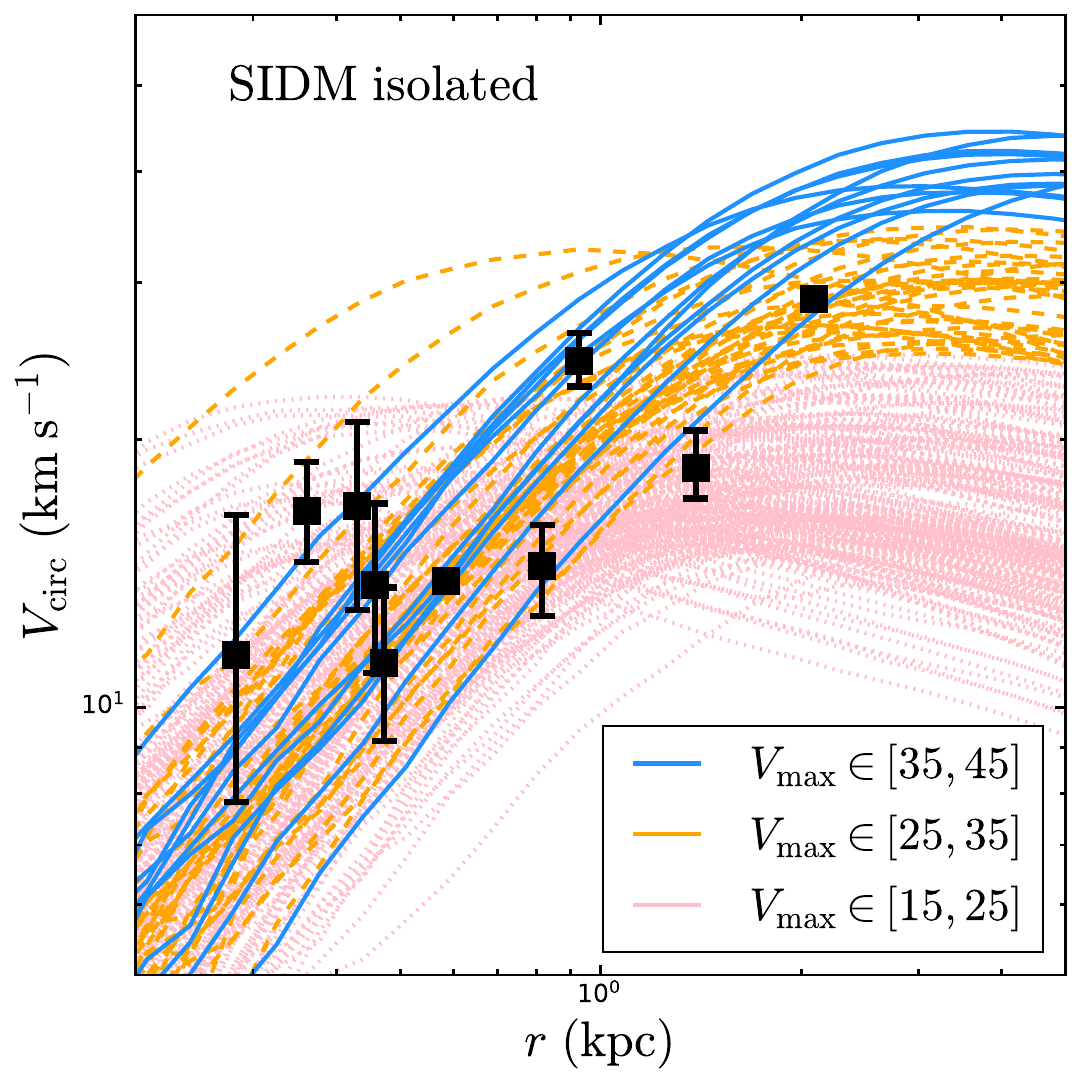}
  \caption{\label{fig:vcirclocal} Circular velocities of isolated halos in CDM (left) and SIDM (right) with distances between $300$ and $2000$ kpc from the MW host halo center. The curves for halos with $V_{\rm max}=35\textup{--}45~\kms$, $25\textup{--}35~\kms$, and $25\textup{--}15~\kms$ in light-blue, orange, and pink, respectively. Data points correspond to circular velocities, measured at the half-light radius, for a sample of local field dwarf galaxies compiled in \cite{Elbert14121477}. The halos of these galaxies have $V_{\rm max}\approx40~{\rm km/s}$ according to the abundance matching relation.
  }
\end{figure*}

We expect that the inclusion of a central disk (or disk potential) will affect this relation by reducing the densities of subhalos with small pericenters due to enhanced tidal stripping.~\cite{Robles190401469} showed that the inner density of CDM and core-forming SIDM subhalos becomes relatively lower for a smaller pericenter due to the presence of a disk, and one would expect a positive correlation between $\rho_{\rm 150}$ and $r_{\rm p}$, instead of an anticorrelation as indicated by the observational data. On the other hand, if core-collapsed subhalos are more resilient to the tidal stripping associated with the Galactic disk than their CDM (or core-forming) counterparts, our strong velocity-dependent SIDM model could produce an even stronger anticorrelation between $\rho_{150}$ and $r_{\rm p}$ in the presence of a disk compared to our current DM-only simulation. We will explore this intriguing possibility in future work. 

\subsection{Isolated Halo Rotation Curves}

Figure~\ref{fig:vcirclocal} shows rotation curves for isolated halos with $V_{\mathrm{max}}=35\textup{--}45~\kms$ (light-blue curves) with distances between $300$ and $2000$ kpc from the MW host halo center, roughly chosen to resemble the sample of 10 observed field dwarf galaxies in the Local Group as compiled in \cite[][black error bars]{Elbert14121477}. According to standard abundance matching relations, the host halos of these galaxies have $V_{\rm max}\approx40~{\rm km/s}$ \citep{Garrison-Kimmel14045313}. In the CDM panel, we find a discrepancy similar to the TBTF problem for satellites, here for field galaxies. This discrepancy has been studied previously (e.g., \citealt{Garrison-Kimmel14045313}). Interestingly, DM self-interactions alleviate this issue without the need to invoke strong baryonic feedback because the isolated halos with $V_{\mathrm{max}}\approx 40~\kms$ are predicted to reside in the core-forming phase (Figure~\ref{fig:vmaxRvmax1}). Our results are broadly consistent with those in~\cite{Elbert14121477}, where they simulated isolated SIDM dwarf halos with constant cross sections in the range $\sigma/m=0.1\textup{--}{50}~\kms$ and found that the TBTF problem in the local field can be alleviated for $\sigma/m>0.5~{\kms}$. 

From Figure~\ref{fig:vcirclocal}, we also see that, for lower masses $V_{\rm max}=25\textup{--}35~\kms$ (orange), and $V_{\rm max}=15\textup{--}25~\kms$ (pink), many of our simulated halos are in the core-collapse phase, resulting in high inner circular velocities; the population of core-collapsed halos increases as $V_{\rm max}$ decreases. For CDM halos, the inner circular velocity at a given radius decreases with $V_{\rm max}$, but the trend can be reversed in SIDM because core collapse occurs predominantly for low-mass halos in our SIDM model. This behavior is similar to our findings for SIDM subhalos, as discussed in the previous subsection. Thus, our SIDM model predicts the existence of dense core-collapsed isolated halos in the Local Group that could host unexpectedly dense galaxies. The correlation between the gravothermal behavior of subhalos and isolated halos is inevitable for models that lead to sufficiently rapid core collapse. 

As shown in~\cite{Papastergis14074665}, there is also a TBTF problem in the field beyond the Local Volume, which could also be difficult to solve in CDM with baryonic feedback~\citep{Papastergis1511087411}. On the other hand,~\cite{Schneider161109362} demonstrated that velocity-dependent SIDM models (e.g., $\sigma/m=14~{\rm cm^2~g^{-1}}$, at $V_{\rm max}=30~{\rm km/s}$) may alleviate or solve the problem. It would be interesting to reassess the SIDM solution in light of our simulations, where both core-forming and core-collapsed halos are populated. We expect that a careful forward-modeling procedure (e.g., based on the subhalo $V_{\mathrm{max}}$ and $V_{\mathrm{peak}}$ functions) is necessary for a detailed comparison aimed at constraining these galaxies' halo properties.

\subsection{Diversity of Isolated and Splashback Halos}
\label{sec:udg}

\begin{figure*}[htbp]
  \centering
  \includegraphics[width=7.2cm]{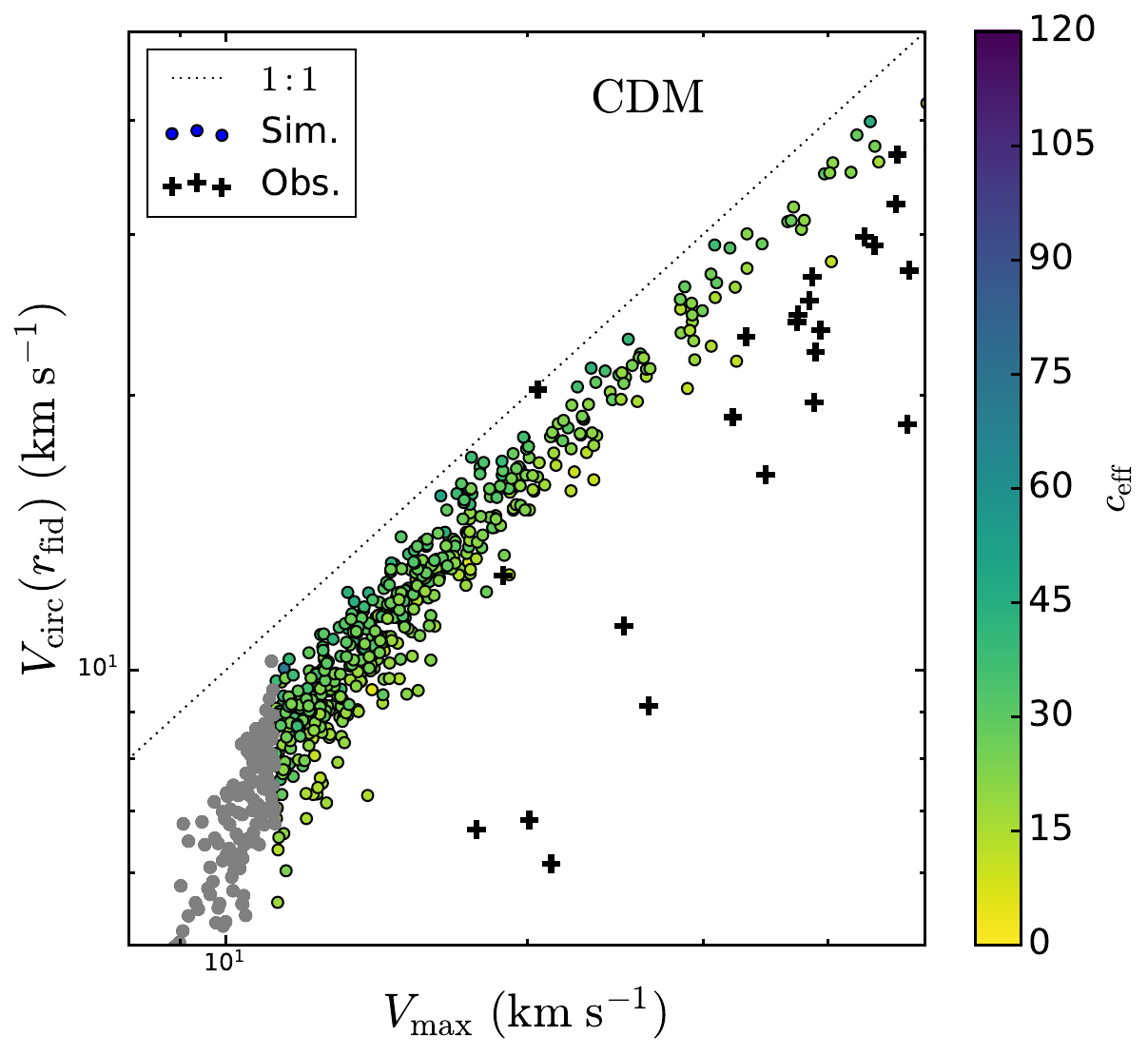}
  \includegraphics[width=7.2cm]{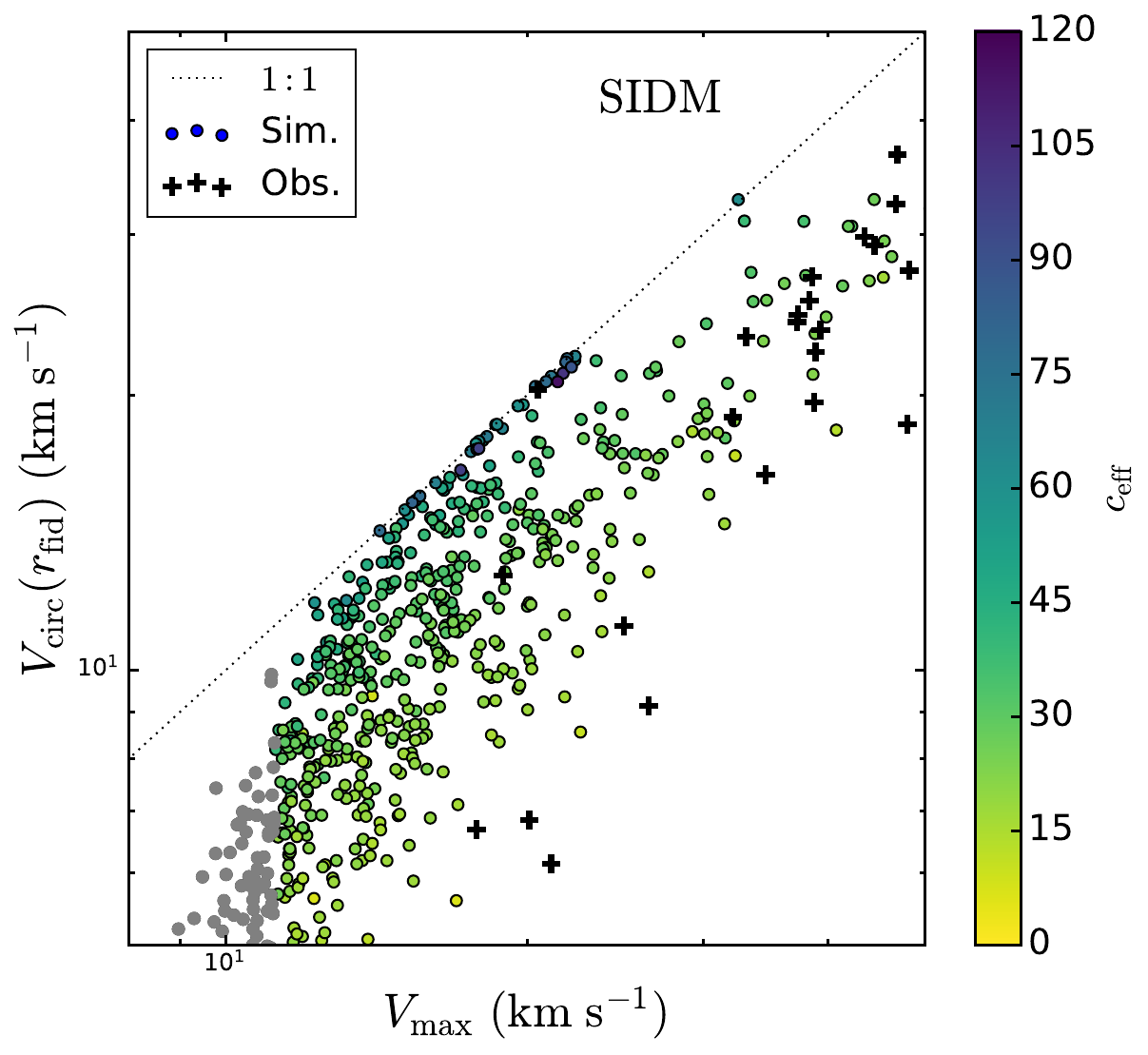}\\
  \includegraphics[width=7.2cm]{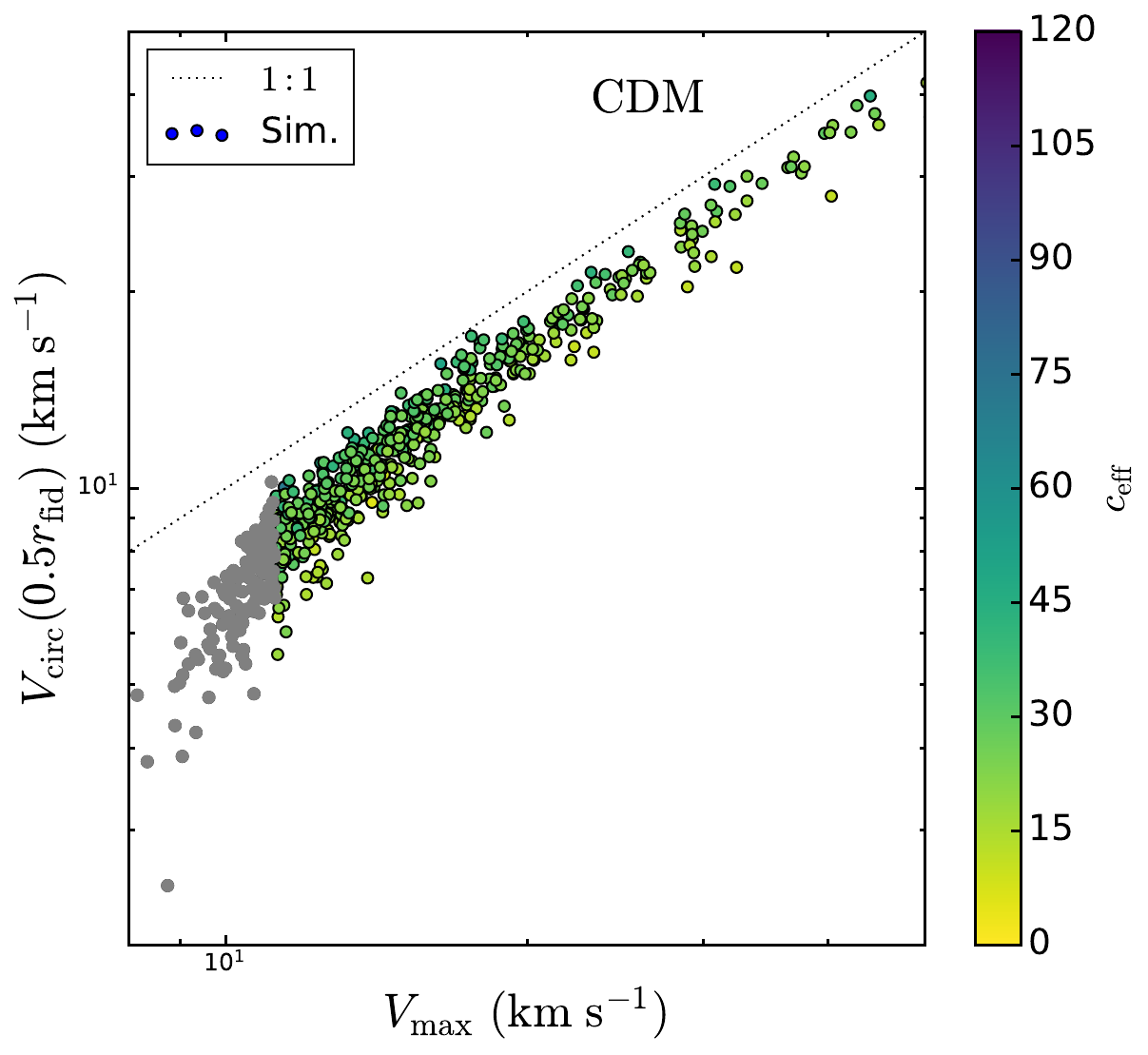}
  \includegraphics[width=7.2cm]{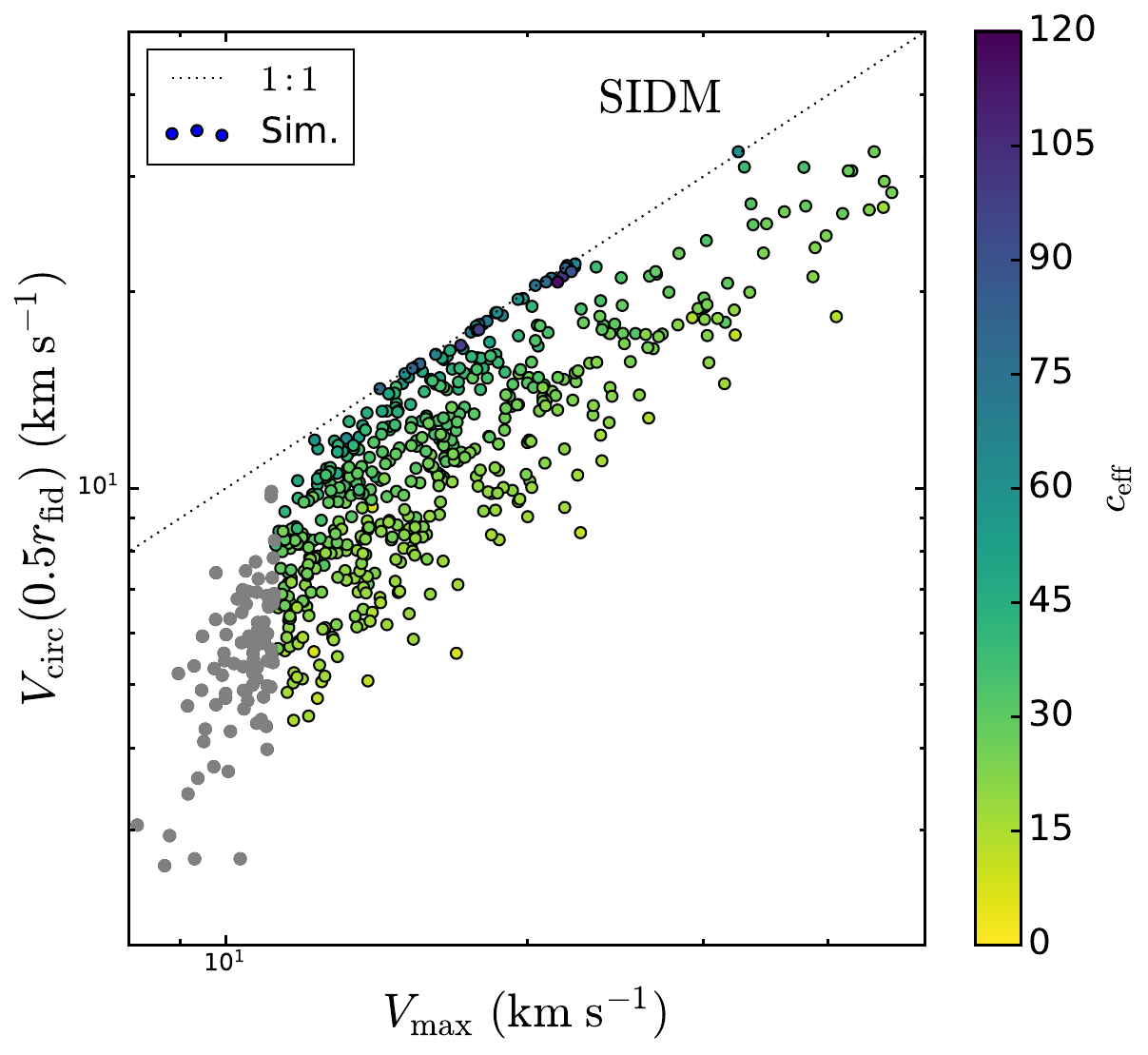}
  \caption{\label{fig:vmaxVcirc2} Top: the $ V_{\rm circ}(r_{\rm fid})\textup{--}V_{\rm max}$ distribution for isolated halos in our CDM (left) and SIDM (right) simulations. The observational points, denoted as crossed black markers, are taken from~\cite{Santos-Santos191109116}. Bottom: the $V_{\rm circ}(0.5 r_{\rm fid})\textup{--}V_{\rm max} $ distribution for isolated halos in our CDM (left) and SIDM (right) simulations. For all panels, simulated halos are colored according to their effective concentration. If $r_{\rm fid}$ is smaller than the convergence radius, the halo is shaded in gray. 
  }
\end{figure*}
 
To further study the effects of SIDM on isolated halos, we follow \cite{Santos-Santos191109116} by computing halo circular velocities evaluated at a fiducial radius,
\begin{equation}
   V_{\rm circ}(r_{\mathrm{fid}})  = \sqrt{\frac{G M(r<r_{\rm fid})}{r_{\rm fid}}},
\end{equation}
where $r_{\rm fid} \equiv 2 V_{\rm max}/(70 ~\rm km/s)~\rm kpc$. This quantity captures the amplitude of the rotation curve halos' inner regions; thus, the $V_{\rm circ}(r_{\rm fid})\textup{--}V_{\rm max}$ relation probes both the shape and magnitude of rotation curves; see also~\cite{Oman150401437} for a similar approach based on a $V_{\rm circ}(2~{\rm  kpc})\textup{--}V_{\rm max}$ relation. 

In the top panels of Figure~\ref{fig:vmaxVcirc2}, we show the $V_{\rm circ}(r_{\rm fid})$--$V_{\rm max}$ distribution from our CDM (left) and SIDM (right) simulations, color-coded by the effective halo concentration. For halos that have $r_{\rm fid}$ smaller than the our spatial resolution limit, we shade them in gray. For comparison, we include the data points of observed galaxies compiled in~\cite{Santos-Santos191109116}. Since our simulated halos mainly populate the low-$V_{\mathrm{max}}$ region of this parameter space, we focus on comparing to observed galaxies with inferred values of $V_{\rm max}<50~\kms$. Our CDM simulation results are consistent with those in \cite{Santos-Santos191109116}, and exhibit much smaller scatter, compared to the observed data, a manifestation of the so-called ``diversity problem'' of galactic rotation curves \citep{deNaray09123518,Oman150401437,Ren180805695,Santos-Santos191109116}. In particular, the median $V_{\rm circ}(r_{\rm fid})$--$V_{\rm max}$ is close to the 1:1 line in CDM, meaning that there are more rapidly rising rotation curves in the simulation compared to the sample of dwarf galaxies shown in Figure~\ref{fig:vmaxVcirc2}. For dwarf galaxies with $V_{\rm max}\lesssim 50~{\rm \kms}$, the effects of baryonic feedback in changing the inner density profiles begin to diminish~(e.g.,~\citealt{Bullock170704256}).

On the other hand, the isolated halos in our SIDM simulation exhibit a much larger scatter in the $V_{\rm circ}({r_{\rm fid}})\textup{--}V_{\rm max}$ plane, compared to CDM, in better agreement with the data points.\footnote{Note that secondary halo properties are also nontrivially correlated with the resulting UDG properties in CDM  (e.g., \citealt{Benavides210901677,Benavides220907539}).} In particular, core-forming halos have shallower rotational curves in their inner halo region and reside toward the lower right of the median relation, while core-collapsed halos have faster-rising rotational curves and shift upward from the median. Some of the SIDM halos are deeply in the collapse phase, and they are placed on the 1:1 line. If rotation curves are measured at smaller radii, e.g., $0.5r_{\rm fid}$, this separation between core-forming and core-collapsed halos is even more clear, as shown in the bottom panels of Figure~\ref{fig:vmaxVcirc2} for our CDM (left) and SIDM (right) simulations. 

Overall, we therefore find that our strong, velocity-dependent SIDM model produces both core-forming and core-collapsed isolated halos and significantly increases the diversity of halos' inner densities. Hence, this potentially provides a better match to the observed diversity of field dwarf galaxy rotation curves. Previous studies on the SIDM solution to the galactic diversity problem mainly focused on the core-formation regime (e.g. \citealt{Creasey161203903,Kamada161102716,Ren180805695,Santos-Santos191109116,Zentner220200012}), and showed that, compared to CDM, SIDM can produce a larger spread in the inner circular velocity profile after accounting for scatter in halo concentration and the impact of baryons on SIDM halo profiles~\citep{Kaplinghat13116524}. We show that the spread can be further amplified in our SIDM model due to core collapse. Thus, we expect outlier galaxies with $V_{\rm circ}(r_{\rm fid})/V_{\rm max}\sim1$ while being low baryon concentration as highlighted in~\cite{Santos-Santos191109116} can be easily accommodated in our SIDM model. 

As in previous sections, we note that our comparisons are not carried out in a full, forward-modeling framework capable of including all relevant observational and theoretical systematics. For example, rotation curve measurements are known to be biased toward recovering artificially flat velocity profiles and cored density profiles, purely due to projection effects, noncircular motions, and dynamical equilibrium assumptions (e.g., \citealt{Read170104833,Oman170607478,Genina191109124,Chang2009001613,Roper220216652}). In addition, the diverse rotation curves are also correlated with the stellar surface densities of spiral galaxies, and hence understanding the origin of baryon distributions is needed to fully address the problem. Here, we have highlighted the trends imprinted by SIDM on the field halo population, which can form the basis of future comparison~studies.

\section{Discussion}
\label{sec:discussion}

We now discuss our main results in the context of previous SIDM studies, focusing on the diversity of SIDM halo populations in Section~\ref{sec:diversity}, comparisons to previous simulations in Section~\ref{sec:sim_comparison}, and areas for future work in Section~\ref{sec:future}.

\subsection{The Diversity of Isolated Halos and Subhalos in SIDM}
\label{sec:diversity}

One of the most distinctive features of SIDM is its ability to diversify the properties of DM halos at roughly fixed mass, depending on their secondary properties (e.g., \citealt{Kaplinghat190404939,Kahlhoefer190410539,Ren180805695,Sameie190407872}). Although the extent to which larger diversity than predicted by CDM is necessary, particularly in the presence of baryons (e.g., \citealt{Kaplinghat191100544,Santos-Santos191109116,Zentner220200012}), upcoming observational facilities will considerably improve measurements of individual galaxies' dynamics, sharpening this test \citep{Chakrabarti220306200}. Thus, it is crucial to achieve self-consistent predictions for the full diversity of SIDM halo populations.

In this context, our results are the first to self-consistently explore the connection between the diversity of SIDM isolated halo and subhalo populations in a realistic cosmic environment. Specifically, the high-resolution region of our zoom-in simulation extends to distances of $3~\mpc$ from the MW center, between the size of the Local Group and Local Volume. Although our simulation is not constrained to match the overdensity of the MW's actual local environment (which may itself be rare; e.g., see \citealt{Neuzil191204307}), we expect it to accurately capture the correlated nature of isolated halo and subhalo populations in our SIDM model. Moreover, our results bridge commonly studied subhalo and isolated halo populations by explicitly including halos in the splashback regime and a realistic LMC analog system.

Consistent with previous studies, we find that a sizable population of MW subhalos likely undergoes core collapse in the presence of strong, velocity-dependent DM self-interactions (e.g., \citealt{Turner201002924}). Furthermore, many isolated and splashback halos show similar core-collapse signatures, albeit at a smaller rate than that for subhalos, likely due to tidal acceleration of gravothermal evolution (consistent with the findings of, e.g., \citealt{Nishikawa190100499,Sameie190407872,Zeng211000259}). We recover a well-known correlation between halo concentration and the efficiency of core collapse at fixed halo mass (e.g., \citealt{Essig180901144,Kaplinghat190404939}), and we quantify how gravothermal evolution can be captured using the halo properties $V_{\mathrm{max}}$ and $R_{\mathrm{max}}$. An analytic gravothermal model is able to predict shifts in the $R_{\mathrm{max}}$--$V_{\mathrm{max}}$ plane relative to CDM fairly well, hinting that the detailed effects of SIDM captured by cosmological simulations can likely be captured by appropriate (semi) analytic modeling as a function of the SIDM cross section (e.g., see \citealt{Jiang220612425} for an example along these lines). We intend to explore this avenue in future work.

A particularly intriguing correlation revealed by our simulations can be phrased as follows. If the densest ultra-faint MW satellite galaxies with small pericenters are explained via core collapse of SIDM subhalos and the low densities of the brightest MW satellites are explained by core formation (see Figure~\ref{fig:vcirc}), then two corresponding populations of field dwarf galaxies exist: (1) a bright field dwarf population that occupies cored DM halos (e.g., corresponding to the rotation curves with high $V_{\mathrm{circ}}$ at small radii in the bottom right panel of Figure~\ref{fig:vcirc}), and (2) a rarer population of faint and ultra-faint field dwarf galaxies that occupy core-\emph{collapsed} halos. 
These correlated predictions are sensitive to the SIDM cross section, implying that a joint fit to these observables would provide a stringent test of SIDM physics.

This latter population has not been remarked on in previous studies, which largely used idealized, noncosmological simulations to model gravothermal evolution and find long core-collapse timescales of isolated halos. Meanwhile, because our simulation self-consistently resolves structure formation in a cosmic environment, gravothermal evolution is often significantly accelerated for the halos that end up as isolated objects at $z=0$ due to major mergers or orbital evolution around the MW (or other nearby hosts) throughout their histories; thus, we predict that a population of MW splashback galaxies also occupy core-collapsed halos. Observationally, these dense, isolated dwarf galaxies predicted in our SIDM model are potentially reminiscent of the Tucana dwarf galaxy, which may have an unusually high inner density as a splashback galaxy of Andromeda (\citealt{Gregory190207228,Santos-Santos220702229}; however, see \citealt{Taibi200111410}).

Finally, we remark on the LMC analog in our SIDM simulations and the potential observational implications of its subhalo population. Although only one LMC-associated subhalo in our SIDM simulation shows signs of core collapse, there are not a large number of LMC-associated subhalos above our conservative halo mass limit. Thus, the implied core-collapse rate of $\approx 10\%$ for LMC-associated subhalos is consistent with that for our isolated halo population. A lack of accelerated core collapse is perhaps expected, because our LMC analog system falls into the MW at very late times and thus does not experience significant and prolonged tidal stripping, unlike most MW subhalos. Furthermore, it appears that tidal evolution of LMC-associated subhalos above our $z=0$ mass threshold, as they orbit the LMC before the LMC falls into the MW, does not play a major role in accelerating their gravothermal evolution.\footnote{On the other hand, \cite{Nadler210912120} found that evaporation due to subhalo--host halo interactions in velocity-independent SIDM models may be enhanced for LMC-associated subhalos, highlighting the sensitivity of halo evolution to SIDM physics.} Conversely, there may be a population of extremely low-mass halos below our resolution limit that were stripped by the LMC at early times and fall into the MW near $z=0$.

Our results therefore highlight that the impact of SIDM on particular halo populations is dependent on both environmental factors and the SIDM cross section. Observationally, as the number of known MW satellite galaxies---including LMC-associated satellites---continues to increase (e.g., \citealt{Kallivayalil180501448,Drlica-Wagner191203302,Patel200101746}), the discovery of a population of very dense ultra-faint dwarf satellite galaxies would be an interesting hint of SIDM-induced core collapse that is difficult to recreate via baryonic effects or other nonstandard DM physics.

\subsection{Comparison to Previous Simulations}
\label{sec:sim_comparison}

Several authors have simulated SIDM models similar to that considered in this paper. Here, we compare our results to these studies, focusing on comparisons between cosmological simulations. A detailed comparison between the gravothermal evolution of halos in idealized, noncosmological simulations (e.g., \citealt{Yang220503392}) versus those found in realistic cosmic environments is an interesting area for future work that will benefit the development of semianalytic models (e.g., \citealt{Jiang220612425,Outmezguine220406568,Yang220502957}).

The most direct comparison we can draw is with the cosmological zoom-in simulations presented in \cite{Zavala190409998,Turner201002924}. In particular, the ``vd100'' cross section model presented in \cite{Zavala190409998} and studied in detail by \cite{Turner201002924}, \cite{Meshveliani221001817} is nearly identical to our SIDM model, and their zoom-in simulation setup is similar to ours, focusing on an MW-mass host halo at similar resolution. The resulting MW subhalo population studied in \cite{Turner201002924} is also broadly similar to ours; for example, the subhalo mass function is not significantly altered relative to a corresponding CDM simulation. Interestingly, SIDM subhalos that show signs of core collapse according to the criteria in \cite{Turner201002924} outnumber uncollapsed subhalos by a factor of $\approx 3$, which implies a much larger core-collapse rate than our fiducial result of $\approx 20\%$, at face value. This is likely due to our very conservative criteria for identifying core-collapsed halos, which effectively requires them to be $\gtrsim 3\sigma$ outliers compared to the CDM $R_{\mathrm{max}}$--$V_{\mathrm{max}}$  relation. The true rate of core-collapsed subhalos in our simulation may therefore be significantly higher, in agreement with the results of \cite{Turner201002924}.

We also note that \cite{Lovell220906834} applied an analytic model based on SIDM scaling relations derived from the \cite{Turner201002924} vd100 simulation to the COCO-CDM simulation \citep{Hellwing150506436} to predict SIDM subhalo population statistics. This technique is, by construction, less accurate than a full SIDM simulation that captures the correlated, environmentally dependent effects of self-interactions on halo and subhalo populations; on the other hand, it is significantly faster because it avoids running a new simulation for each SIDM system of interest. Quantitatively, the results of \cite{Lovell220906834} corroborate our finding that core collapse is necessary to explain the full diversity of inferred MW satellite inner densities in an SIDM scenario (also see, e.g., \citealt{Kim210609050}).

Importantly, our MW host halo stands out from those used in many previous SIDM simulations, including those described above \citep{Zavala190409998,Turner201002924,Lovell220906834}, because (1) its early merger history resembles that of the MW, and (2) it hosts a massive LMC analog on recent infall \citep{Nadler191203303}. This host has been used to fit the MW satellite galaxy population in CDM and alternative models \citep{Nadler191203303,Nadler200800022,Mau220111740}, and it has been resimulated in velocity-independent SIDM models \citep{Nadler210912120}. These properties add to the realism of our MW subhalo population predictions, and allow us to study the LMC-associated subhalo population in a scenario that includes gravothermal core collapse for the first time.

We also highlight an important difference between our \emph{analysis} and that of many previous SIDM simulation studies, including \cite{Zavala190409998,Turner201002924}. In particular, we leveraged the entire high-resolution volume of our zoom-in region to self-consistently study the effects of SIDM on isolated halos, splashback halos, and MW and LMC-associated subhalos. This is crucial because, as we have shown, the gravothermal evolution and other properties of isolated halos and subhalos are intrinsically linked in an SIDM model like ours. This allows us to make novel predictions concerning the population of field dwarf galaxies hosted by core-forming and core-collapsed halos, including in the ultra-faint regime, and assuming that the densest observed MW satellite galaxies are interpreted as core-collapsed systems. Imminent observational advances, e.g.\ delivered by the Rubin Observatory Legacy Survey of Space and Time \citep{IvezicLSST}, will increase the census of faint dwarf galaxies known in the local universe (e.g., \citealt{Mutlu-Pakdil210501658}). Combined with spectroscopic follow-up (including from next-generation observational facilities; \citealt{Chakrabarti220306200}), these measurements will significantly sharpen the power of dwarf galaxy internal dynamics as a probe of SIDM physics.

Finally, note that \cite{Correa220611298} also studied the vd100 SIDM model in a large-volume cosmological simulation. These authors studied both isolated halos and subhalos in a $\sim 25~\mpc$ volume, which includes several massive halos and thus, in principle, provides excellent statistical power for unified studies of isolated halos' and subhalos' gravothermal evolution. However, compared to zoom-in simulations, there is necessarily a tradeoff in resolution when simulating such a large volume; for example, the mass resolution and softening used by \cite{Correa220611298} were $1.44\times 10^6~\msun$ and $650~\pc$, which are factors of $\sim 25$ and $5$ larger than in our simulation, respectively. Our study is thus highly complementary to that of \cite{Correa220611298}; in particular, these authors did not study the effects of SIDM on very-low-mass halos with masses of $\lesssim 10^9~\msun$, where we the effects of our SIDM model are most pronounced. Investing in simulations that combine the statistical power of large-volume simulations with the resolution of zoom-ins is therefore important for future SIDM studies.

\subsection{Future Work}
\label{sec:future}

As discussed above, our study has several unique advantages relative to recent SIDM work. However, as we have emphasized throughout, several aspects of the simulations and analysis presented in this paper are simplified. We now summarize the most important areas for improvement, and avenues for future work aimed at more realistic modeling of halo and galaxy population in SIDM. 

First, we have presented DM-only simulations. Galaxy formation and baryonic feedback is known to affect halo properties in a manner that can be degenerate with SIDM physics (e.g., \citealt{Pontzen11060499,Brook14103825,Burger210807358}). However, simulations using the FIRE-2 galaxy formation model show that SIDM density cores are much more robust to the inclusion of baryonic physics compared to CDM on dwarf galaxies~\citep{Robles170607514}; we therefore do not expect core formation of low-mass halos in our SIDM simulation to be strongly affected by baryonic feedback (e.g., see \citealt{Ren180805695}). In future work, it will be particularly interesting to perform hydrodynamic simulations in the presence of core collapse. We expect that the cuspy density profiles of collapsed halos remain resilient to feedback as self-interactions can redistribute energy rapidly (e.g., see \citealt{Sameie210212480,Rose220614830} for studies of MW-mass SIDM halos). This may provide a mechanism to form compact dwarf galaxies in SIDM even when feedback is \emph{strong}.

Second, and related to the previous point concerning baryonic physics, our MW host halo is expected to host a massive central galaxy in reality. The central baryonic component, and particularly the Galactic disk, is known to have a major impact on the evolution of subhalos that pass sufficiently close to the host halo center, potentially reducing the abundance of surviving subhalos relative to a corresponding CDM-only simulation by $\approx 50\%$ (e.g., \citealt{Garrison-Kimmel170103792,Nadler171204467,Kelley1811112413,Richings181112437,Samuel190411508}; however, also see \citealt{Webb200606695,Green211013044}). The interplay between the effect of the disk and SIDM has only received limited attention, e.g.\ in \cite{Robles190401469}, and is likely nontrivial. For example, SIDM responds to the gravitational potential of the MW galaxy efficiently (e.g., \citealt{Kaplinghat13116524}), implying that the central core of our MW host halo likely remains a cusp in the presence of a disk, while many MW subhalos are cored in our SIDM scenario, increasing the efficiency of tidal disruption as they orbit the central potential (e.g., \citealt{Errani221001131}).

On the other hand, the low-mass core-collapsed subhalos produced by our strongly velocity-dependent SIDM model are likely \emph{more} resilient to tidal disruption than their CDM counterparts, particularly in the presence of a central galaxy. Taken together, these effects can potentially enhance the diversity of MW subhalos in our SIDM model even further. In the presence of the Galactic disk, a particularly intriguing possibility is that the combination of enhanced tidal disruption relative to CDM for massive, core-forming subhalos and increased resilience for low-mass, core-collapsed subhalos \emph{amplifies} the anticorrelation between subhalos' central densities and pericentric distances studied in the bottom panels of Figure~\ref{fig:vcirc}. We therefore plan to pursue SIDM simulations with embedded, analytic disk potentials to model the tidal effects of the MW galaxy in future work.

Third, we have not attempted to forward-model the galaxies compared to those in Section~\ref{sec:population_stats}, including the MW satellite population or the populations of local field dwarf galaxies presented there. A forward-modeling study would affect the comparisons we have presented in several ways. For example, a careful abundance matching procedure that accounts for uncertainties in the global and satellite luminosity functions, as well as the (uncertain) intrinsic scatter in the faint-end stellar mass--halo mass relation, is necessary to robustly associated halos and subhalos in our simulations with observed galaxies, in a statistical fashion (e.g., see \citealt{Nadler180905542,Nadler191203303} for examples in the context of MW satellite galaxies). Furthermore, for many of the halo and subhalo properties we have studied, observational selection effects and systematics play an important role when inferring observed systems' halo properties. For example the difficulty of accurately inferring field galaxy rotation curves was discussed in Section~\ref{sec:udg}; meanwhile, MW satellite galaxies are subject to significant selection effects as a function of their distance, size, and luminosity (e.g., \citealt{Drlica-Wagner191203302}), and all of these properties are correlated with the expected distribution of underlying halo properties, in a given DM model. Thus, follow-up studies that forward-model properties of observable galaxies predicted to inhabit halos and subhalos in our simulations are worth pursuing. \\

\section{Conclusions}
\label{sec:conclusions}

We explored the impact of strong, velocity-dependent DM self-interaction cross section (Figure~\ref{fig:bmxs}) on halo populations within and surrounding the MW. In particular, we compared high-resolution CDM and SIDM cosmological zoom-in simulations focused on a host halo with a mass, merger history, and LMC analog that resembles the MW. 
Crucially, we analyze high-resolution halos out to $\sim 3~\mpc$ from the MW center, allowing us to self-consistently analyze the gravothermal evolution of SIDM halos in different environments, including MW and LMC-associated subhalos, splashback halos of the MW, and isolated halos. 
Our main findings are summarized below:
\begin{itemize}
    \item The stage of SIDM halos' gravothermal evolution is correlated with their concentration, formation history, and environment, and imprints as a shift toward larger $V_{\mathrm{max}}$ and smaller $R_{\mathrm{max}}$ for halos in the core-collapse phase (Figure~\ref{fig:vmaxRvmax1}).
    \item Environmental effects lead to subtle differences among the gravothermal evolution of subhalos and isolated halos; at a fixed mass, the fraction of core-collapsed subhalos is $\approx 2$ times higher for subhalos, likely due to tidal evolution (Figure~\ref{fig:hmaxcum}).  
    \item Mass functions for subhalos and isolated halos in our CDM and SIDM simulations are nearly identical (Figure~\ref{fig:shmf}). Thus, the evaporation due to subhalo--host halo interactions, which can reduce the amplitude of the subhalo mass function for SIDM models with large cross sections at the MW host halo's infall velocity scale, is less important in our model because of its strong velocity-dependence and possibly because some subhalos are more resilient to tidal disruption due to core collapse.
    \item Our SIDM model yields diverse inner DM distributions for \emph{both} subhalos isolated halos (top right panel, Figure~\ref{fig:vcirc}; also see Figure \ref{fig:vcirclocal}). If dense, ultra-faint MW satellite galaxies are hosted by core-collapsed subhalos, then we predict an undiscovered population of dense, isolated faint dwarf galaxies in the field, which are compelling observational targets.
    \item Strong self-interactions produce a significant anticorrelation between MW subhalos' inner densities and pericentric distance (bottom right panel, Figure~\ref{fig:vcirc}). This may help explain the observed trend for MW satellites; this comparison is limited by uncertainties in these galaxies' inferred DM profiles, which will be reduced by upcoming observational facilities.
\end{itemize}

We conclude with a few speculative remarks hinted by our results. First, the matched evolution of SIDM halos and subhalos that show signs of core collapse (Figure~\ref{fig:bmlmc} and Appendix~\ref{sec:additional_benchmark}) suggests that halos' detailed merger histories are correlated with their gravothermal evolution. In particular, even the isolated core-collapsed halos we identify undergo early tidal evolution and/or major mergers, which seem to facilitate a rapid approach toward core collapse relative to the core-forming isolated halos of similar masses. It is not clear whether this relationship is causal, in the sense that merger dynamics or tidal evolution accelerate core collapse, or due to more rapid evolution of higher-concentration remnants of early mergers. In either case, comparing the gravothermal evolution we measured in a realistic cosmic environment to idealized simulations and predictions from (semi) analytic gravothermal predictions and isothermal Jeans models is an important area for future study. If core collapse strongly correlates with certain features of halos' merger histories, this would have interesting implications for how SIDM physics may manifest in observable galaxy population statistics.

Second, our analysis in Section \ref{sec:sidm_identification} and Figure~\ref{fig:tcbarCDM} indicates that core-collapse timescales estimated using an analytic effective cross section model predict the impact of SIDM on simulated halos' $V_{\rm max}$ and $R_{\rm max}$ surprisingly well. This is compelling because it suggests that SIDM effects resulting from detailed, cosmological simulations can be predicted analytically---to a reasonable degree of accuracy---as a function of the effective cross section. Thus, we can use our formalism to predict SIDM signatures for cross section models that have not yet been simulated in order to analytically \emph{transform} the results of any CDM simulation into a corresponding SIDM run. It will be interesting to test how well such analytic transformations recover the gravothermal evolution predicted by simulations in as-yet untested regions of SIDM parameter space.


\acknowledgments

We are grateful to Susmita Adhikari, Arka Banerjee, Andrew Benson, Nitya Kallivayalil, Manoj Kaplinghat, Victor Robles, Laura Sales, and Shengqi Yang for helpful discussions. We thank Mike Grudi\'{c} for assistance with \textsc{meshoid}, Philip Mansfield for sharing the version of \textsc{ROCKSTAR} used in this work, and Yao-Yuan Mao for sharing the initial conditions from \cite{Mao150302637}.

The computations presented here were conducted through Carnegie's partnership in the Resnick High Performance Computing Center, a facility supported by Resnick Sustainability Institute at the California Institute of Technology. This work was supported by the John Templeton Foundation under grant ID \#61884 (D.Y., H.-B.Y.), and the U.S. Department of Energy under grant No.\ de-sc0008541 (H.-B.Y.). The opinions expressed in this publication are those of the authors and do not necessarily reflect the views of the John Templeton Foundation.

\bibliographystyle{yahapj2}
\bibliography{reference}


\appendix

\section{Cosmological SIDM Implementation}
\label{sec:appSIDM}

The SIDM simulation performed in this paper is based on a module that extends the public \textsc{GADGET-2} \citep{Springel0505010} program to incorporate elastic DM scatterings. 
It adopts the parallelization implementation in \cite{Robertson160504307}, with a few modifications. 
For isolated halo simulations, details of the program have been introduced in \cite{Yang220503392}. 
For cosmological simulations, the evaluation of differential cross section and scattering probability needs to be updated to incorporate the expansion of the universe. 

The extension from isolated to cosmological simulation is based on the time integration scheme of \textsc{GADGET-2} \citep{Springel0505010}, where the kick and drift of positions and momenta are implemented as follows: 
\begin{eqnarray}
\label{eq:kad}
\vect{x}_i &\to& \vect{x}_i + \dfrac{\vect{q}_{i}}{m} \int_{t_0}^{t_1} \dfrac{d t}{a^2} = \vect{x}_i + \dfrac{\vect{q}_{i}}{m} \int_{a_0}^{a_1} \dfrac{d a}{H a^3}, \\ \nonumber
\dfrac{\vect{q}_{i}}{m} &\to& \dfrac{\vect{q}_{i}}{m} + \int_{a_0}^{a_1} \dfrac{d a}{H a^2} a^3 \left( \ddot{\vect{x}}_i + 2 H \dot{\vect{x}}_i \right), \\ \nonumber
                        &=&  \dfrac{\vect{q}_{i}}{m} + \int_{a_0}^{a_1} \dfrac{d a}{H a^2} \left( - G \sum_{j\neq i} \dfrac{m_j \vect{x}_{ij}}{|\vect{x}_{ij}|^3}  + \dfrac{H_0^2}{2} \Omega_{M,0} \vect{x}_i \right),
\end{eqnarray}
where $x_i$ refers to the components of comoving coordinates, $q_i\equiv a^2 \dot{x}_i m$ refers to the canonical momentum, and $H_0$, $\Omega_{M,0}$ are the Hubble and relative matter density today.  
The program uses the comoving coordinates, and a velocity variable $\vect{v}_{\mathrm{code}} \equiv \vect{q}/m$, which is consistently translated into the peculiar velocity when evaluating a velocity-dependent cross section. 

The scattering probability between a pair of neighboring particles is evaluated in a single time step of displacement $\Delta x$ as follows: 
\begin{equation}
\label{eq:probability}
{\cal P}_{ij} = \frac{\sigma(v_{ij})}{2 a^2} W(x_{ij},h) (\Delta x),  
\end{equation}
where $\sigma(v_{ij})$ is a (velocity-dependent) cross section, $x_{ij}$ is the separation between the two particles, $W(x_{ij},h)$ is a weighting kernel, and the factor $1/2$ removes double counting from looping over the particle indices $i,j$. 
We choose the kernel function to be the same as the smoothing kernel in~\textsc{GADGET-2}~\citep{Springel0505010}; also see the discussion in \cite{Yang220503392}.

A halo virialized in a static background maintains virialized in a comoving background.
We make use of this feature to test if an isolated virialized halo evolves similarly under SIDM in a cosmological (comoving) versus static background. 
We perform this test for a dwarf-sized halo with a constant cross section $\sigma/m=5~\rm cm^2~g^{-1}$ and obtained almost identical results in physical coordinates, which verified our cosmological implementation of the SIDM module.  

\section{Convergence tests}
\label{sec:convergence}

\begin{figure*}[htbp]
  \centering
  \includegraphics[width=7.2cm]{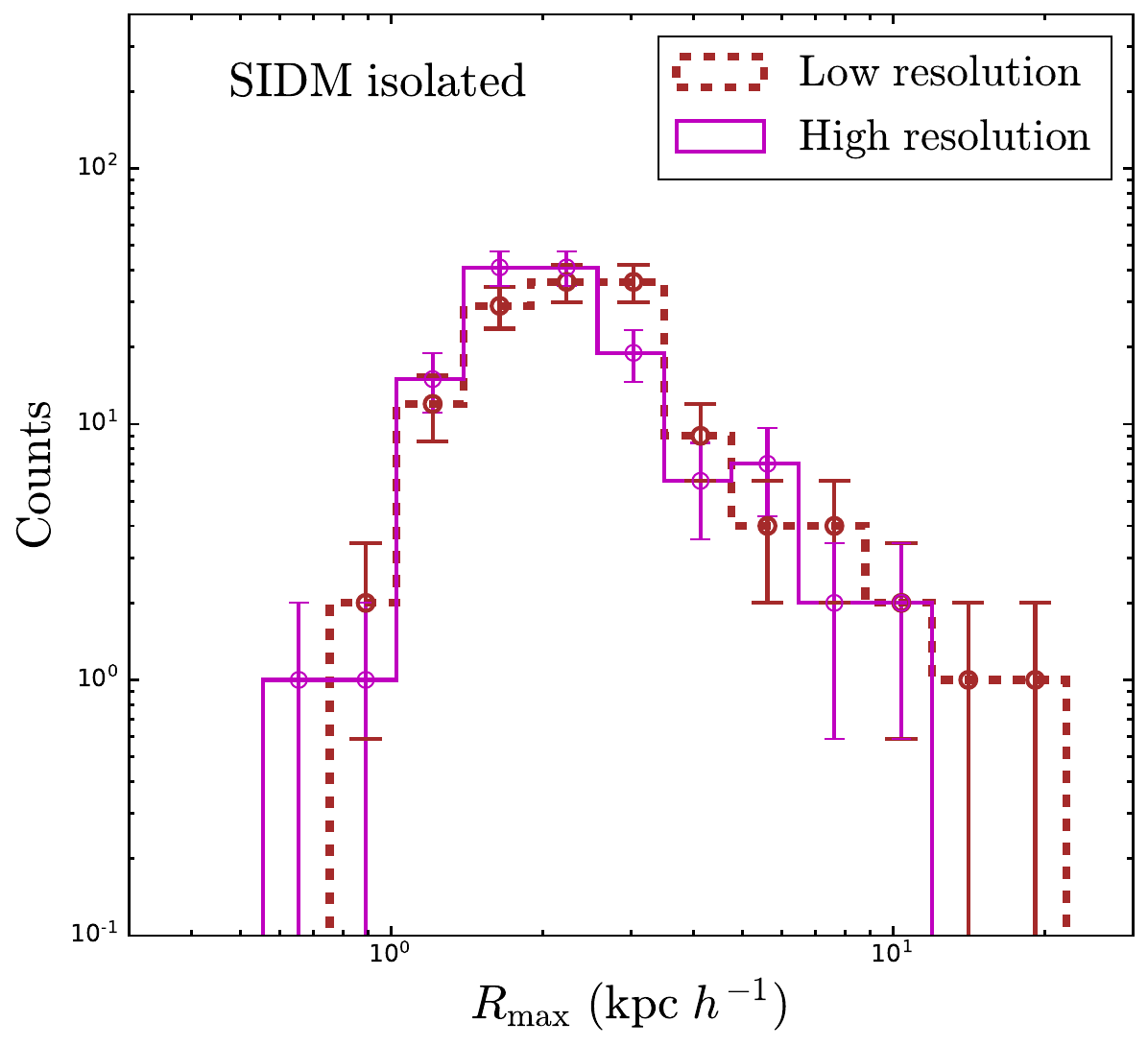}
  \includegraphics[width=7.2cm]{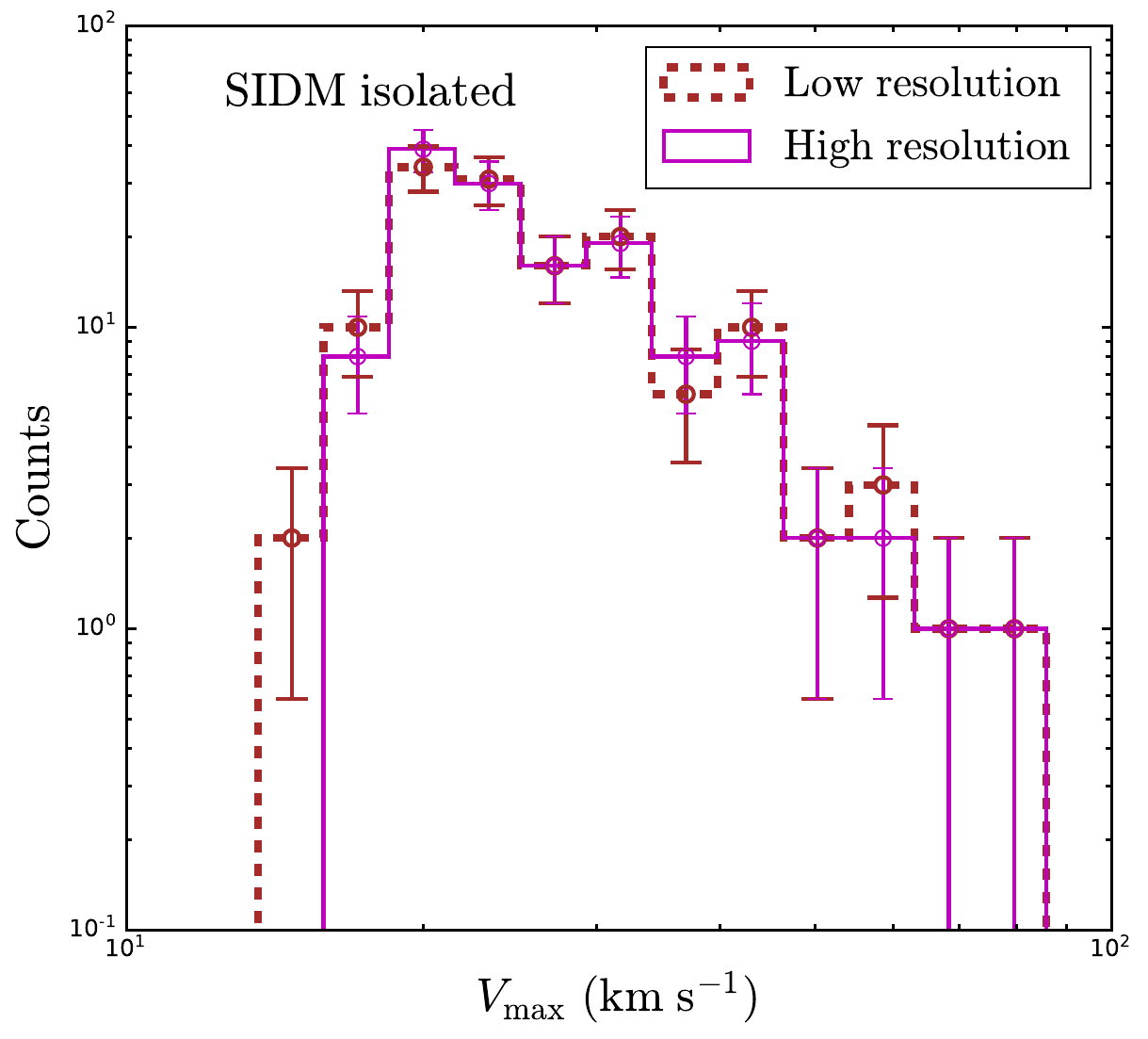}
  \caption{\label{fig:conv0} 
Distribution of $R_{\rm max}$ (left) and $V_{\rm max}$ (right) in our fiducial, high-resolution SIDM simulation (solid) and in a lower-resolution resimulation (dashed) using $8$ times more massive particles and $2$ times larger softening.}
\end{figure*}

\begin{figure*}[htbp]
  \centering
  \includegraphics[height=4.6cm]{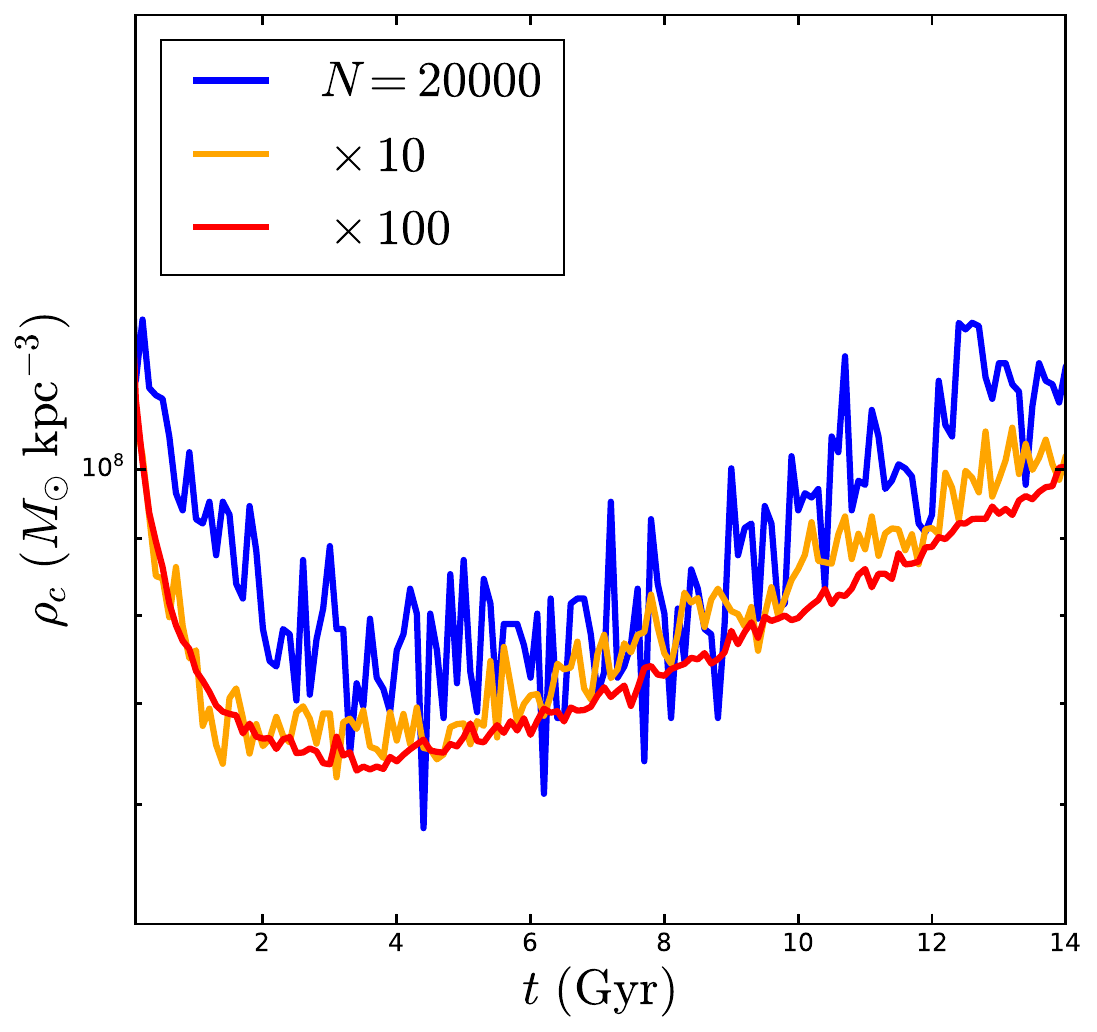}
  \includegraphics[height=4.6cm]{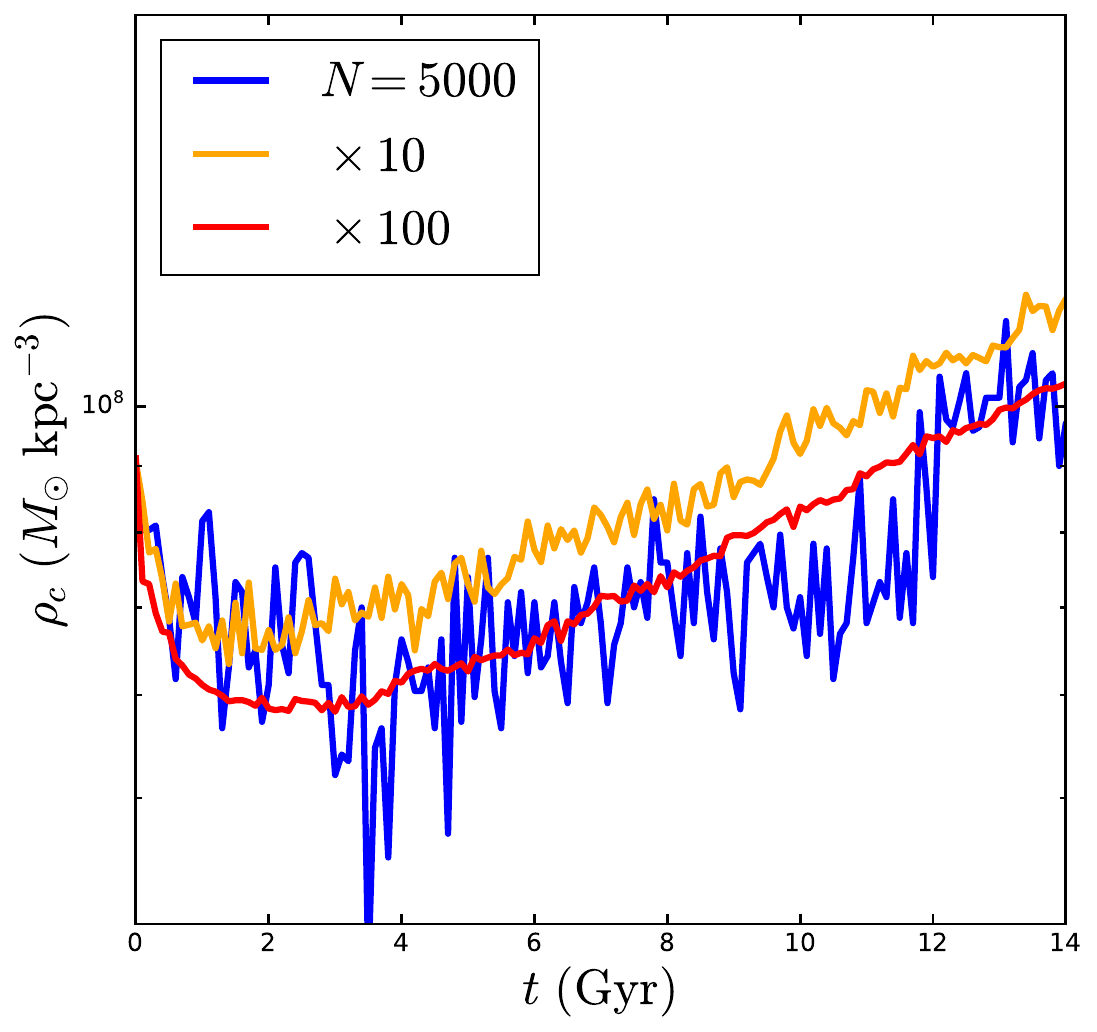}
  \includegraphics[height=4.6cm]{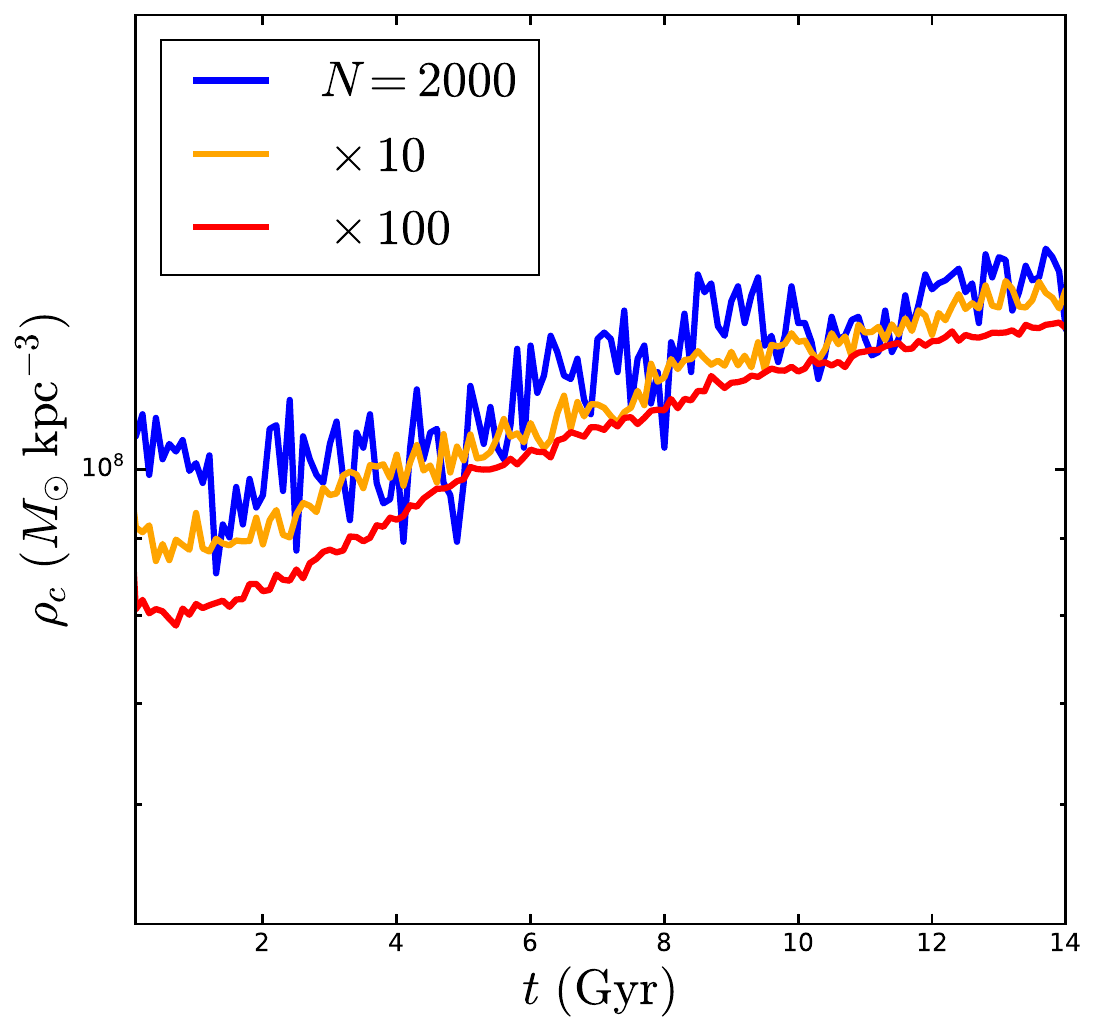} \\
  \includegraphics[height=4.6cm]{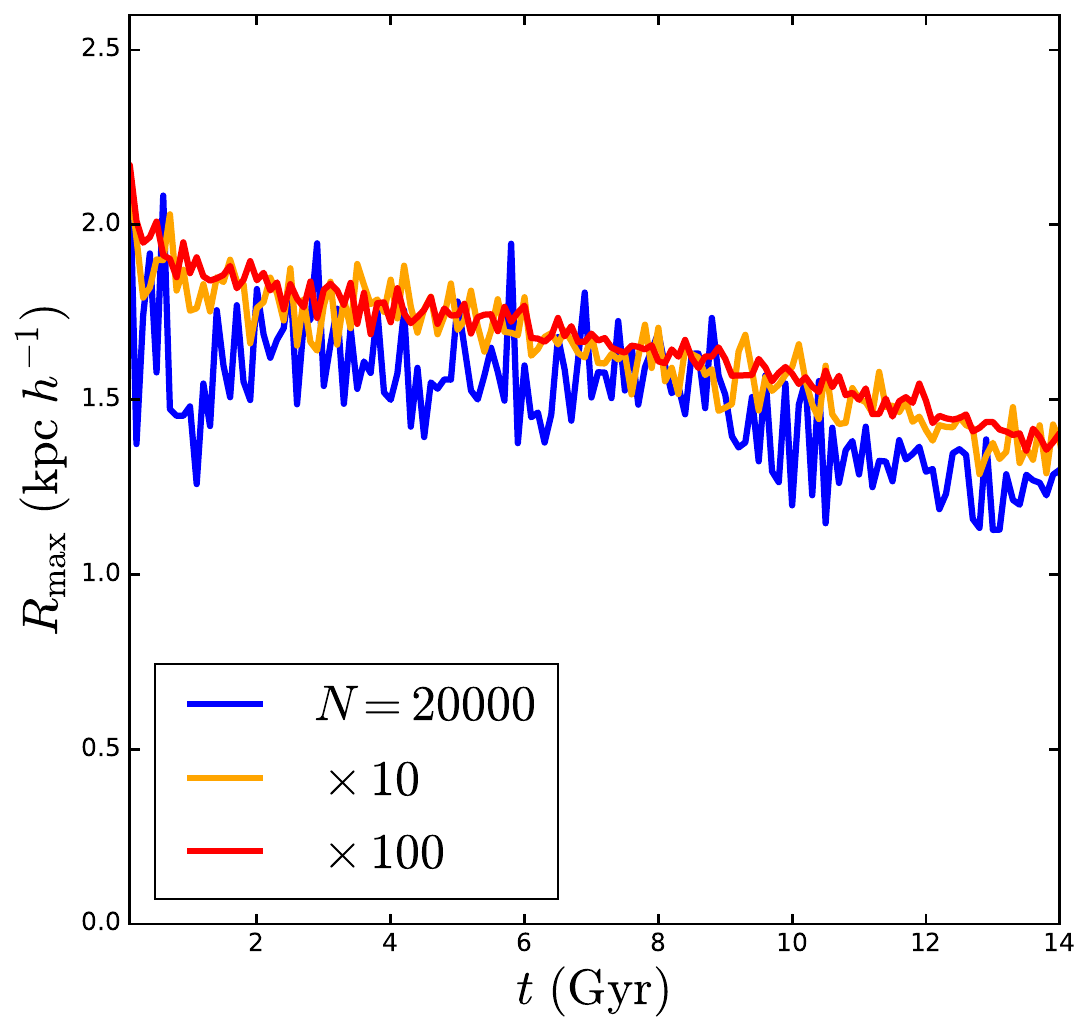}
  \includegraphics[height=4.6cm]{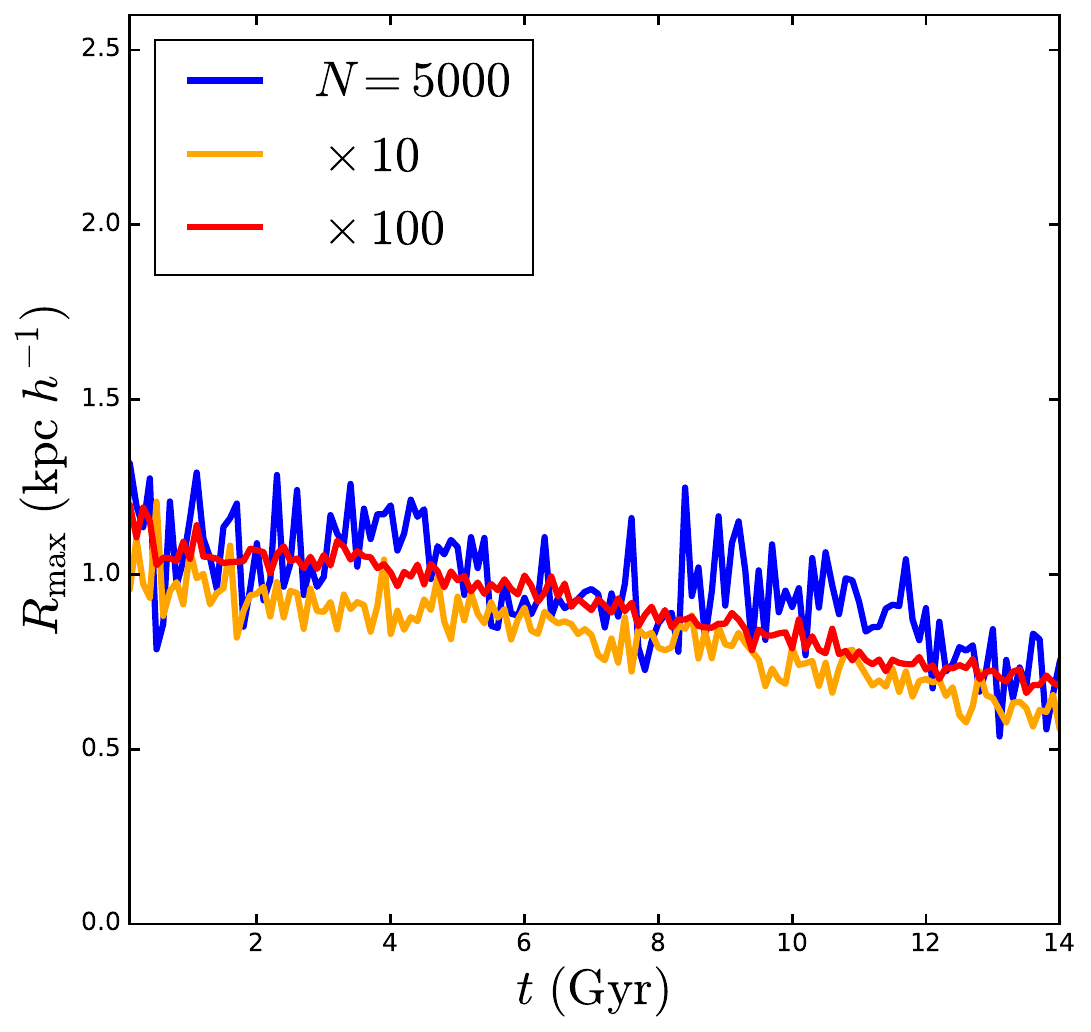}
  \includegraphics[height=4.6cm]{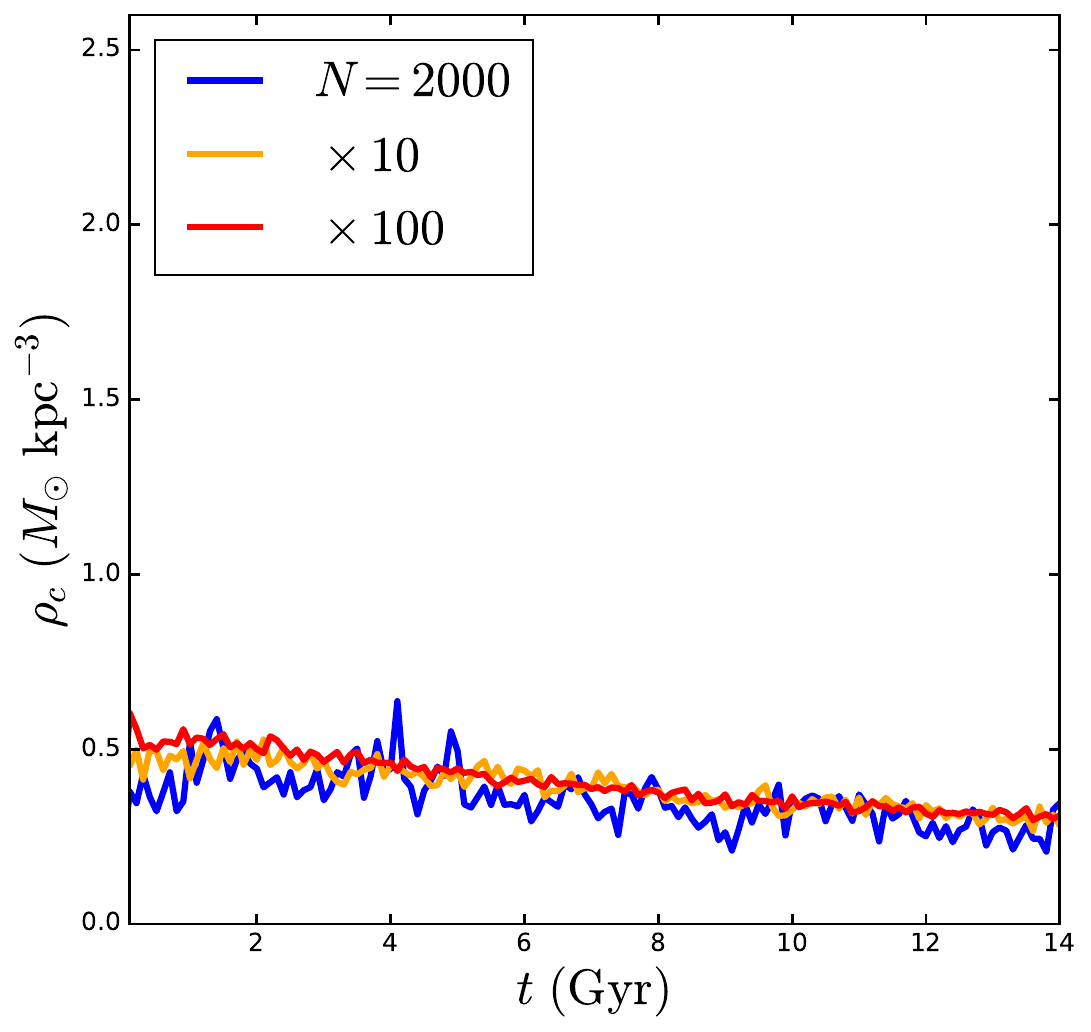} \\
  \includegraphics[height=4.6cm]{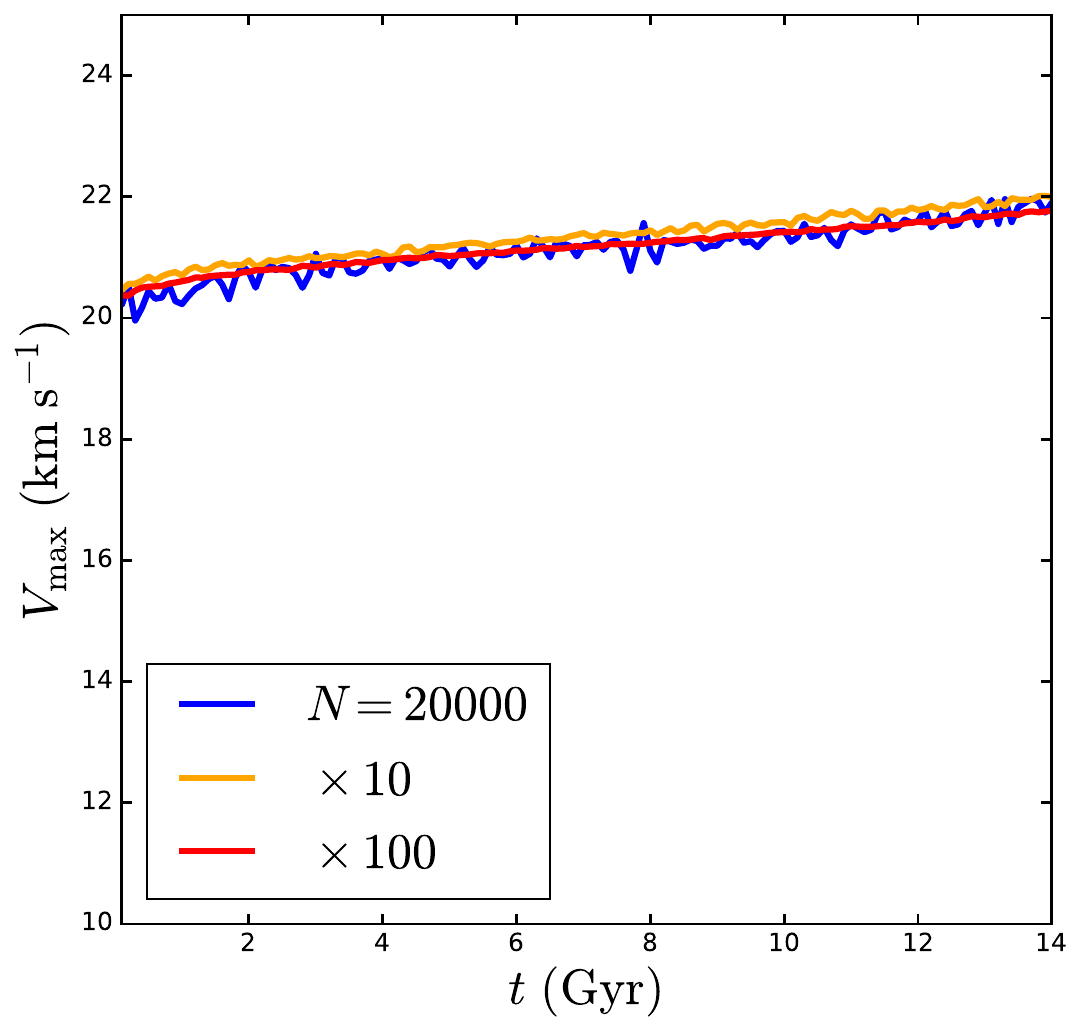}
  \includegraphics[height=4.6cm]{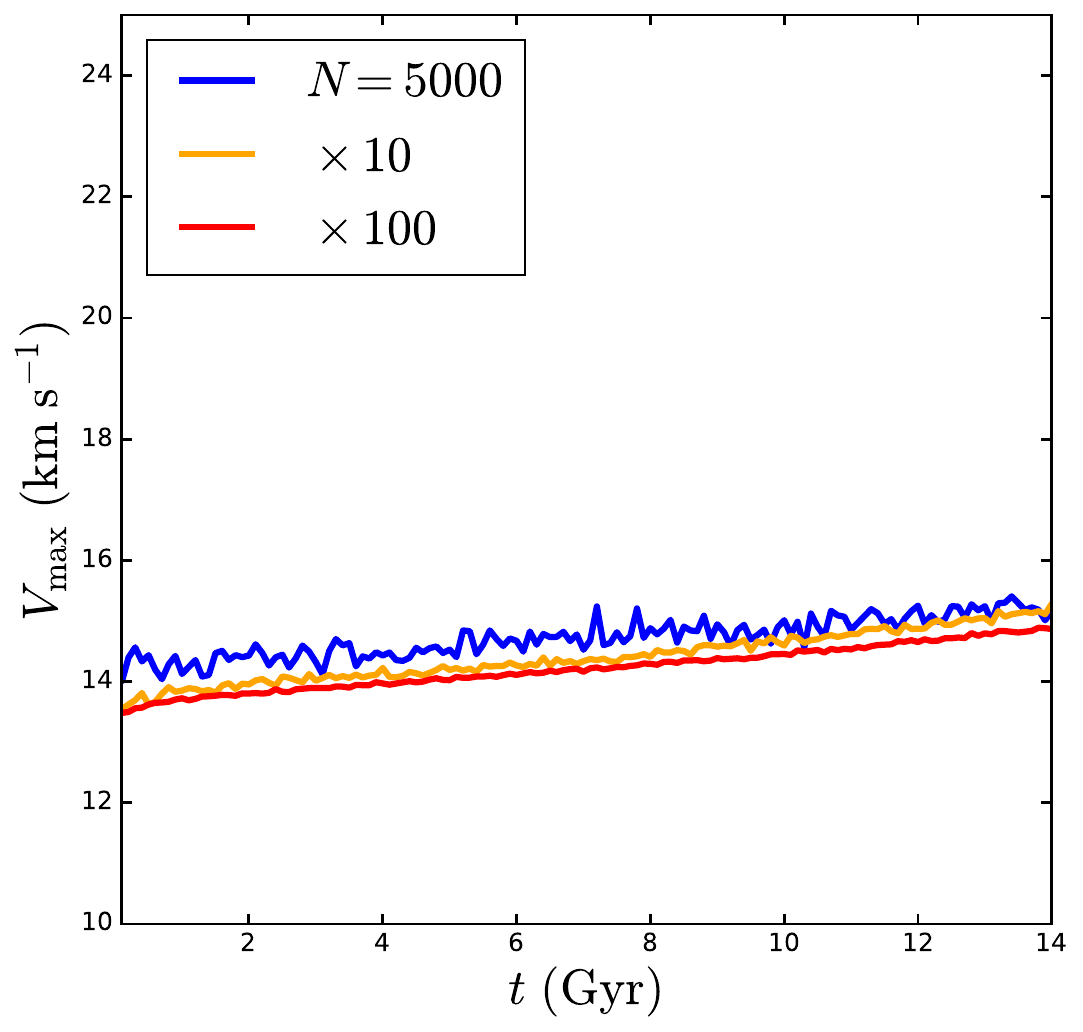}
  \includegraphics[height=4.6cm]{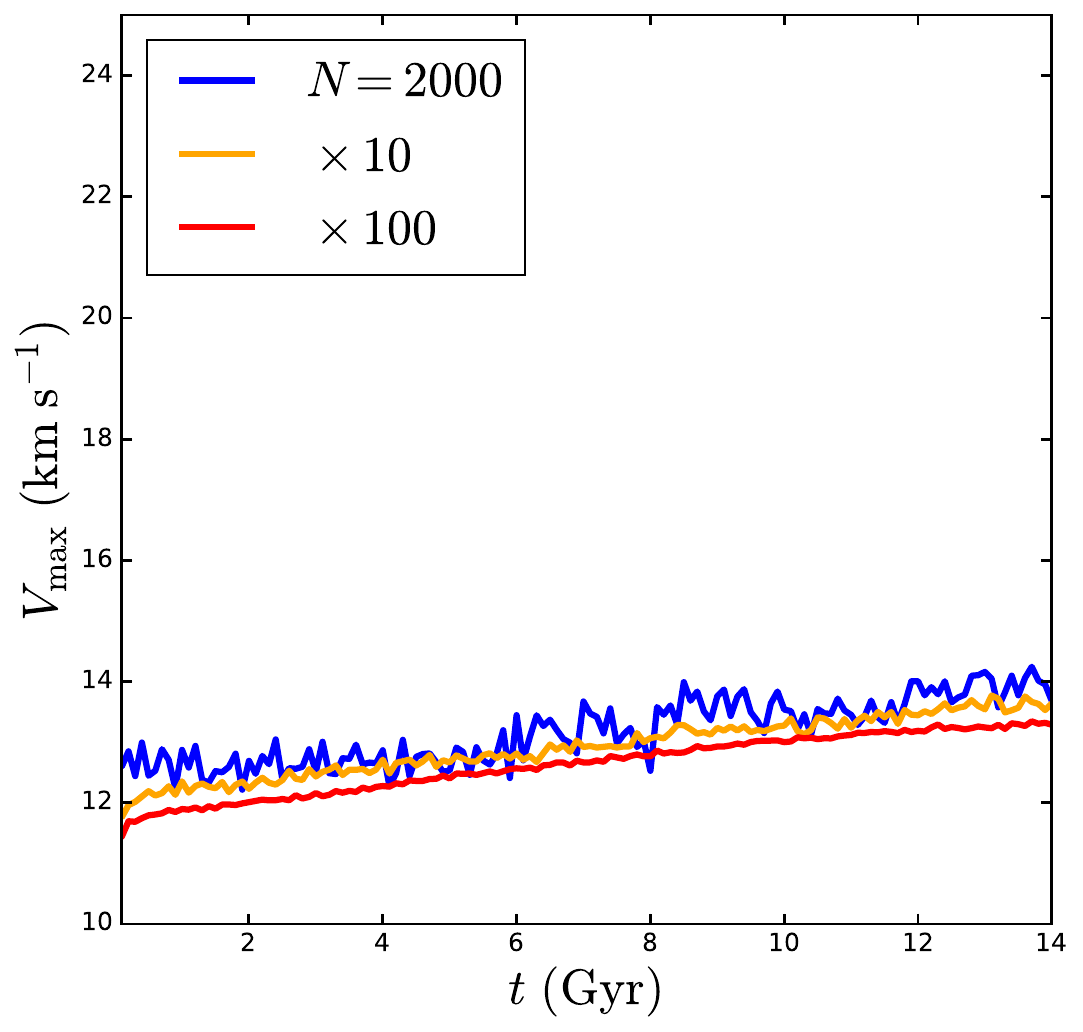}
  \caption{\label{fig:conv} Convergence test for three representative core-collapsed halos using idealized (noncosmological) simulations. 
From left to right, the simulated halos correspond to an isolated halo (Isolated 1), a splashback halo (Splashback), and a low-mass isolated halo (Isolated 2), with initial properties listed in Table~\ref{tab:conv}. From top to bottom, panels show evolution of the inner density, $V_{\mathrm{max}}$, and $R_{\mathrm{max}}$. Runs with particle masses close to the fiducial resolution in our cosmological simulation are shown in blue; orange and red lines show results for runs with $10\times$ and a $100\times$ the fiducial particle counts, respectively.
}
\end{figure*}

We perform two tests to evaluate the convergence properties of our halos in our SIDM simulation. 
As shown in the main text, the core-forming halos in our simulation are typically more massive than the core-collapsed halos; we therefore expect better convergence for core-forming halos, in general. 
In Figure~\ref{fig:conv0}, we show the distribution of $R_{\mathrm{max}}$ and $V_{\mathrm{max}}$ for isolated SIDM halos that are more massive than $8\times 10^8~M_{\odot}$, in the fiducial high-resolution simulation used in the paper, and in a lower-resolution simulation that uses an $8$ times larger particle mass and $2$ times larger softening, with an equal time-stepping criteria. 
This mass threshold selects the halos in the lower-resolution simulation containing at least $2000$ simulation particles; hence, the selected halos are expected to have converged properties, which is confirmed by Figure~\ref{fig:conv0}. The minor shifts we find in these distributions are not systematic, and may be caused by, for example, subtle differences in \textsc{ROCKSTAR}'s reconstruction of halo properties at varying resolution.

To test the convergence properties of core-collapsed halos, which have relatively low particle counts even in our high-resolution simulation, we take a different approach. 
In particular, we perform isolated, noncosmological simulations of the three representative core-collapsed halos from our cosmological simulation. 
We generate initial conditions using the \textsc{SpherIC} code~\cite{Garrison-Kimmel13013137} and assuming the NFW profile parameters similar to those reported by \textsc{ROCKSTAR} at $z=0$. 
Simulation parameters of the three halos are collected in Table~\ref{tab:conv}. 
In all cases, the force softening lengths are fixed to be $\ell=280~$pc.
For an extreme test, we take the third benchmark to be an isolated core-collapsed halo close to the mass threshold of $10^8~\msun~h^{-1}$ used throughout the analysis, corresponding to $2500$ high-resolution particles.  

We simulate the gravothermal evolution of these halos using the same simulation particle mass in our fiducial high-resolution simulation; to test for convergence, we also simulate the same systems with $10$ and $100 \times$ smaller particle masses and larger particle counts. 
The results are shown in Figure~\ref{fig:conv}. From top to bottom rows, the evolution of the central density $\rho_c$ (i.e., the average density within the force softening length), $R_{\rm max}$, and $V_{\rm max}$, are shown for the three different resolutions.
In all cases, we obtained reasonably converged results, with differences at the sub $20\%$ level at most times for $\rho_c$, comparable to the size of statistical fluctuations given the particle count. 
Convergence is particularly good in terms of $V_{\rm max}$, with differences at the $\approx 1\%$ level among our convergence runs, and $R_{\rm max}$ is converged at the $\approx 10\%$ level. We have also checked that the convergence is as good as shown in Figure 15 for the central density evaluated within $2\ell$ and $3\ell$.

\begin{table}
\begin{center}
\begin{tabular}{c|c|c|c|c}
\hline
\hline
Name & $M_{\rm vir}$ ($\msun$) & $R_{\rm vir}~\rm (kpc)$ & $V_{\rm max}~(\kms)$  & $R_{\rm max}~(\kpc)$ \\
\hline
Isolated 1 & $10^{9}$  & $26$  & $20$ & $2.0$ \\ 
Splashback       & $2.6\times 10^{8}$   & $17$ &  $14$  &  $1.1$ \\ 
Isolated 2& $10^8$    & $15$   & $12$ & $0.41$ \\ 
\hline
\hline
\end{tabular}
\caption{Properties of convergence test halos.  
\label{tab:conv} }
\end{center}
\end{table}

\section{Fitting the $R_{\rm max}\textup{--}V_{\rm max}$ relation}
\label{sec:fitting}

Here, we describe our method for fitting the $R_{\rm max}\textup{--}V_{\rm max}$ relation and the results. Assuming that $\log_{10}R_{\rm max}$ is at most a first-order polynomial in $\log_{10} V_{\rm max}$, we derive the median $R_{\rm max}\textup{--}V_{\rm max}$ relation by minimizing the median absolute error. 
To quantify the deviation from the median, we measure statistical fluctuations based on the residual distributions and evaluate the percentiles of upward and downward fluctuations at $68.27\%$ ($1\sigma$) and $95.45\%$ ($2\sigma$) coverage probabilities. 
This yields the following: 
\begin{eqnarray}
R_{\rm max} &=& 10^{+0.16}_{-0.14} 0.0610 V_{\rm max}^{1.12} ~~\text{  ($1\sigma$ band, isolated halos)},\nonumber \\ \nonumber
            &=& 10^{+0.38}_{-0.29} 0.0610 V_{\rm max}^{1.12} ~~\text{  ($2\sigma$ band)}, \\ \nonumber
R_{\rm max} &=& 10^{+0.17}_{-0.13} 0.0477 V_{\rm max}^{1.06} ~~\text{  ($1\sigma$ band, MW subhalos)}, \\ \nonumber
            &=& 10^{+0.48}_{-0.27} 0.0477 V_{\rm max}^{1.06} ~~\text{  ($2\sigma$ band)}. 
\end{eqnarray}

In Figures~\ref{fig:vmaxRvmax1} and \ref{fig:tcbarCDM}, we show the median relation and $\pm 0.6$~dex uncertainties to emphasize the differences between the CDM and SIDM results. 
For this purpose, we show the same curves in all $R_{\rm max}\textup{--}V_{\rm max}$ planes, which is based on the CDM simulation. For completeness, we also quantify the statistical deviations of the $\pm 0.6$~dex lines from the CDM median: the $+0.6$~dex line corresponds to a $0.6/0.16\approx 3.8\sigma$ ($0.6/0.17\approx 3.5\sigma$) upward fluctuation in the isolated halo (subhalo) case, while the $-0.6$~dex line corresponds to a $0.6/0.14\approx 4.3\sigma$ ($0.6/0.13\approx 4.6\sigma$) downward fluctuation in the isolated halo (subhalo) case. 

\section{Benchmark Isolated and Splashback Halo Histories}
\label{sec:additional_benchmark}

\begin{figure*}[htbp]
  \centering
  \includegraphics[height=4.6cm]{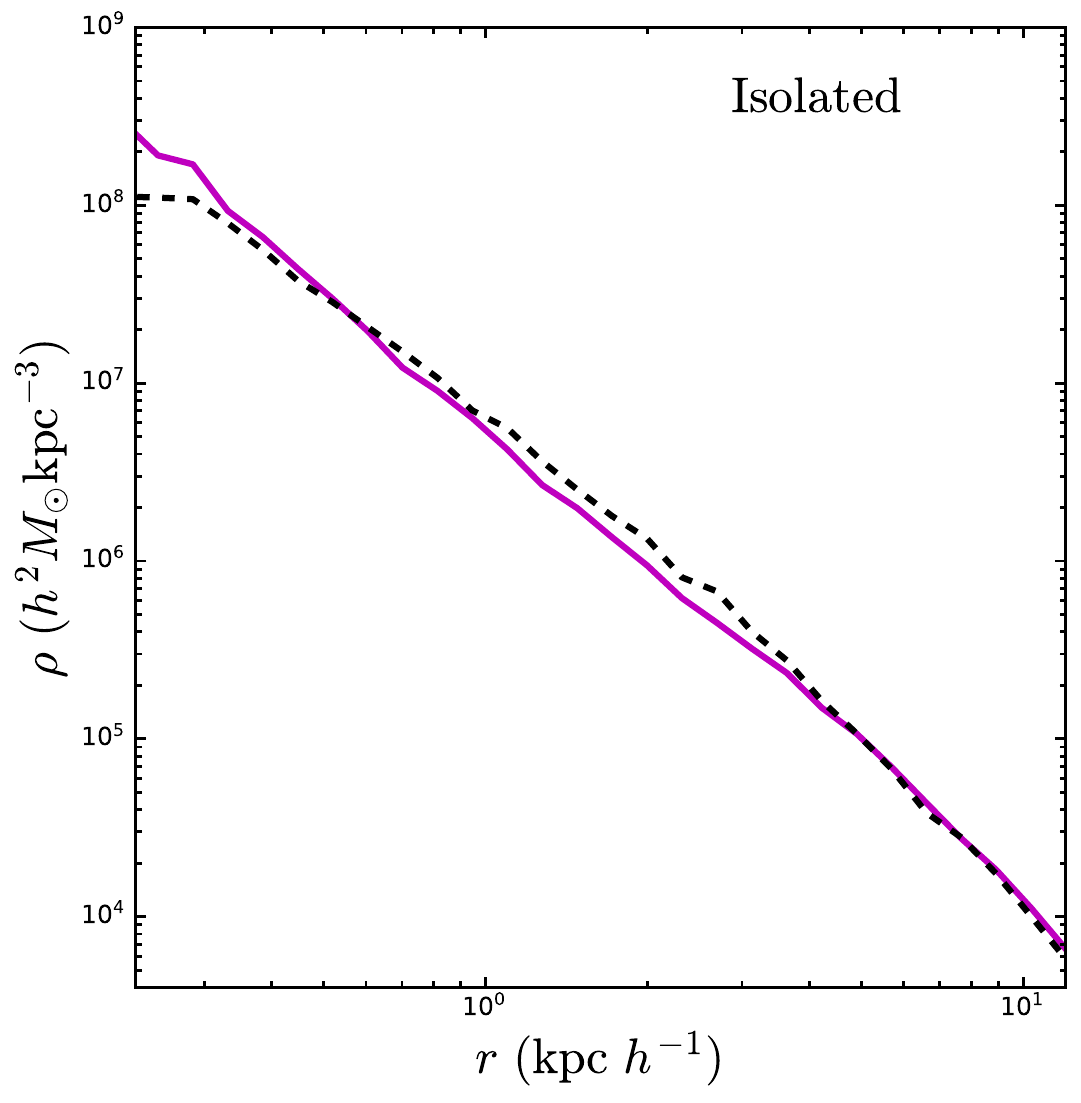}
  \includegraphics[height=4.6cm]{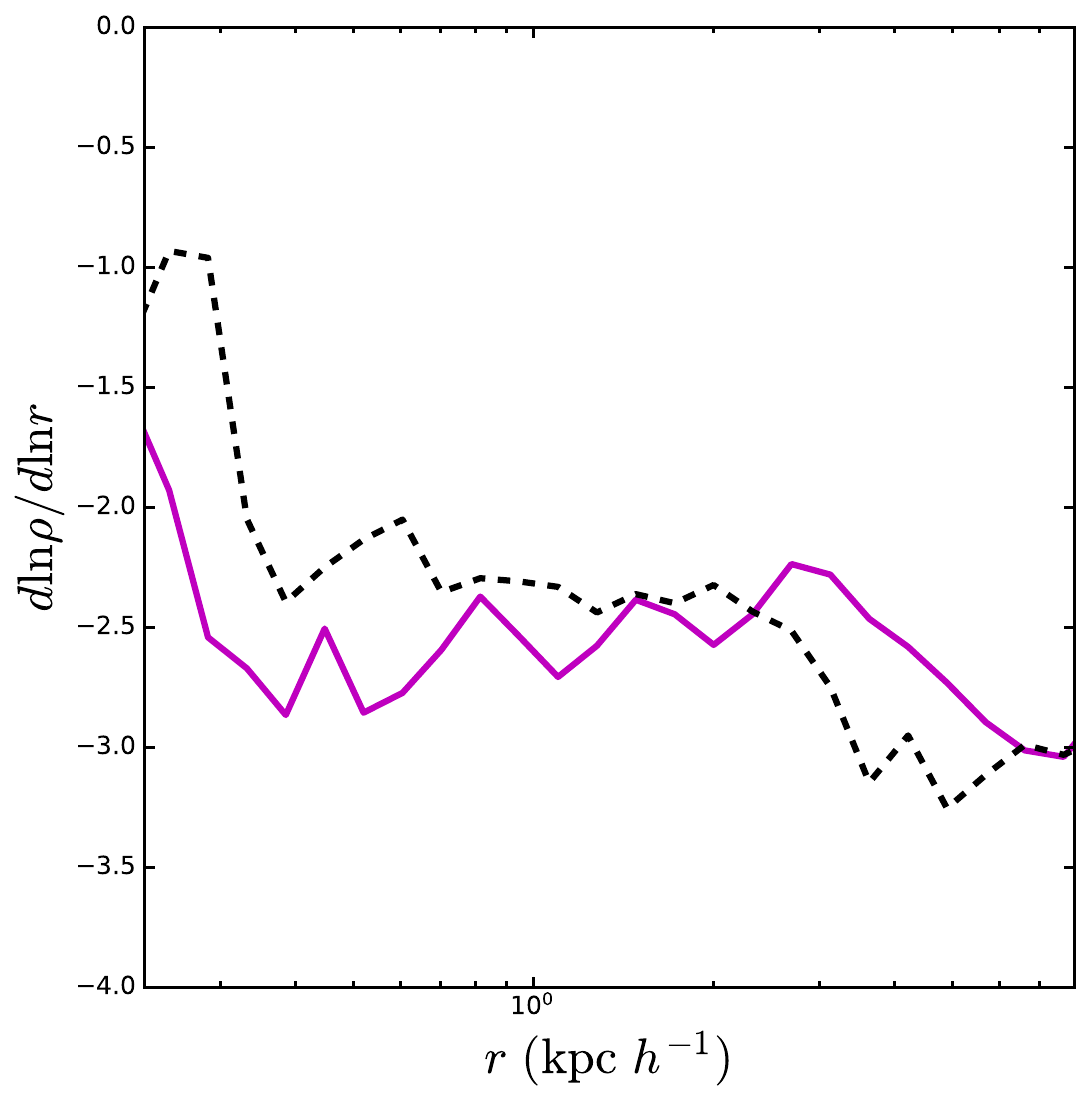}
  \includegraphics[height=4.6cm]{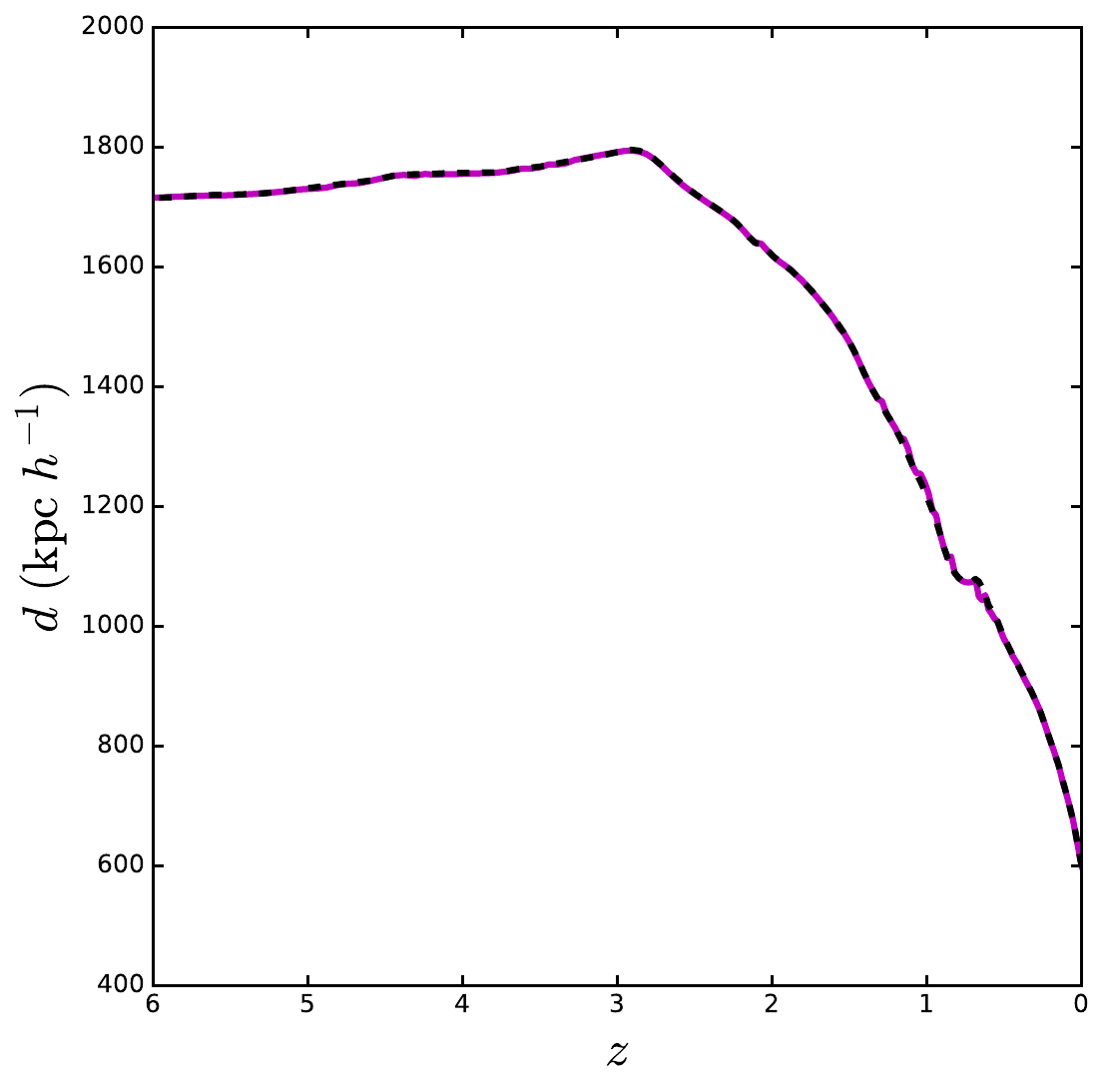} \\
  \includegraphics[height=4.6cm]{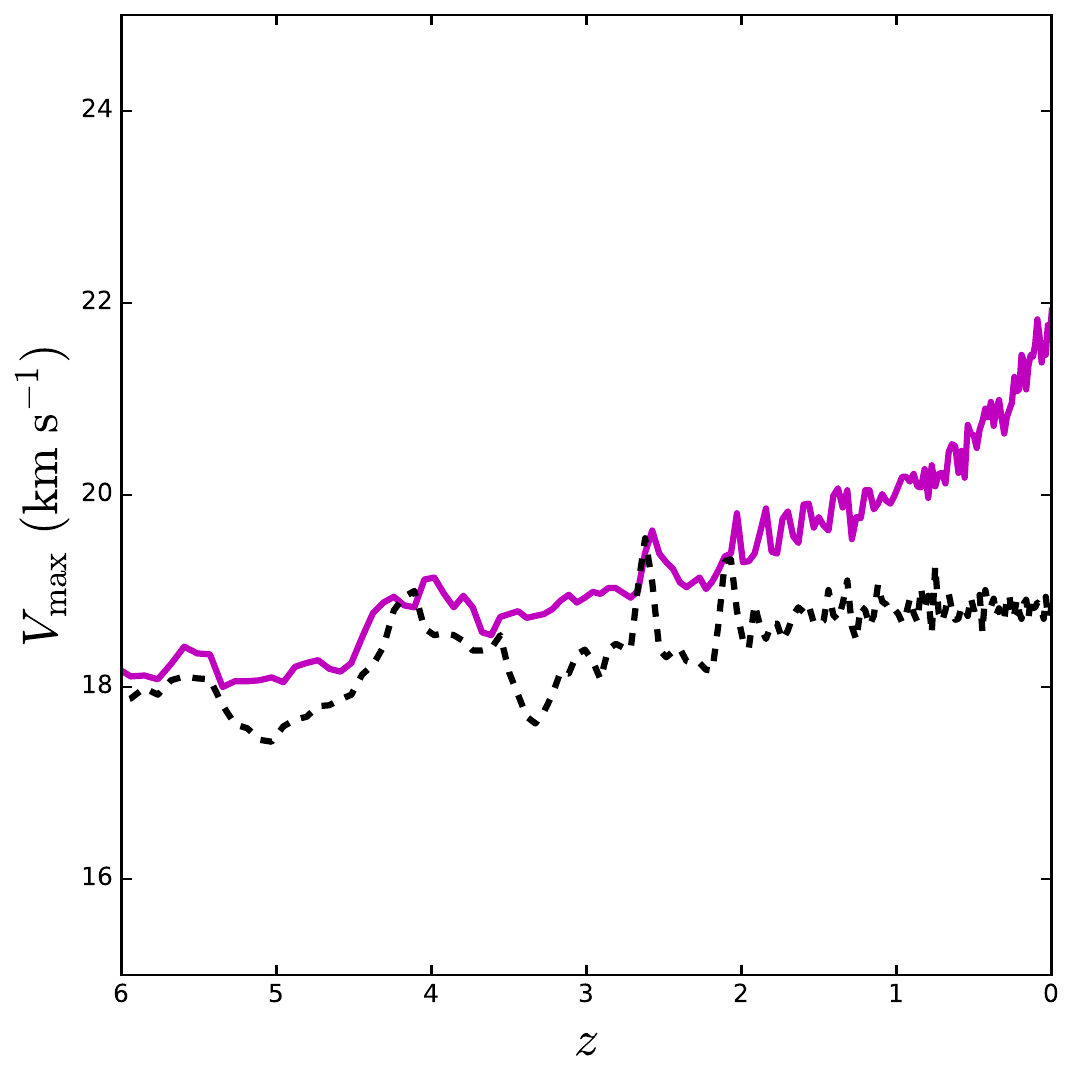}
  \includegraphics[height=4.6cm]{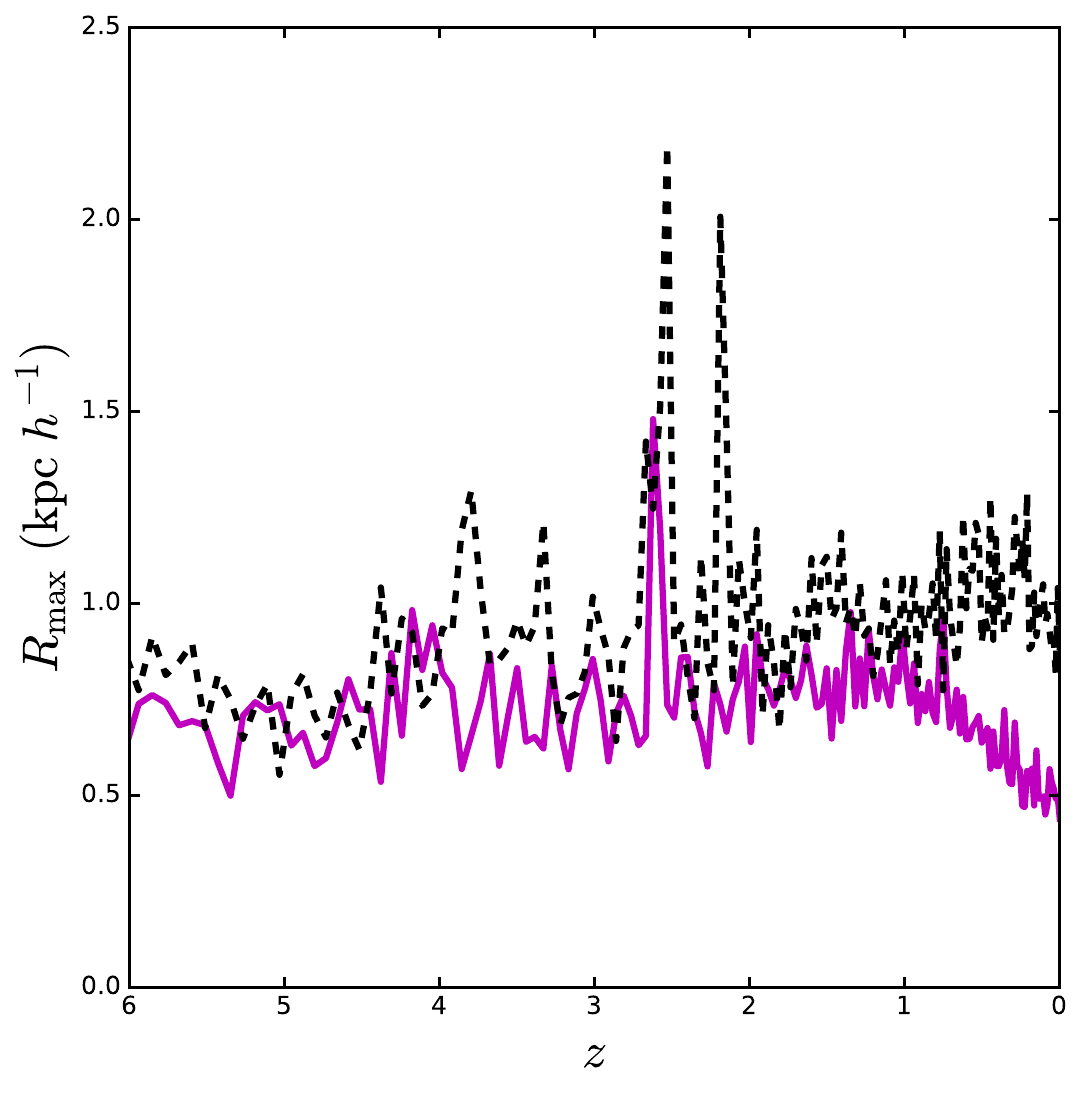}
  \includegraphics[height=4.6cm]{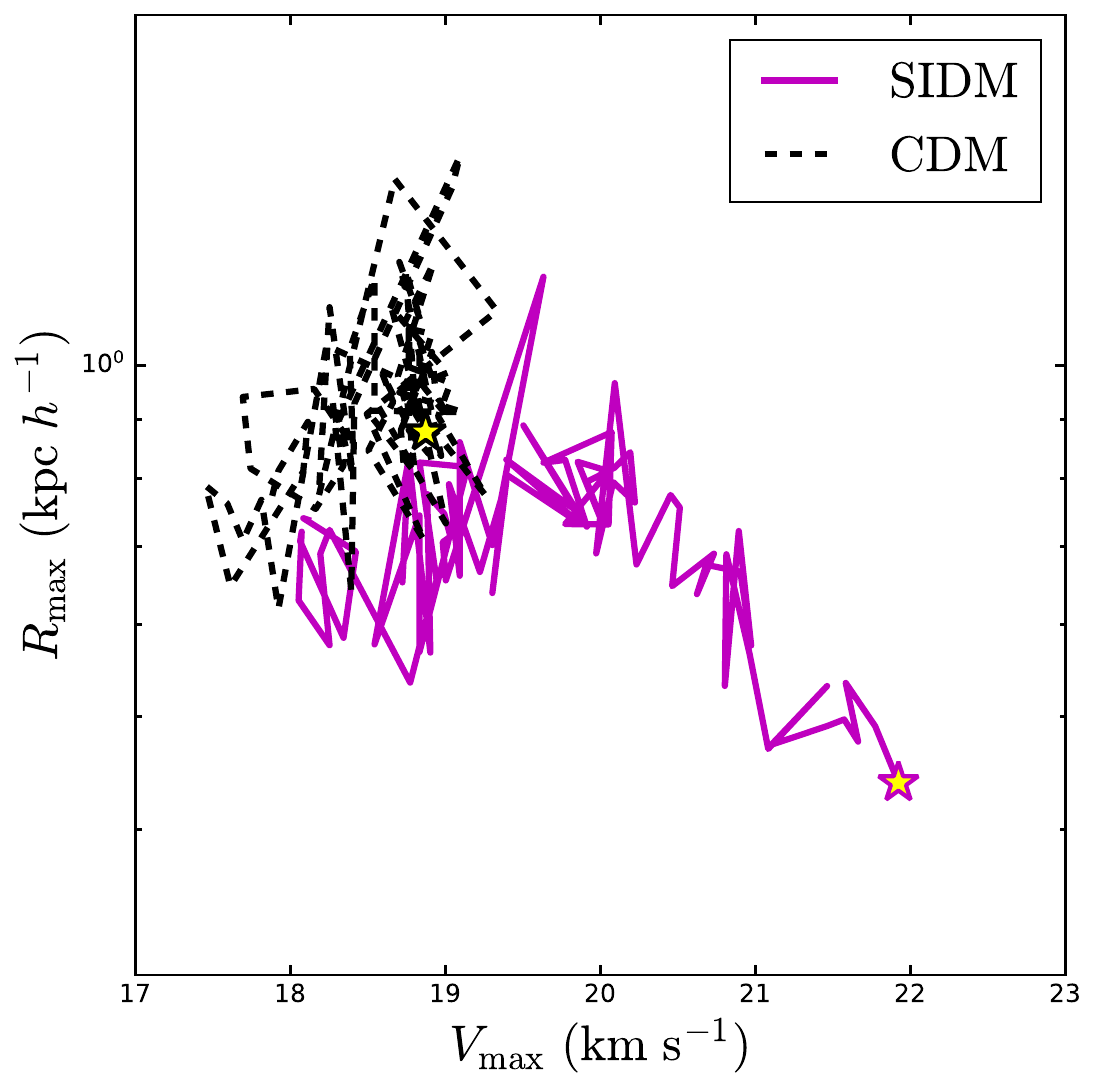}
  \caption{\label{fig:bmiso}
Benchmark isolated core-collapsed halo evolution. The top row shows the density profile, density profile slope, and distance from the MW center as a function of redshift, from left to right.
The bottom row shows evolution of $V_{\rm max}$, $R_{\rm max}$, and the trajectory in the $R_{\rm max}$--$V_{\rm max}$ plane as functions of redshift, from left to right.}
\end{figure*}

\begin{figure*}[htbp]
  \centering
  \includegraphics[height=4.6cm]{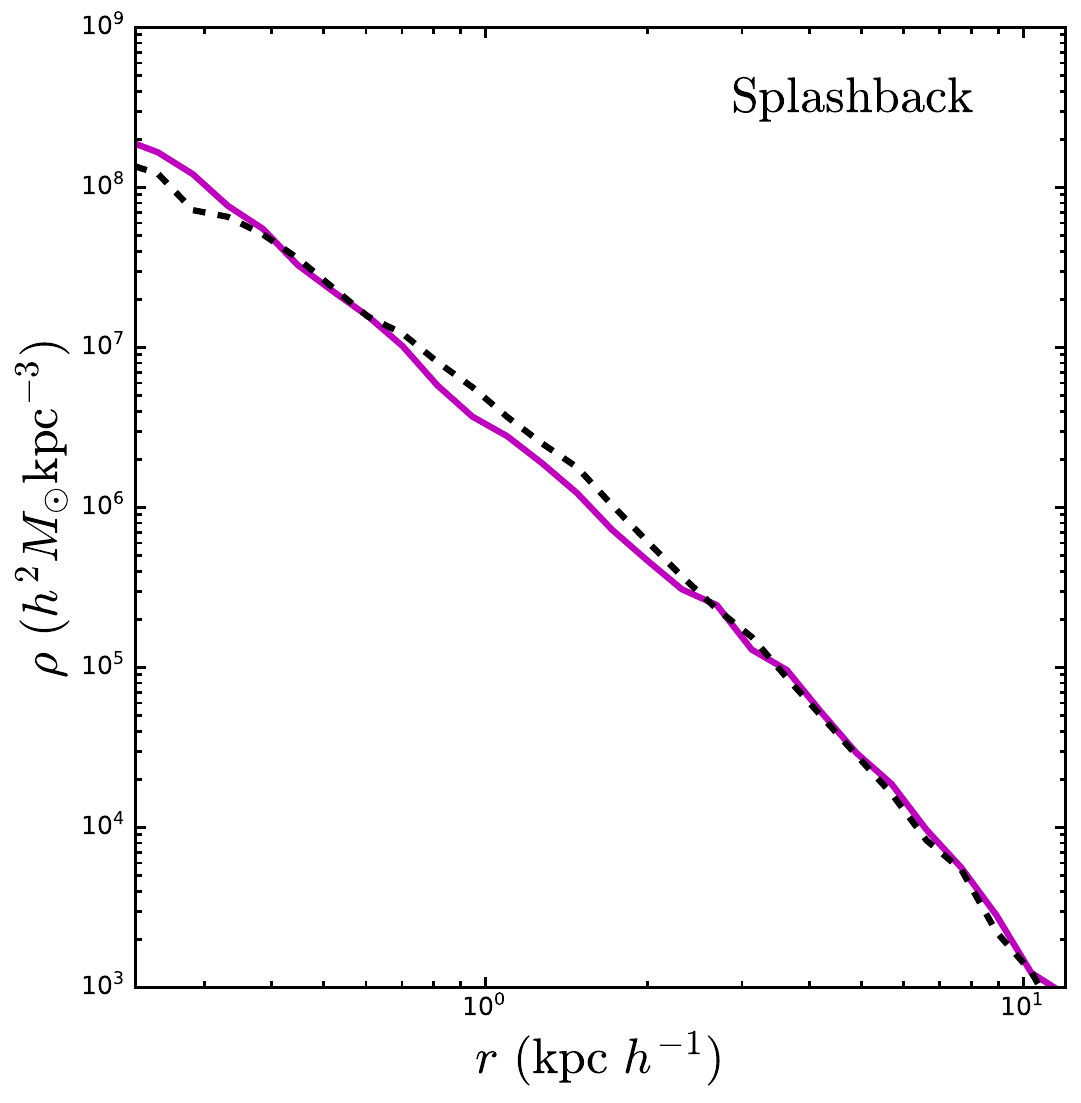}
  \includegraphics[height=4.6cm]{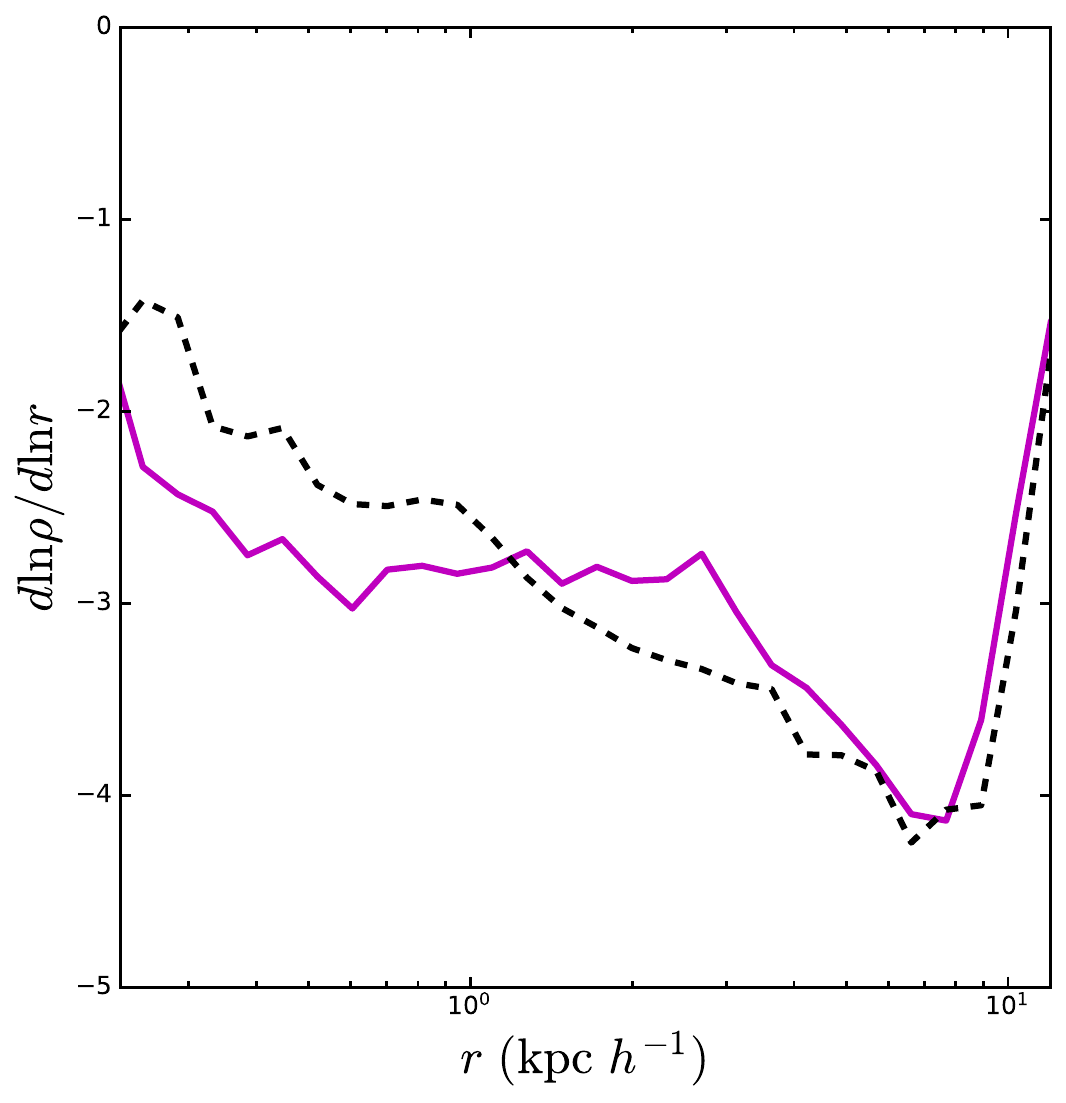}
  \includegraphics[height=4.6cm]{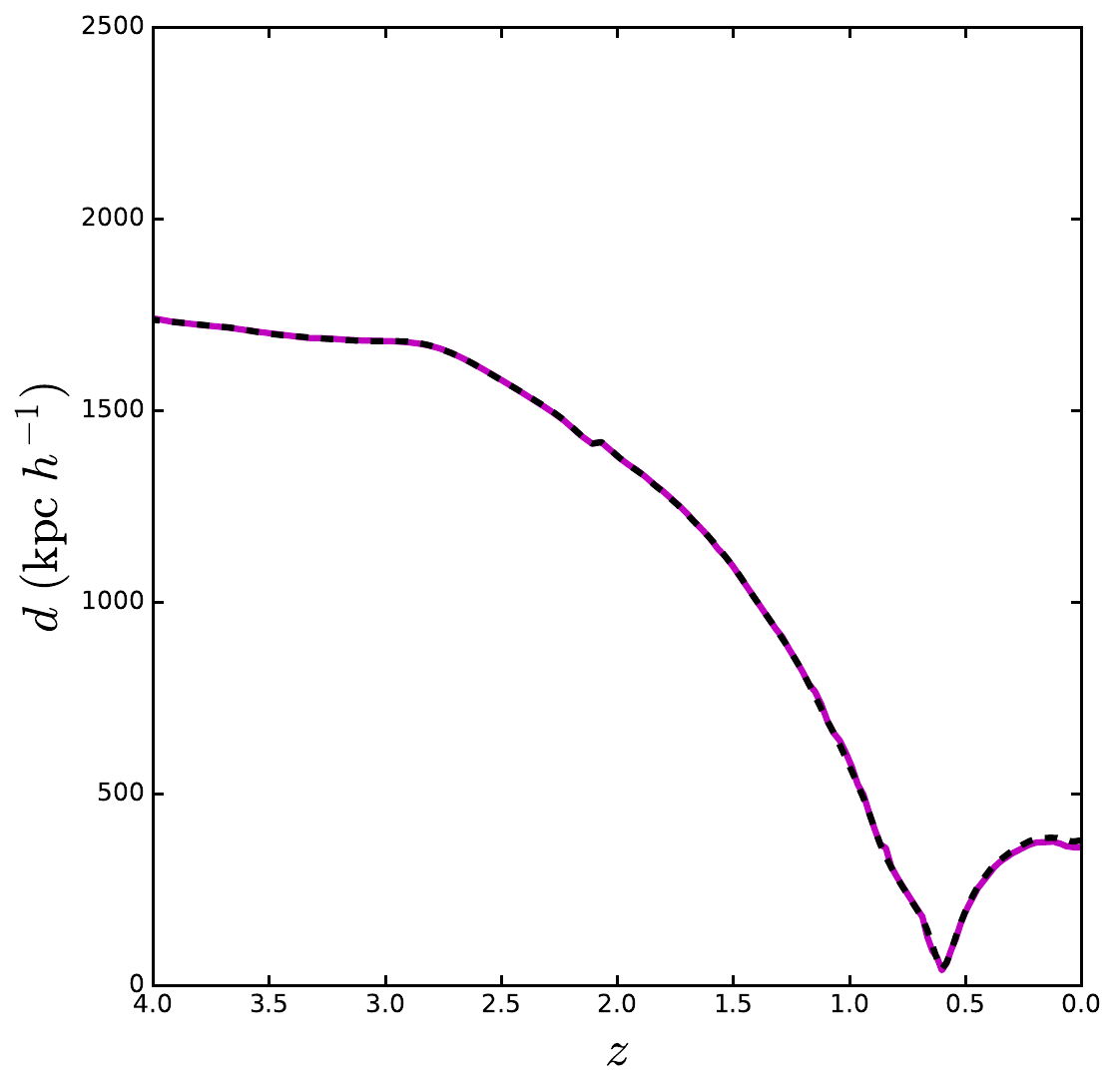}
  \includegraphics[height=4.6cm]{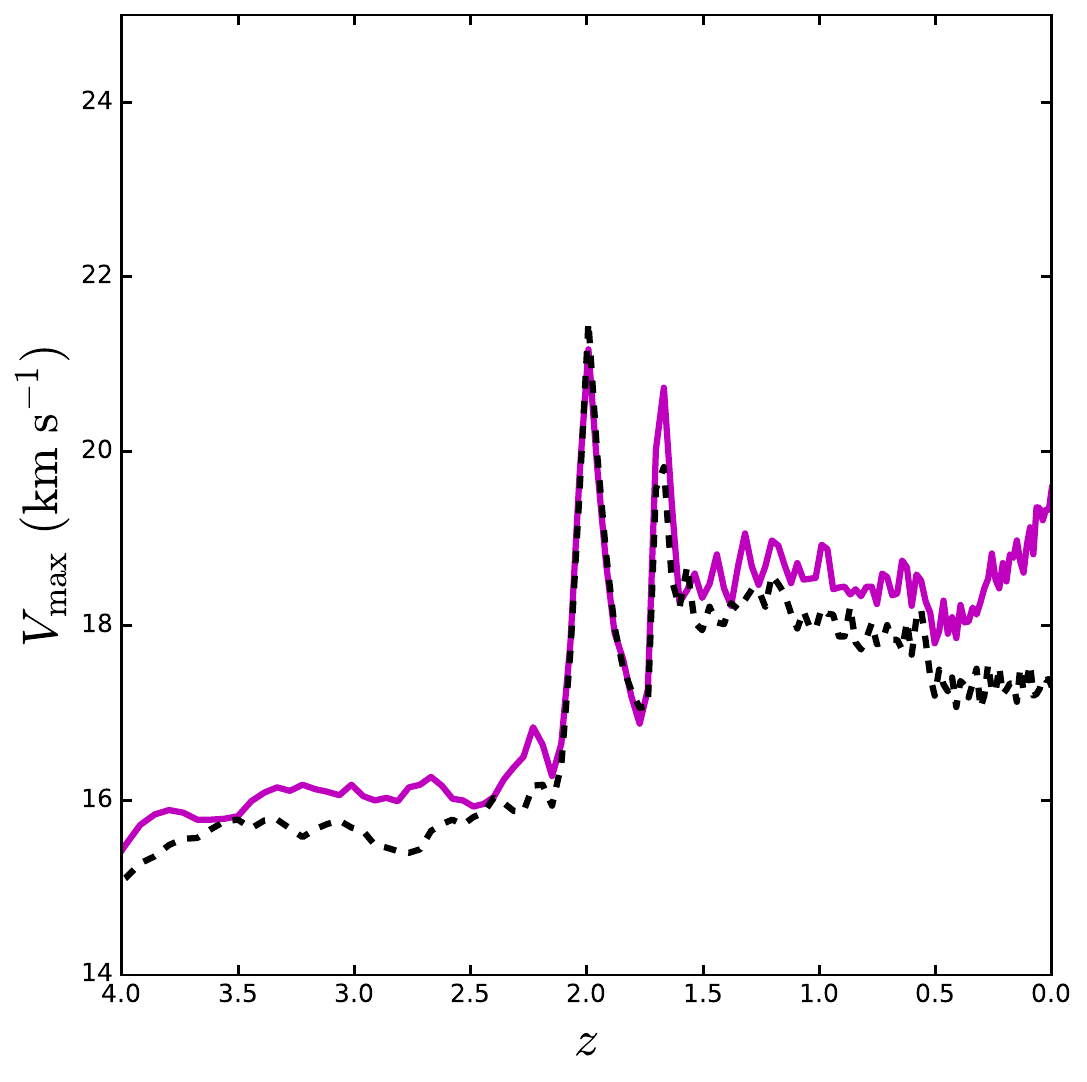}
  \includegraphics[height=4.6cm]{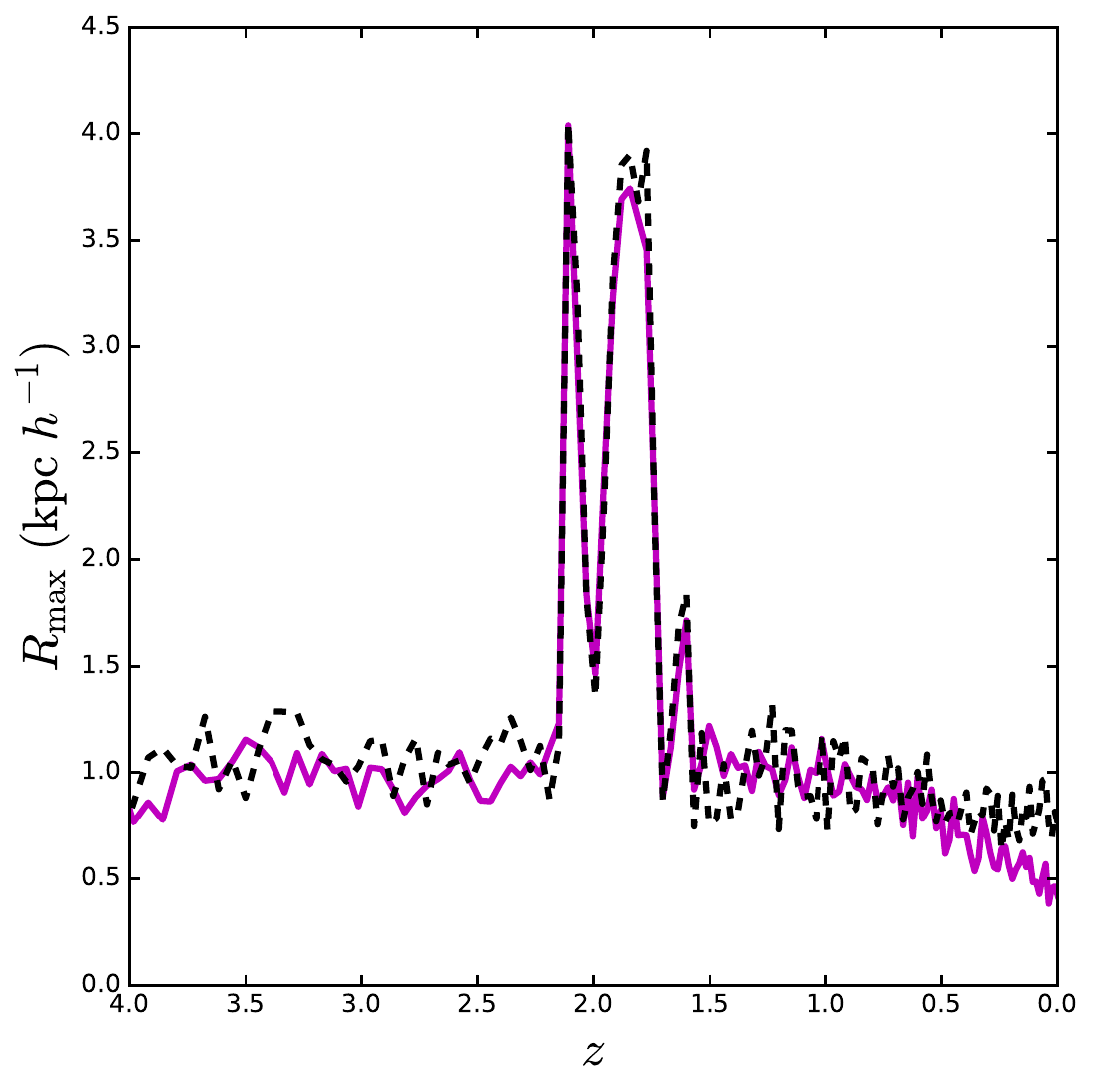}
  \includegraphics[height=4.6cm]{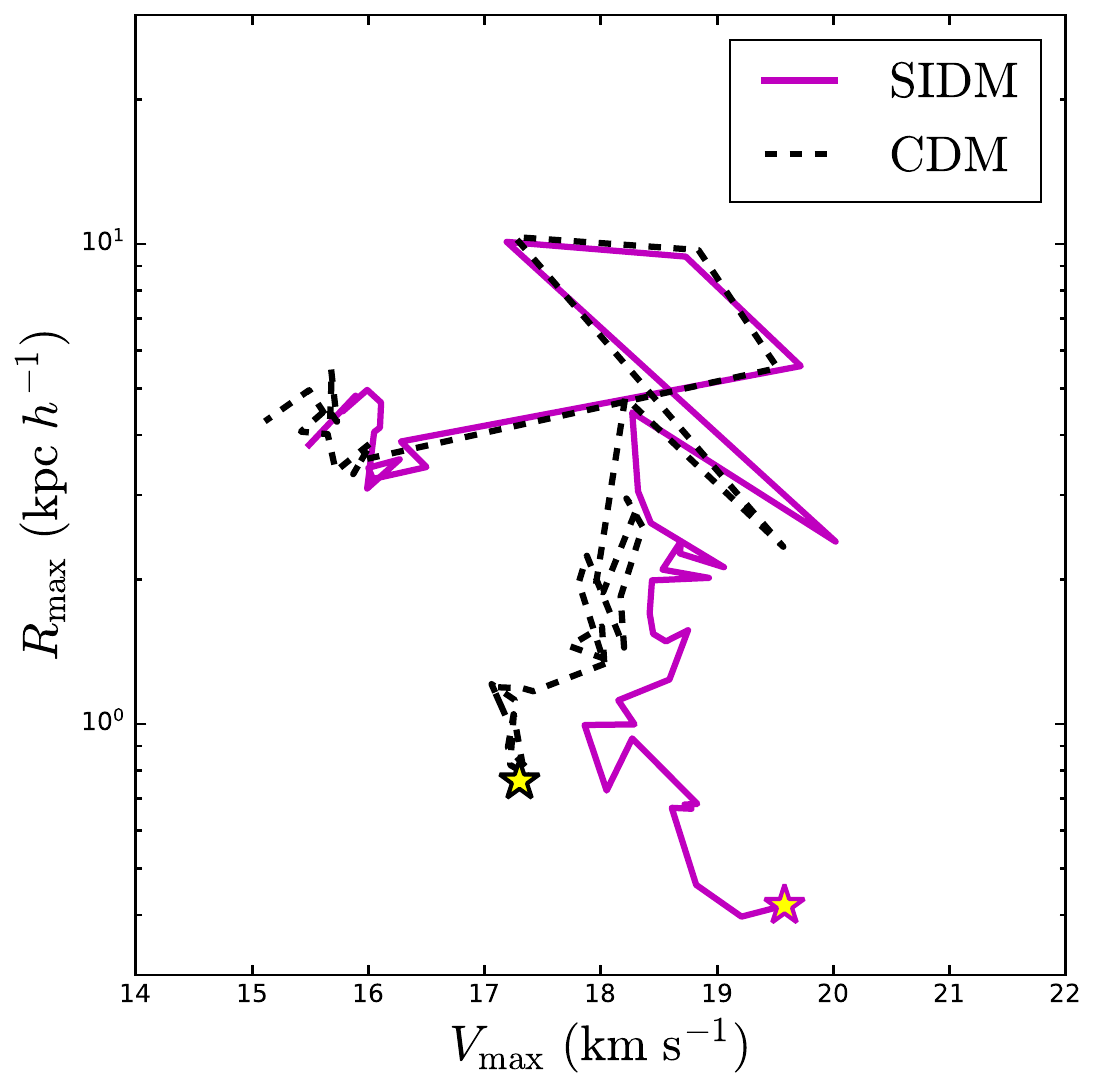}
  \caption{\label{fig:bmspb}
Same as Figure~\ref{fig:bmiso}, for our benchmark splashback core-collapsed halo.
}
\end{figure*}

We present the evolution histories of the isolated and splashback benchmark core-collapsed halos from Table~\ref{tab:bm} in Figures~\ref{fig:bmiso} and \ref{fig:bmspb}, respectively. Comparing the isolated benchmark with the MW subhalo benchmark from Figure~\ref{fig:bmmw}, we see that the isolated one has a large distance to the MW center at all times but does not go beyond $3~$Mpc due to our distance cut. 
This halo has a high concentration of $c_{\mathrm{eff}}\approx 37$ in the CDM case, which is $3\sigma$ higher than the median concentration mass relation from ~\cite{2014MNRAS.441.3359D} at $z=0$.  
Limited by the resolution, the density profile inside the NFW scale radius is not very well resolved.
Aside from that, its density profile is consistent with an NFW profile, as can be seen from the density slope plot in Figure~\ref{fig:bmiso}, where the slope increases from $-3$ to $-1$. 

The splashback halo is characterized by a single pericenter passage that is close to the MW center, with $R_{\rm vir} =17~\kpc$. 
From the evolution of its $V_{\rm max}$ and $R_{\rm max}$, we see that its SIDM evolution differs from its CDM evolution significantly only at late times, after its pericentric passage. 
Thus, this halo shows a rather extreme case of accelerated gravothermal evolution due to tidal effects. 

\end{document}